\documentclass[11pt]{article}
\usepackage{graphicx} %

\usepackage{mathpazo}
\usepackage{pdfpages}
\usepackage{bm}
\usepackage[margin=1in]{geometry}
\usepackage{amsmath,amssymb}

\usepackage[utf8]{inputenc}
\usepackage{color,amsmath,amsfonts,fullpage,amsthm,mathrsfs,fancyhdr,verbatim,framed,hyperref,xspace,mdframed,tabularx,booktabs,enumitem,comment,float}
\usepackage{cleveref}
\usepackage[dvipsnames]{xcolor}
\usepackage{graphicx}
\usepackage{amsthm}
\usepackage{mathpazo}
\usepackage{amsmath}
\usepackage{amssymb}
\usepackage{mathtools}

\usepackage{thmtools,thm-restate}
\usepackage{fec} 
\usepackage{soul}

\usepackage{physics}
\usepackage{makecell}
\usepackage{array}
\usepackage{subcaption}
\usepackage{authblk}

\usepackage{amssymb}

\newcommand{\diff}{\mathrm{d}}
\newcommand{\tildetau}{\widetilde{\tau}}

\usepackage[style=alphabetic,backref=true]{biblatex}
\bibliography{biblio}

\title{From Quantum Tunneling to Classical Optimization through Instantons}

\author[1]{Jiayu Shen\thanks{\texttt{jiayu.shen@jpmchase.com}}}
\author[1]{Dylan Herman}
\author[1]{Guneykan Ozgul}
\author[1]{Akshay Seshadri}
\author[1]{Junhyung Lyle Kim}
\author[1]{Shree Hari Sureshbabu}
\author[1]{Shouvanik Chakrabarti\thanks{\texttt{shouvanik.chakrabarti@jpmchase.com}}}
\author[1]{Niraj Kumar\thanks{\texttt{niraj.x7.kumar@jpmchase.com}}}

\affil[1]{Global Technology Applied Research, JPMorganChase, New York, NY 10001, USA}

\date{}

\begin{document}

\maketitle

\begin{abstract}

    We use an \emph{instanton}-based analysis to show that \emph{Path Integral Stochastic Gradient Descent} (PISGD), a classical optimization algorithm instantiated from path integral Monte Carlo (PIMC), can inherit the quantum tunneling mechanism. Quantum tunneling is the phenomenon in which a quantum particle passes through a potential barrier that a classical trajectory cannot cross in the same way.
    This mechanism can contribute to the performance of quantum optimization algorithms, since a common bottleneck in optimization is that the solution becomes trapped in a local minimum, separated from better solutions by a potential barrier. Classical noise-driven optimizers, such as stochastic gradient descent (SGD), have escape times that scale exponentially with the potential barrier height. For quantum tunneling, the escape time is instead exponential in the bounce action, which equals twice the \emph{Agmon distance} between the two basins. When the barrier is tall but thin, quantum tunneling can allow escaping local minima faster than such classical methods. Remarkably, this scaling is not exclusive to algorithms with genuine quantum dynamics. For PISGD applied to classical nonconvex global optimization, we adapt Kramers' escape-rate theory to path space and obtain the same exponential scaling as quantum tunneling, in terms of the bounce action rather than the barrier height. We confirm this scaling numerically by implementing PISGD for both separable and nonseparable potentials. That said, PISGD retains only incoherent tunneling, since coherence among multiple tunneling paths is absent. Our analysis therefore does not rule out a potential quantum advantage for quantum-tunneling-based algorithms over classical ones in general.
\end{abstract}

\tableofcontents
\newpage

\section{Introduction}
\subsection{Motivation}
Quantum tunneling is one of the defining features of quantum mechanics, with no evident counterpart in ordinary classical mechanics. Informally, a quantum particle can pass through a barrier that classical mechanics forbids it from crossing, or at least reach the other side by a route unavailable to a classical trajectory. The phenomenon is of broad physical interest because it is universal: it arises whenever the system has structure such as a potential barrier, from the classic one-dimensional example of tunneling through a potential wall to the quantum field theories that describe fundamental high-energy physics~\cite{Coleman:1977py,callan1977fate,vauinshteuin1982abc}. The apparent
absence of tunneling in classical mechanics suggests that quantum algorithms exploiting it may hold an advantage over classical ones for solving certain tasks~\cite{isakov2016understanding,7782986}.

Finding the global minimum of a nonconvex function is one of the main applications of the quantum algorithms that exploit the tunneling phenomenon, since the typical challenge is that the function landscape can contain many local minima separated by barriers. This problem is known as \emph{nonconvex optimization}, and it is described as follows: given a differentiable objective function $V\colon\mathbb{R}^d\to\mathbb{R}$, find a point $\tilde{x}\in\mathbb{R}^d$ 
\begin{equation}
    \norm{\tilde{x} - x^\star} \le \varepsilon
\end{equation}
where $x^\star = \argmin_{x\in\mathbb{R}^d} V(x)$ is the \emph{global minimum} and $\varepsilon > 0$ is a prescribed tolerance. 

\begin{figure}[!htbp]
    \centering
    \begin{subfigure}[t]{0.47\linewidth}
        \centering
        \includegraphics[width=\linewidth]{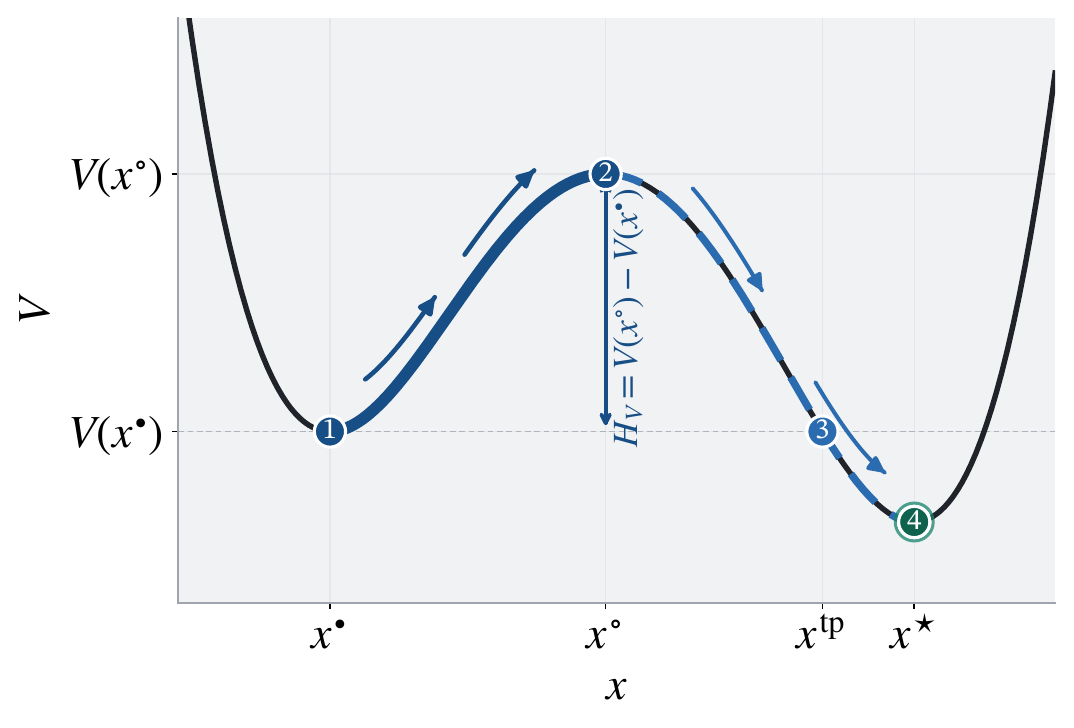}
        \caption{SGD: Escape over the potential barrier.}
        \label{fig:sgd_potential_barrier}
    \end{subfigure}
    \hfill
    \begin{subfigure}[t]{0.47\linewidth}
        \centering
        \includegraphics[width=\linewidth]{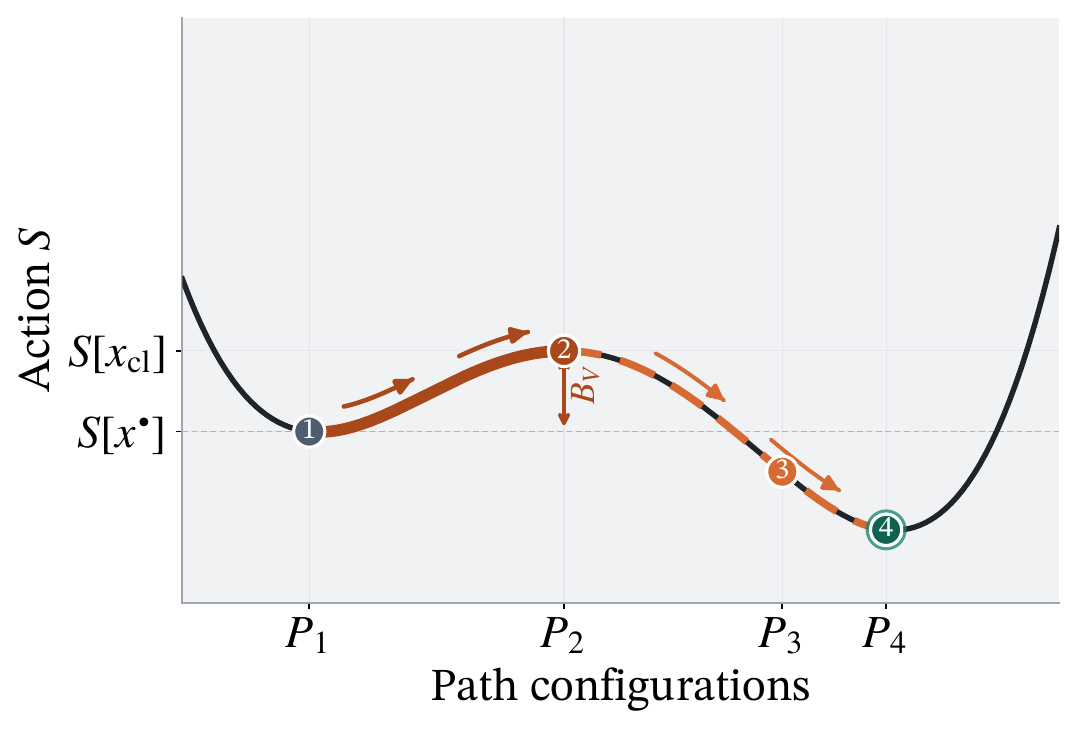}
        \caption{PISGD: Escape over the action barrier.}
        \label{fig:pisgd_action_barrier}
    \end{subfigure}

    \vspace{0.75em}

    \begin{subfigure}[t]{0.98\linewidth}
        \centering
        \includegraphics[width=\linewidth]{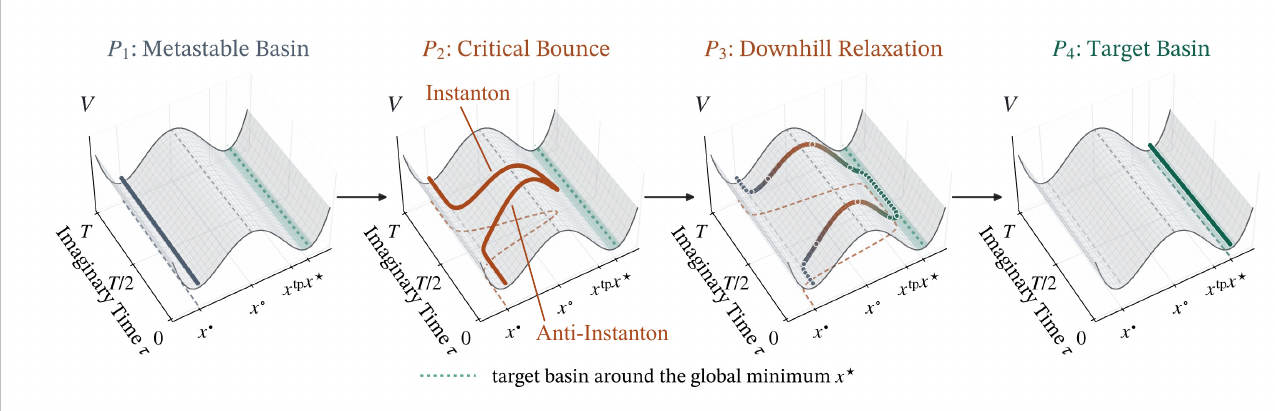}
        \caption{Path evolution during an escape in PISGD.}
        \label{fig:pisgd_bounce_sequence}
    \end{subfigure}
    \caption{Barrier-escape mechanisms of SGD and PISGD for a one-dimensional potential $V$. In (a), SGD evolves a single point $x$ and must climb a potential barrier of height $H_V$. In (b), PISGD instead evolves an entire path. Along the schematic slice of the path-integral action shown here, the relevant barrier is the bounce action $B_V$. The four path configurations marked on this action profile are shown explicitly in (c). Here, $P_1$ is the constant path at the metastable local minimum $x^\bullet$, while $P_2$ is the critical bounce, the labeled instanton--anti-instanton pair, at the top of the action barrier. After this barrier is crossed, $P_3$ undergoes downhill relaxation, and $P_4$ is the constant path at the target global minimum $x^\star$.}
    \label{fig:sgd_pisgd_escape_mechanisms}
\end{figure}

When the objective function $V$ is nonconvex, a local minimum $x^\bullet$ and the global minimum $x^\star$ is typically separated by a bottleneck saddle point $x^\circ$ (see Figure~\ref{fig:sgd_potential_barrier} for an illustration). When the barrier is high,
a local algorithm can be trapped in the basin containing $x^\bullet$ and spend very long time before finding a new local minimum. This challenge provides a very natural motivation for some sort tunneling mechanism to pass through such barriers for faster optimization algorithms.

However, the need for a quantum computer to simulate such a tunneling mechanism is less clear-cut than it seems. In the context of studying quantum physics, well before the appearance of quantum computers and quantum algorithms, there already existed classical algorithms to simulate quantum systems of physical interest and to compute their physical quantities. An important class of these is the quantum Monte Carlo algorithms~\footnote{We refer here to classical algorithms for quantum systems, rather than to quantum algorithms for sampling on quantum computers, which are in some contexts also called Quantum Monte Carlo.} or Path Integral Monte Carlo (PIMC), which use Monte Carlo sampling from the path space to extract useful information about quantum systems. This algorithm is based on the path integral formalism of quantum mechanics, a fundamental construction of quantum mechanics and quantum field theory: one assigns a probability interpretation to the integrand of the path integral, so that a classical algorithm can sample from the resulting distribution to estimate the certain properties of the quantum systems. It is widely used, with applications ranging from condensed matter systems~\cite{suzuki1976relationship,ceperley1995path} to quantum chromodynamics~\cite{wilson1974confinement,creutz1980monte}, where it is used to study the strong force in particle physics. As a classical algorithm that in a sense simulates quantum systems, path integral Monte Carlo captures some key features of quantum mechanics.

Because of this, path integral Monte Carlo may inherit quantum effects, possibly including tunneling itself~\cite{isakov2016understanding,jiang2017scaling}, the very mechanism behind the expected advantage of quantum optimization algorithms. This possibility has been exploited for discrete-variable optimization with path integral Monte Carlo and its variants~\cite{crosson2021rapid,jiang2017scaling}. In~\cite{jiang2017scaling}, for example, a mean-field description, approximating the discrete system by a continuous one, is applied to the Hamiltonian of the quantum Monte Carlo algorithm for a discrete optimization problem, yielding results in the Wentzel--Kramers--Brillouin (WKB) form traditionally used to study tunneling rates in the semiclassical limit. For continuous variables, path integral Monte Carlo is even more natural to analyze: the time to escape from a local minimum to the global minimum is again associated with the bounce action $B_V$ and is strongly connected to quantum tunneling~\cite{shen2023vacuum}, and the analysis via WKB and instantons is more direct than in the discrete case, requiring no mean-field approximation.

This raises the algorithmic question that motivates this paper: can path integral Monte Carlo be turned into a practical optimizer for continuous global optimization that, like standard first-order methods, uses only gradients of the objective, yet it also exploits the tunneling behavior existing in quantum algorithms to get accelerated rates?
We answer this affirmatively by instantiating path integral Monte Carlo as \emph{path integral stochastic gradient descent} (PISGD), a gradient-based optimizer acting on paths rather than on points. 

To understand how PISGD behaves differently than the standard classical algorithms, we first need to highlight some behavior of classical and quantum dynamics from an algorithmic perspective. The most natural classical algorithm is Stochastic Gradient Descent (SGD), a widely used optimization algorithm in machine learning and various other fields. For a fixed objective function the noise is introduced artificially in order to help escape local minima and the mechanism that allows to overcome the barriers is purely statistical. That is, the algorithm needs to inject a noise at one point to climb the barrier instead of converging to the local minimum. Its runtime can analyzed through the Langevin stochastic differential equation~\cite{herau2011tunnel,shi2023learning}. The dominant exponential scaling of the runtime is given by $\exp(\widetilde{\lambda}_{\mathrm{SGD}} H_V)$, where $H_V = V(x^\circ) - V(x^\bullet)$ is the barrier height and $\widetilde{\lambda}_{\mathrm{SGD}} > 0$ is a scaling constant treated as a hyperparameter. In some contexts~\cite{shi2023learning}, it can be seen as a inverse learning rate.

On the quantum side, the tunneling time instead scales as $\exp(\widetilde{\lambda}_{\mathrm{QT}} B_V)$, and the two exponents measure different things. In SGD, the noise must carry the iterate over the top of the barrier, so $H_V$ measures only how \emph{tall} the barrier is. On the other hand, $B_V$ averages the barrier height along the path that connects the local minima, as the definition of the bounce action shows:
\begin{equation*}
    B_V = 2 \inf_{\gamma} \int_{\gamma} \abs{\diff x} \, \sqrt{2 \max \{ V(x) - V(x^\bullet), \, 0 \}},
\end{equation*}
where the infimum runs over paths $\gamma$ from the local minimum to the far side of the barrier. Hence $B_V$ depends on both the height and the width of the barrier. For a barrier that is high but thin, $B_V$ is small while $H_V$ is large, and tunneling is the cheaper route by an exponential factor. This is the regime in which quantum optimizers are expected to outperform, consistent with the runtime analysis of the quantum tunneling walk algorithm in~\cite{liu2023quantum}.

This mechanism, however, is not only accessible to quantum algorithms or quantum systems. Since solving the Schr\"{o}dinger equation for real-time quantum dynamics is very challenging, the tunneling rate in the semiclassical limit is already accessible through \emph{instantons}, classical trajectories constructed from the theory that serve as the starting point for semiclassical analysis. Instantons are the standard route to tunneling rates across a wide range of quantum systems~\cite{Coleman:1977py,callan1977fate,vauinshteuin1982abc,andreassen2017precision}. In PISGD, each iterate is a discretized path rather than a single point, and SGD is applied to the corresponding path-integral action, so that tunneling can be exploited by a gradient-based method. The path can cross a potential barrier through instanton-like configurations, providing a classical analogue of the tunneling mechanism. Motivated by this, we give a systematic study of PISGD for continuous global optimization from an instanton perspective and obtain concrete rates for solving optimization problems.

Figure~\ref{fig:sgd_pisgd_escape_mechanisms} contrasts the SGD and PISGD escape mechanisms for a one-dimensional potential and illustrates the path evolution during a PISGD escape. SGD must cross the potential barrier of height $H_V$ in the original $x$-space, as shown in Figure~\ref{fig:sgd_potential_barrier}, whereas PISGD crosses the action barrier $B_V$ in path space, as shown in Figure~\ref{fig:pisgd_action_barrier}, through the path configurations in Figure~\ref{fig:pisgd_bounce_sequence}. This replacement of $H_V$ by $B_V$ is the central difference between the exponential scalings in~\eqref{eqn:main_results_single_barrier}.

\subsection{Overview of Results}
\paragraph{Scaling for the single barrier case.}
Our first result is to characterize the runtime of the PISGD algorithm for nonconvex functions using Kramers' escape rate formalism~\cite{langer1969statistical}. For simplicity, we consider nonconvex functions with a single barrier for explicit characterization of the escape rate; however, this is done without loss of generality since in multiple barrier case, the asymptotic runtime is dominated by the worst-case tunneling rate between two local minima. We discuss the extension to multiple barriers case in Section~\ref{sec:pisgd_beyond_a_single_barrier}. In particular, we establish that the runtime $\mathcal{T}_{\mathrm{PISGD}}$ is characterized by the bounce action similar to the quantum tunneling rate:

\begin{equation}
\begin{alignedat}{3}
    &\text{SGD}        &\quad \mathcal{T}_{\mathrm{SGD}}   &\sim \exp(\widetilde{\lambda}_{\mathrm{SGD}} H_V)   &\quad& \text{\cite{herau2011tunnel,shi2023learning}} \\
    &\text{Quantum tunneling} &\quad \mathcal{T}_{\mathrm{QT}} &\sim \exp(\widetilde{\lambda}_{\mathrm{QT}} B_V) &\quad& \text{\cite{Coleman:1977py,liu2023quantum}} \\
    &\text{PISGD}      &\quad \mathcal{T}_{\mathrm{PISGD}} &\sim \exp(\widetilde{\lambda}_{\mathrm{PISGD}} B_V) &\quad& \text{Eq.~\eqref{eqn:T_PISGD}}.
\end{alignedat}
\label{eqn:main_results_single_barrier}
\end{equation}
PISGD, although classical, thus shares the exponent of quantum tunneling: the bounce action $B_V$ rather than the barrier height $H_V$.  We show further that the choices of $\widetilde{\lambda}_{\mathrm{QT}}$ and $\widetilde{\lambda}_{\mathrm{PISGD}}$ also agree, so that the leading scalings of quantum tunneling and PISGD coincide in this case. Like quantum tunneling, PISGD can thus hold an advantage over SGD whenever $B_V$ is smaller than $H_V$. Finally, the intermediate configurations visited during PISGD iterations exhibit instanton-like shapes, evidence that PISGD captures the quantum tunneling mechanism in classical optimization through instantons.

We note that the scaling factor~\eqref{eqn:main_results_single_barrier} is not a free parameter and it is determined by the required concentration properties of the stationary distribution of the corresponding algorithm. For example, SGD with isotropic Gaussian noise converges to Gibbs distribution with log-density $\sim \widetilde{\lambda}_{\text{SGD}} V(x)$ so that increasing $\widetilde{\lambda}_{\text{SGD}}$ concentrates the final samples around the global minima. However, as implied by~\eqref{eqn:main_results_single_barrier}, the mixing time can increase exponentially with these scale factors.

Escape rate and concentration requirements therefore pull $\widetilde{\lambda}$ in opposite directions\footnote{In practice, a learning-rate-decay schedule~\cite{shi2023learning}, which for a fixed objective function leads to a decaying noise strength, can relieve this tension: a large noise strength early on helps escape, and a small one later helps concentration near the target. We consider only the fixed-noise setting, for the reasons given in Section~\ref{sec:fixed-noise_sgd_convergence}.}, and the best choice is the smallest $\widetilde{\lambda}$ at which this distribution still concentrates within distance $\varepsilon$ of $x^\star$. This threshold grows with the dimension $d$ and with $1/\varepsilon^2$. For SGD, it is set by the curvature of $V$ at $x^\star$. For PISGD, whose stationary distribution approaches the quantum ground-state density (Section~\ref{sec:distribution_equivalence}), it is set instead by the zero-point spread of the ground state, which depends only on the square root of that curvature; it therefore coincides with the threshold for quantum tunneling. With $\widetilde{\lambda}$ chosen at these thresholds, quantum tunneling and PISGD retain the same optimal escape-time scaling, with the bounce action $B_V$ in place of the barrier height $H_V$ of SGD (Sections~\ref{sec:sgd_runtime} and~\ref{sec:runtime}). This replacement is the central theme of this paper.

\paragraph{Specialization to separable case.}
Since in the most general case, it is very challenging to make the characterization of these parameters tracktable, we consider $1$-$D$ potentials and also high dimensional potentials with coordinate separability, i.e. $V(x) = \sum_{i=0}^{d-1} V_i(x_i)$, where each $V_i$ a one-dimensional function of the coordinate $x_i$.  For non-separable case, we use numerical experiments to demonstrate the similar mechanism and separation between PISGD and SGD (Section~\ref{sec:numerics}). 

For the separable case, for SGD, quantum tunneling, and PISGD alike, our analysis shows that the worst-case escape time is then governed by the slowest coordinate. When all components are identical, each with curvature $(\omega^\star)^2$ at its global minimum, and $\widetilde{\lambda}$ is chosen at the concentration threshold, the optimal escape times are
\begin{equation}
\begin{alignedat}{3}
    &\text{SGD}        &\quad \mathcal{T}_{\mathrm{SGD}}^{\mathrm{opt}}   &\sim \exp\!\left( \frac{1}{\varepsilon^2} \frac{d}{(\omega^\star)^2} H_V \right)   &\quad& \text{Sec.~\ref{sec:sgd_runtime}} \\
    &\text{Quantum tunneling} &\quad \mathcal{T}_{\mathrm{QT}}^{\mathrm{opt}} &\sim \exp\!\left( \frac{1}{2 \varepsilon^2} \frac{d}{\omega^\star} B_V \right) &\quad& \text{Sec.~\ref{sec:runtime}} \\
    &\text{PISGD}      &\quad \mathcal{T}_{\mathrm{PISGD}}^{\mathrm{opt}} &\sim \exp\!\left( \frac{1}{2 \varepsilon^2} \frac{d}{\omega^\star} B_V \right) &\quad& \text{Sec.~\ref{sec:runtime}}
\end{alignedat}
\label{eqn:main_results_separable}
\end{equation}
where $H_V$ and $B_V$ now denote the barrier height and the bounce action of the common one-dimensional component. All three escape times grow exponentially with the dimension $d$, because the concentration threshold for $\widetilde{\lambda}$ grows linearly with $d$; they differ only in the coefficient of $d$. Comparing these coefficients, PISGD has a scaling advantage over SGD whenever 
\begin{equation}
\label{eqn:pisgd_advantage_condition}
  B_V < 2 H_V / \omega^\star  
\end{equation}
a condition favored by high, thin barriers and by a flat global basin~\footnote{Interestingly, Appendix~\ref{sec:quartic_functions} shows that the asymmetric quartic double-well potential---a standard and particularly simple model of quantum tunneling---does not satisfy this condition. Reproducing the tunneling mechanism in PISGD thus does not by itself guarantee faster optimization: for this family of potentials, PISGD has a worse asymptotic scaling than SGD in the setting analyzed here.}.

\begin{figure}[!htbp]
    \centering
    \includegraphics[width=0.6\linewidth]{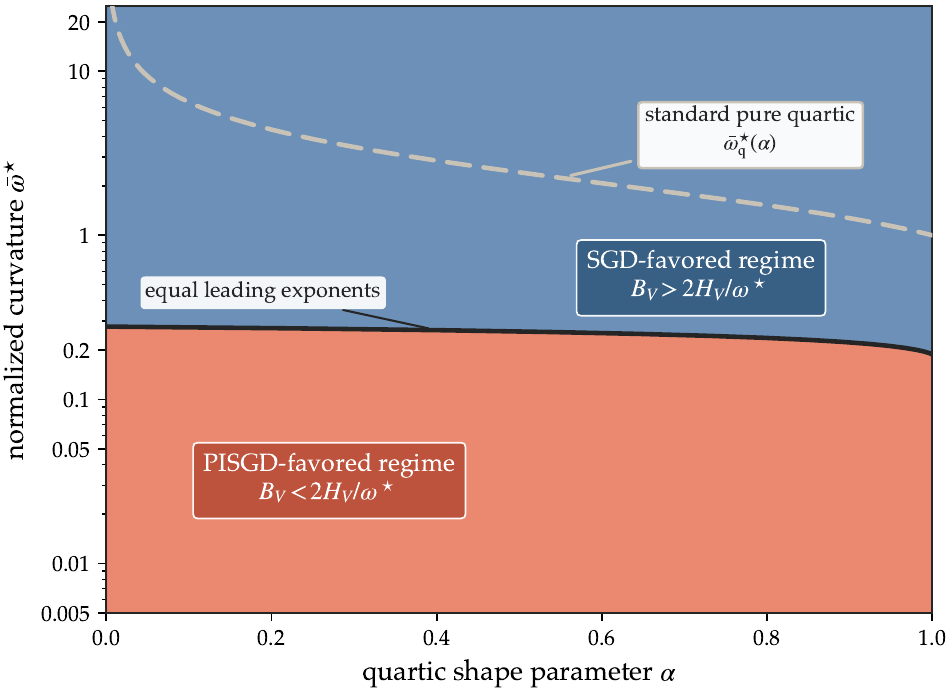}
    \caption{Concentration-adjusted PISGD- and SGD-favored parameter regimes for the separable variant quartic potential.
    In the orange PISGD-favored region, $B_V<2H_V/\omega^\star$, so PISGD has the smaller leading exponent. In the blue SGD-favored region, $B_V>2H_V/\omega^\star$, so SGD has the smaller leading exponent.
    The small open circle marks the setting $(\alpha,\bar{\omega}^\star)=(0.9,0.1)$ used in our variant-quartic numerical experiments.
    The dashed gray curve is the normalized-curvature curve $\bar{\omega}_{\mathrm{q}}^\star(\alpha)$ of the standard pure quartic and lies entirely in the SGD-favored regime. Although the pure quartic is not literally a quartic branch joined to a quadratic branch, the same comparison applies.}
    \label{fig:variant_quartic_pisgd_sgd_phase_diagram}
\end{figure}

\paragraph{Explicit cases where PISGD outperforms SGD.} We construct explicit families of nonconvex separable functions where PISGD outperforms SGD.  We first start with a $d$ dimensional function with each coordinate given by the asymmetric quartic double well 
\begin{equation}
    U(z)=c\,q_\alpha(z),
\end{equation}
with $q_\alpha(z):=\frac{1}{2}z^2-\frac{1}{2}z^3+\frac{1}{8}\alpha z^4$, $0<\alpha<1$, and $c>0$, is a standard model of quantum tunneling~\cite{Coleman:1977py,dunne2008functional}, yet it violates~\eqref{eqn:pisgd_advantage_condition} for every $\alpha$. The condition is met, however, once the curvature of the global basin is decoupled from the barrier. Next, we modify this to construct \emph{variant quartic} to keep the quartic part up to its turning point $z^{\mathrm{tp}}=2(1-\sqrt{1-\alpha})/\alpha$, where $q_\alpha(z^{\mathrm{tp}})=q_\alpha(0)=0$, and continues it by a quadratic branch of tunable curvature,
\begin{equation}
    U(z)=
    \begin{cases}
        c\,q_\alpha(z), & z\leq z^{\mathrm{tp}}, \\
        c\,q_\alpha'(z^{\mathrm{tp}})(z-z^{\mathrm{tp}})+\dfrac{c}{2}(\bar{\omega}^\star)^2(z-z^{\mathrm{tp}})^2, & z>z^{\mathrm{tp}},
    \end{cases}
\label{eqn:variant_quartic_overview}
\end{equation}
which is continuously differentiable. The barrier lies entirely on the quartic branch, so $H_V$ and $B_V$ are those of the quartic, whereas the global minimum lies on the quadratic branch with curvature $(\omega^\star)^2=c(\bar{\omega}^\star)^2$. By condition in~\eqref{eqn:pisgd_advantage_condition}, PISGD then has the smaller exponent whenever the normalized curvature $\bar{\omega}^\star$ is below the critical value $\bar{\omega}_{\mathrm{c}}^\star(\alpha):=2H_V/(\sqrt{c}\,B_V)$, which depends only on $\alpha$. In this case, PISGD outperforms SGD. Figure~\ref{fig:variant_quartic_pisgd_sgd_phase_diagram} maps the resulting PISGD- and SGD-favored regimes in more detail on this toy model.

The advantage is even more pronounced for tall, thin barriers, such as a quadratic well split by a narrow Gaussian spike,
\begin{equation}
    U(z)=c\left(\frac{1}{2}(\bar{\omega}^\star)^2z^2+\frac{h}{\sqrt{2\pi w^2}}\exp\!\left(-\frac{(z-\mu)^2}{2w^2}\right)\right),
\label{eqn:spike_overview}
\end{equation}
with width $w$, weight $h$, and center $\mu<0$. For explicit choice of $w, h, \mu$ and $c$, we showed in Section~\ref{sec:numerics_spiky_func}, the SGD exponent is about $8.5$ times larger than the PISGD exponent.

\begin{figure}[!htbp]
    \centering
    \centering
    \begin{subfigure}[t]{0.31\linewidth}
        \centering
        \includegraphics[width=\linewidth]{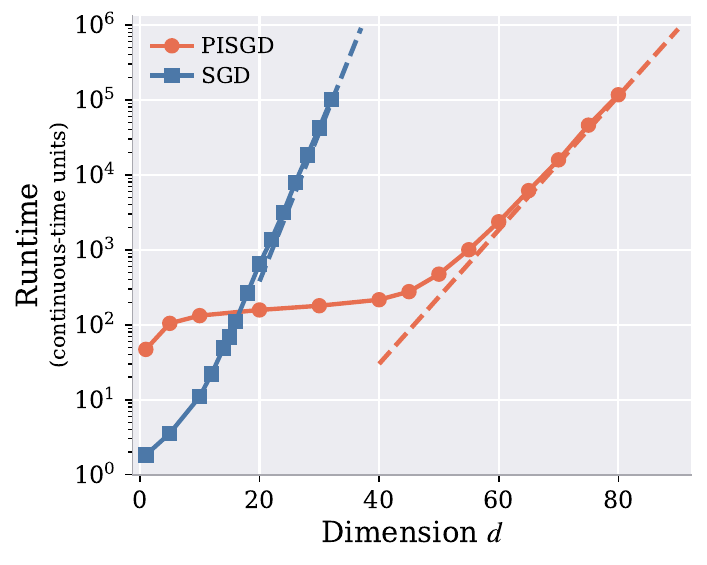}
        \caption{Separable variant quartics.}
        \label{fig:comparison_variant_quartic}
    \end{subfigure}
    \hfill
    \begin{subfigure}[t]{0.31\linewidth}
        \centering
        \includegraphics[width=\linewidth]{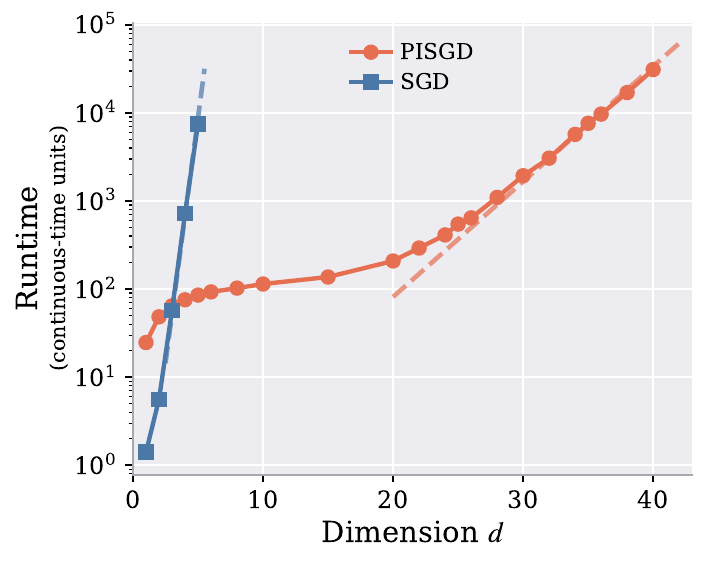}
        \caption{Separable continuous spikes.}
        \label{fig:comparison_spike}
    \end{subfigure}
    \hfill
    \begin{subfigure}[t]{0.31\linewidth}
        \centering
        \includegraphics[width=\linewidth]{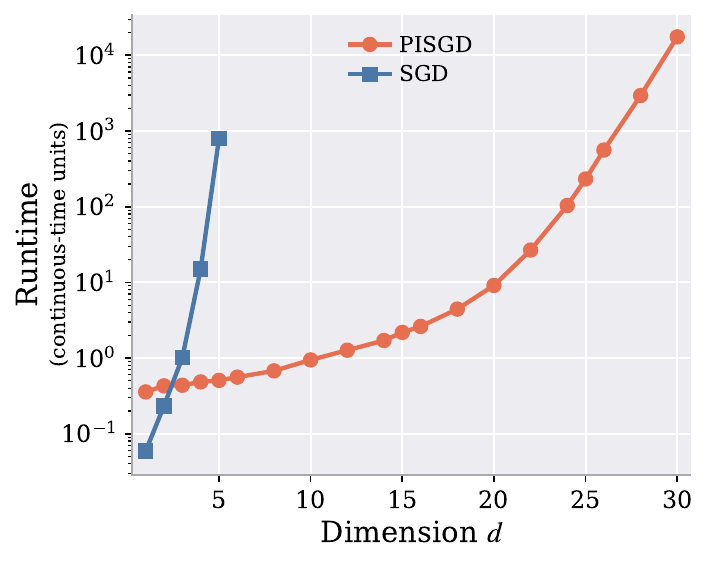}
        \caption{Nonseparable squared quartics.}
        \label{fig:comparison_squared_quartic}
    \end{subfigure}
    \caption{Comparison of PISGD and SGD to solution for different potentials across different dimensions $d$, measured in terms of the number of gradient steps times the learning rate. Figure~\ref{fig:comparison_main_results}(a) is for the single component of a quartic potential in equation~\eqref{eqn:variant_quartic_overview}. Figure~\ref{fig:comparison_main_results}(b) is for the Gaussian spike in equation~\eqref{eqn:spike_overview}. Figure~\ref{fig:comparison_main_results}(c) is for a non-separable example detailed in Section~\ref{sec:nonseparable_potentials}. For separable potential cases, the dashed lines show the theoretically predicted scalings in the large $d$ asymptotic limit.}
    \label{fig:comparison_main_results}
\end{figure}

Figures~\ref{fig:comparison_variant_quartic} and~\ref{fig:comparison_spike} compare SGD and PISGD numerically on two such potentials and confirm the scalings in~\eqref{eqn:main_results_separable}, and Figure~\ref{fig:comparison_squared_quartic} reports the same comparison for a nonseparable potential whose stationary points have a comparable structure (Sections~\ref{sec:numerics_variant_quartic_func}, \ref{sec:numerics_spiky_func}, and~\ref{sec:nonseparable_potentials}). For separable potentials, we also derive the PISGD escape time beyond its leading exponential factor (Section~\ref{sec:runtime}). The resulting prefactor makes the dependence on the remaining hyperparameters explicit, allows a direct comparison with these numerical runs, and lets us check whether the omitted factors introduce any further exponential scaling.

We should note that our results should not be read as ruling out quantum advantage for tunneling-based algorithms. In certain landscapes outside of our scope, a genuine quantum system can exploit coherence among several tunneling paths, whereas a Monte Carlo sampler such as PISGD can suffer from a topological obstruction~\cite{hastings2013obstructions,andriyash2017can}. We will return to this point in Section~\ref{sec:related_work}. Furthermore, in this paper, we compare PISGD only with the algorithm exploiting tunneling mechanisms rather than all quantum dynamical algorithms. Section~\ref{sec:related_work} briefly discusses why PISGD does not capture the mechanisms of ~\cite{herman2025mechanisms,leng2023quantum}, and we leave a full comparison to future work.

\subsection{Related Work}
\label{sec:related_work}

Finding the global optimum of a nonconvex function in general is very difficult and even a quadratic nonconvex case is known to be NP-hard~\cite{murty1987some,pardalos1991quadratic}.
First-order methods such as GD and SGD are guaranteed to converge to
approximate stationary points~\cite{ge2015escaping, carmon2020lower}, 
but finding a global minimum is a fundamentally harder task.  Tight
bounds on the spectral gap of Langevin diffusions show that escape
from metastable basins requires time exponential in the barrier
height~\cite{langer1969statistical, digesù2015smallnoisespectralgap,
brooks2019sharptunnelingestimatesdoublewell}.  Raginsky et
al.~\cite{raginsky2017nonconvex} analyzed the non-asymptotic
convergence of SGD (Langevin dynamics) for nonconvex optimization and
showed that the mixing time depends exponentially on certain
properties of the landscape.  Shi, Su, and
Jordan~\cite{shi2023learning} connected the learning-rate schedule of
SGD to the spectral theory of Schr\"odinger operators, a connection
that also underlies PISGD.

Simulated annealing~\cite{kirkpatrick1983sa} and its variants
(dual annealing, basin-hopping, parallel tempering, hardware
annealers~\cite{aramon2019fphy, mohseni2022ising}) overcome
metastability by varying the temperature: running at high temperature
enables barrier crossing, then cooling concentrates the distribution
near the global minimum.  PISGD takes a complementary approach: rather
than changing the temperature schedule, it changes the search space by
lifting single points to worldlines, so that the barrier-crossing cost
is reduced at any fixed temperature.

Quantum algorithms for optimization exploit tunneling and coherence to
navigate nonconvex landscapes.  Quantum adiabatic
optimization~\cite{farhi2002quantum,farhi2014quantum} and its gate-model variants
prepare the ground state of a problem Hamiltonian by slowly
deforming a simple initial Hamiltonian.  Liu, Su, and
Li~\cite{liu2023quantum} introduced quantum tunneling walks and showed that
tunneling rates are governed by the Agmon distance rather than the
barrier height, yielding polynomial or exponential speedups over
classical random walks on certain landscapes.  Herman et
al.~\cite{herman2025mechanisms} identified
mechanisms for provable quantum advantage and demonstrated
polynomial-time quantum optimization for broad families of nonconvex
functions via a real-space adiabatic algorithm.  Leng et
al.~\cite{leng2023quantum} proposed Quantum Hamiltonian Descent, a
quantum analogue of gradient descent driven by Schr\"odinger
evolution. 
PISGD is motivated by the same operator correspondence that underlies
these quantum algorithms but operates entirely classically. Moreover, the practicality of quantum optimization algorithms is less clear for data-driven objectives than for fixed objective functions. As a classical algorithm, PISGD is in principle compatible with such objectives, although its benefits in that setting also remain unknown.

The Suzuki--Trotter decomposition~\cite{suzuki1976relationship} maps a
$d$-dimensional quantum system to a $(d{+}1)$-dimensional classical
system, enabling classical simulation via Path Integral Monte Carlo
(PIMC).  When used as an optimizer, this approach is known as
simulated quantum annealing
(SQA)~\cite{martonak2002quantum}.  Crosson and
Harrow~\cite{7782986} proved that SQA can be exponentially faster
than classical simulated annealing.  Isakov et
al.~\cite{isakov2016understanding} demonstrated that quantum Monte
Carlo reproduces instanton-mediated tunneling rates, and Jiang et
al.~\cite{jiang2017scaling} provided a detailed scaling analysis
connecting instantons to thermally assisted tunneling in QMC.
Denchev et al.~\cite{denchev2016computational} studied the
computational value of finite-range tunneling in the context of
quantum annealing hardware.  Bravyi~\cite{bravyi2014monte} and
Bergamaschi~\cite{bergamaschi2020simulated} analyzed the complexity of
Monte Carlo simulation for stoquastic Hamiltonians and showed
efficient mixing for specific problem classes. PISGD departs from
these Metropolis-based PIMC methods by using gradient-based updates
(SGD) on the discretized action, which enables parallel updates of all
time slices.
That said, other PIMC algorithms that sample according to the same action may exhibit scaling behavior similar to that of PISGD. We adopt SGD update rules partly for theoretical convenience, as they allow us to draw on established results for overdamped Langevin dynamics.

As discussed in~\cite{hastings2013obstructions,andriyash2017can}, quantum Monte Carlo algorithms can suffer from topological obstructions and cannot reproduce coherent quantum tunneling, in which multiple tunneling paths interfere to enhance the tunneling amplitude in genuinely quantum algorithms. As an instantiation of PIMC, PISGD shares these limitations and benefits only from incoherent tunneling.

Beyond optimization, PISGD also admits an application in the reverse direction. PIMC methods were developed to study quantum systems and inherit some of their properties. In particular, the simulation dynamics of PIMC has been proposed as a probe of quantum tunneling and false vacuum decay rates~\cite{shen2023vacuum}, especially their instanton-governed exponential factors. Since our analysis of PISGD extends beyond the leading exponential factor with a more concrete framework for classical numerical experiments on quantum systems and may yield more detailed information about tunneling and decay rates.

The Euclidean path integral and instanton (bounce) solutions are
standard tools in quantum field
theory~\cite{coleman1988aspects, vauinshteuin1982abc,
gattringer2009quantum}.  Kramers' theory~\cite{langer1969statistical,hanggi1990reaction}
and its generalizations provide the framework for computing
barrier-crossing rates in stochastic
systems~\cite{hong2022introduction, herau2011tunnel}.  We use the
Gel'fand--Yaglom method~\cite{dunne2008functional, dunne2005beyond}
for functional-determinant ratios to evaluate the semiclassical
prefactors entering the PISGD escape rate. 
The stochastic dynamics of PISGD can be modeled by the class of stochastic partial differential equations (SPDEs), which utilize a space-time Brownian perturbation to stochastically update a path \cite{goodair2023stochasticcalculusinfinitedimensions}. Specifically, these dynamics under a certain symmetric potential is the well-known Allen-Cahn equation. There has been some previous work on analyzing metastability in these equations, \textit{i.e.}, extending Kramer's law to SPDE case: \textit{e.g.}, \cite{berglund2013sharp} and \cite{barret2015sharp} with focus. 
However, a rigorous spectral theory for a general SPDE setting is not yet available, so our analysis partly follows physics-style derivations rather than fully rigorous proofs.

\paragraph{Organization.}
Section~\ref{sec:algo} derives PISGD from the Euclidean path integral, presents the algorithm and its hyperparameter normalization, and explains its instanton-mediated escape mechanism. Section~\ref{sec:distribution_equivalence} establishes the correspondence between the PISGD stationary marginal distribution and the quantum ground-state density. Section~\ref{sec:sgd_runtime} reviews the escape and concentration analysis for SGD, while Section~\ref{sec:runtime} derives the PISGD runtime, compares it with quantum tunneling, extends the analysis to separable potentials, and evaluates the factors beyond the leading exponential. Section~\ref{sec:numerics} tests the predicted behavior on separable and nonseparable potentials. 
The appendices provide supporting derivations, benchmark potentials, and ablation studies.

\section{PISGD: Algorithm and Mechanism}\label{sec:algo}

\subsection{Background: Quantum Mechanics, Path Integrals, and Partition Functions}

We review the background from quantum mechanics and its path-integral formalism, in particular the Euclidean-time version. This is the foundation of path-integral Monte Carlo algorithms for quantum systems, which we later adapt to optimization.

In quantum mechanics, we consider a time-independent Hamiltonian operator with a potential $V: \mathbb{R}^d \rightarrow \mathbb{R}$ and a scaling
parameter $\lambda > 0$:
\begin{equation}
    \hat{H} = -\frac{1}{2} \Delta + \lambda^2 V(\hat{x}),
\label{eqn:hamiltonian}
\end{equation}
a second-order differential operator acting on $L^2 (\mathbb{R}^d)$. The Hamiltonian governs the
time evolution of the quantum system via the Schr\"odinger equation
\begin{equation}
    i\, \frac{\partial}{\partial t} \ket{\psi (t)} = \hat{H} \ket{\psi(t)}
\end{equation}
where $\ket{\psi (t)} \in L^2 (\mathbb{R}^d)$ is a quantum state at time $t$. Its inner product with the position state $\ket{x}$ is the wavefunction $\psi(x; t) = \braket{x}{\psi(t)}$.
Formally, the Schr\"odinger equation can be solved by the time evolution operator
\begin{equation}
    \hat{U}(t) = \exp(- i \hat{H} t)
    ,
\end{equation}
such that for an initial quantum state $\ket{\psi (0)}$ at time $0$, the time-evolved quantum state at time $t$ is
\begin{equation}
    \ket{\psi (t)} = \hat{U}(t) \ket{\psi(0)}
    .
\end{equation}
The complex-valued transition amplitude from a point $x \in \mathbb{R}^d$ at time $0$ to a point $x' \in \mathbb{R}^d$ at time $t$ is the \emph{propagator} $K(x, x'; t) = \mel{x'}{\hat{U}(t)}{x}$. The propagator reconstructs the solution to the Schr\"odinger equation: for any point $x' \in \mathbb{R}^d$, the wavefunction at time $t$ is
\begin{equation}
    \psi(x'; t) = \braket{x'}{\psi(t)} = \int \diff x \, K(x, x'; t) \braket{x}{\psi(0)} = \int \diff x \, K(x, x'; t) \psi(x; 0)
\end{equation}
given the initial wavefunction $\psi(x; 0)$ for all the points $x \in \mathbb{R}^d$.

As an alternative to the Hamiltonian formalism, the path-integral formalism describes the same quantum dynamics by expressing the propagator as
\begin{equation}
    K(x_i, x_f; t) = \int_{x(0) = x_i}^{x(t) = x_f} \mathcal{D} x \, \exp(i S[x; \lambda])
    ,
\end{equation}
where $\mathcal{D} x$ is a measure on the space of functions $x: [0, t] \rightarrow \mathbb{R}^d$, \textit{i.e.}, on all paths in $\mathbb{R}^d$ parametrized by time, and $S$ is the action, a functional of the path $x$.
The action is
\begin{equation}
    S[x;\lambda]
    = \int_0^t \mathrm{d}t\;\left[
        \frac{1}{2}\!\left(\frac{\mathrm{d}x}{\mathrm{d}t}\right)^{\!2}
        - \lambda^2 V\!\bigl(x(t)\bigr)
      \right].
\end{equation}

The discussion above uses a real time $t$. Physics also considers a Euclidean time $\tau = i t$. In physics, one also considers a Wick rotation that formally rotates the time variable in the complex plane so that $\tau$ becomes real-valued while the original time $t$ becomes imaginary-valued; for this reason $\tau$ is conventionally called the imaginary time even though it is real-valued. After this reparametrization, the action becomes the Euclidean action
\begin{equation}
    S_{\mathrm{E}}[x;\lambda]
    = \int_0^T \mathrm{d}\tau\;\left[
        \frac{1}{2}\!\left(\frac{\mathrm{d}x}{\mathrm{d}\tau}\right)^{\!2}
        + \lambda^2 V\!\bigl(x(\tau)\bigr)
      \right],
\label{eqn:euclidean_action}
\end{equation}
and the Euclidean propagator, which describes the Euclidean-time dynamics, is
\begin{equation}
    K_{\mathrm{E}}(x_i, x_f; T) = \int_{x(0) = x_i}^{x(T) = x_f} \mathcal{D} x \, \exp(- S_{\mathrm{E}}[x; \lambda])
    ,
\end{equation}
where $K_{\mathrm{E}}(x_i, x_f; T) = \mel{x_f}{\hat{U}_{\mathrm{E}} (T)}{x_i}$ and $\hat{U}_{\mathrm{E}} (T) = \exp(- \hat{H} T)$.
The key difference from the real-time case is that the integrand $\exp(- S_{\mathrm{E}}[x; \lambda])$ is real and positive, so it admits a probabilistic interpretation~\footnote{We use the subscript $\mathrm{E}$ to denote ``Euclidean'', but drop it later when no ambiguity arises from the context.}.

We now introduce the Euclidean partition function, whose probabilistic interpretation leads to path-integral Monte Carlo algorithms.

We impose a periodic boundary condition in Euclidean time over the imaginary-time length $T$, so that $x (\tau) = x(\tau + T)$ for all $\tau$, and restrict attention to $\tau \in [0, T]$. The partition function is then
\begin{equation}
\begin{aligned}
    Z = \tr \left[ e^{- T \hat{H}} \right] = \int \diff x_0 \, \mel{x_0}{e^{- T \hat{H}}}{x_0} = \int \diff x_0 \, K_{\mathrm{E}}(x_0, x_0; T) = \int \mathcal{D} x \, \exp(- S_{\mathrm{E}}[x; \lambda])
\end{aligned}
\end{equation}
where the path-integral measure now runs over functions satisfying the periodic boundary condition.

In quantum physics, once an ensemble of paths is sampled from the distribution proportional to $\exp(- S_{\mathrm{E}}[x; \lambda])$ by a Monte Carlo algorithm, observables can be estimated from the sample to study properties of the quantum system.

This paper instead applies path-integral Monte Carlo to optimization. The idea is that for a sufficiently large scaling parameter $\lambda$, \textit{i.e.}, in the semiclassical limit, an ensemble of paths drawn with probability proportional to $\exp(- S_{\mathrm{E}}[x; \lambda])$ approximately represents the distribution of the quantum ground state, which in this limit concentrates near the global minimum $x^\star$.

\subsection{Algorithm}

The path integral runs over paths $x(\tau)$ parametrized by a continuous imaginary time $\tau$. Following standard practice in physics, we discretize $\tau$ into $N_T$ equally spaced points $\tau_j = j\,a$ on the interval $[0, T]$, where $a = T/N_T$ is the lattice spacing and $j \in [N_T] = \{0, 1, \ldots, N_T - 1\}$. The path is then represented by the collection of time slices $(x_0, x_1, \ldots, x_{N_T-1}) \in \mathbb{R}^{d\times N_T}$. Replacing the derivative in~\eqref{eqn:euclidean_action} by a finite difference and the integral by a sum gives the discretized action
\begin{align}
    S \!\left(\{x_j\}_j;\lambda\right)
    = a \sum_{j=0}^{N_T-1} \left(
        \frac{1}{2a^2}\left(x_j - x_{j+1}\right)^2
        + \lambda^2 V(x_j)
      \right),
\label{eqn:discrete_action_lambda2}
\end{align}
where we have omitted the subscript $\mathrm{E}$ for notational convenience. The periodic boundary condition is imposed by identifying $x_{N_T} \equiv x_0$. Crucially, $S$ is an ordinary real-valued function on $\mathbb{R}^{d\times N_T}$.

Gradient-based sampling requires the gradient of Eq.~\eqref{eqn:discrete_action_lambda2},
\begin{equation}
    \frac{\partial S}{\partial x_{j}}
    = \frac{1}{a}\!\left(2x_j - x_{j-1} - x_{j+1}\right)
      + a\,\lambda^2\,\nabla V(x_j),
\label{eqn:gradient_action}
\end{equation}
a $d$-dimensional vector for each $j \in [N_T]$, with slice indices taken modulo $N_T$.

We assess PISGD and SGD using the same success criterion. A candidate output
$\tilde{x}$ is successful when
\begin{equation}
    \norm{\tilde{x} - x^\star} \le \varepsilon,
\label{eq:nonconvex-optimization-problem}
\end{equation}
where $x^\star$ is the global minimum and $\varepsilon>0$ is the prescribed
tolerance.

PISGD applies the standard noisy gradient-descent update to $S$, treating the entire worldline as the optimization variable.
\begin{algorithm}[H]
\caption{Path Integral Stochastic Gradient Descent}\label{alg:pisgd}
\textbf{Input:} \\
Differentiable function $V$: $\mathbb{R}^d \rightarrow \mathbb{R}$; \\
Initial point $x^{(0)} \in \mathbb{R}^d$; \\
Maximum number of iterations $k_{\mathrm{max}} \in \mathbb{Z}_+$; \\
\textbf{Hyperparameters:} \\
Scaling parameter $\lambda \in \mathbb{R}_+$; \\
Learning rate $s \in \mathbb{R}_+$; \\
Noise strength $\sigma \in \mathbb{R}_+$; \\
Imaginary-time lattice spacing $a \in \mathbb{R}_+$; \\
Number of imaginary time slices $N_T \in \mathbb{Z}_+$; \\
\textbf{Output:} Candidate solution $x_{\mathrm{sol}} \in \mathbb{R}^d$ obtained by postprocessing the worldline $\{x_j\}_{j=0}^{N_T-1}$;
\begin{algorithmic}[1]
    \State Initialize $x_j \gets x^{(0)}$ for all $j \in [N_T]$, and set $k \gets 0$
  \For{$k < k_{\mathrm{max}}$}
        \State \State Compute $g_j \gets \left.\frac{\partial S}{\partial x_{j}}\right|_{\{x_{j'}\}_{j'}}$ for all $j \in [N_T]$ in parallel%
        \State Sample $\xi_j^{(k)} \sim \mathcal{N}(0, I_d)$ independently for all $j \in [N_T]$
        \State Update $x_j \gets x_j - s\, g_j + \sigma \sqrt{s}\, \xi_j^{(k)}$ for all $j \in [N_T]$
    \State $k \gets k + 1$
  \EndFor
    \State \Return $x_{\mathrm{sol}} \gets x_0$
\end{algorithmic}
\end{algorithm}

We now describe the iteration in detail. As with SGD and other optimizers, PISGD starts from an initial point $x^{(0)} \in \mathbb{R}^d$, which is copied to every time slice, $x_j = x^{(0)}$ for all $j \in [N_T]$. At the $k$-th iteration the worldline is updated according to
\begin{equation}
    x^{(k+1)}_j = x^{(k)}_j - s \left.\frac{\partial S}{\partial x_{j}}\right|_{\{x_{j'}\} = \{x^{(k)}_{j'}\}} + \sigma \sqrt{s}\, \xi^{(k)}_j
\label{eqn:update_rule}
\end{equation}
where $\xi^{(k)} \sim \mathcal{N} (0, I_{d \cdot N_T})$ is Gaussian noise, independent across iterations, dimensions, and time slices, and $\xi^{(k)}_j$ is its $j$-th component along the time-slice axis. The learning rate $s$ and the noise strength $\sigma$ are two independent hyperparameters; we revisit this choice of parameterization in Section~\ref{sec:discretization}. As in ordinary SGD, the gradient of this $(d \times N_T)$-dimensional function can in principle be evaluated in parallel across its components. After $k_{\mathrm{max}}$ iterations, the algorithm returns a solution $x_{\mathrm{sol}}$ derived from the final worldline $(x_0, x_1, \ldots, x_{N_T-1})$. Since each $x_j$ is a $d$-dimensional vector, postprocessing is needed to produce a single $d$-dimensional solution, and we take $x_{\mathrm{sol}} = x_0$, \textit{i.e.}, the value of $x$ at $\tau = 0$.~\footnote{
    Alternatives include the slice average $x_{\mathrm{sol}} = \frac{1}{N_T} \sum_j x_j$,
    or the best slice $x_{\mathrm{sol}} = x_{j^*}$ with $j^* = \argmin_j V(x_j)$, determined by evaluating $V(x_j)$ for each $j$. The latter may achieve a shorter hitting time than $x_{\mathrm{sol}} = x_0$, but we adopt $x_0$ because it is more amenable to analysis.
}
Algorithmically, PISGD is simply SGD applied to the function $S$ on $\mathbb{R}^{d\times N_T}$. The per-iteration cost is $N_T$ evaluations of $\nabla V$, one per slice, all of which are independent and can be computed in parallel. Everything that distinguishes PISGD from a naive high-dimensional SGD comes from the specific form of $S$: the nearest-neighbor coupling inherited from the path integral. Section~\ref{sec:instantons} explains why this coupling opens escape routes in the landscape of $S$ that have no counterpart in the landscape of~$V$ itself.

\subsection{Hyperparameter Normalization}\label{sec:discretization}

The four hyperparameters $s$, $a$, $\lambda$, and $\sigma$ are not independent. We show here that one of them is redundant, which yields a more concise parametrization and identifies the physically meaningful combination that controls the dynamics.

Substituting the gradient~\eqref{eqn:gradient_action} into the update rule~\eqref{eqn:update_rule} gives
\begin{equation}
    x^{(k+1)}_j = x^{(k)}_j + \frac{s}{a} \left(x^{(k)}_{j+1} - 2 x^{(k)}_j + x^{(k)}_{j-1}\right) - s a \lambda^2 \nabla V\!\left(x^{(k)}_j\right) + \sigma \sqrt{s} \xi^{(k)}_j
    .
\label{eqn:update_rule_substituted}
\end{equation}

The hyperparameters enter only through the three coefficients $s/a$, $s a \lambda^2$, and $\sigma \sqrt{s}$ multiplying the kinetic, potential, and noise terms. Two hyperparameter choices $(s, a, \lambda, \sigma)$ and $(\widetilde{s}, \widetilde{a}, \widetilde{\lambda}, \widetilde{\sigma})$ therefore generate identical updates precisely when these three coefficients agree,
\begin{equation}
    \frac{s}{a} = \frac{\widetilde{s}}{\widetilde{a}}, \quad s a \lambda^2 = \widetilde{s} \widetilde{a} \widetilde{\lambda}^2, \quad \sigma \sqrt{s} = \widetilde{\sigma} \sqrt{\widetilde{s}}
    .
\label{eqn:hyperparameter_invariants}
\end{equation}
Three constraints on four parameters leave one redundant degree of freedom, which we use to fix the noise strength at a convenient value. Throughout this paper we adopt
\begin{equation}
    \widetilde{s} = \frac{\sigma^2 s}{2}, \quad \widetilde{a} = \frac{\sigma^2 a}{2}, \quad \widetilde{\lambda} = \frac{2 \lambda}{\sigma^2}, \quad \widetilde{\sigma} = \sqrt{2}
    ,
\label{eqn:hyperparameter_normalization}
\end{equation}
which satisfies~\eqref{eqn:hyperparameter_invariants}.

The three surviving hyperparameters play distinct roles. The learning rate $s$ sets the simulation-time discretization and the lattice spacing $a$ sets the imaginary-time discretization; the path-integral dynamics is asymptotically invariant as both are taken to zero, and we report an ablation study supporting this in Appendix~\ref{app:ablation_study}.

The remaining parameter $\widetilde{\lambda}$ is the physically relevant one. It is the combination that governs both the continuous-time dynamics and the stationary distribution, and it is the parameter with respect to which the semiclassical limit is taken.

The normalization $\widetilde{\sigma} = \sqrt{2}$ is chosen so that the stationary distribution takes a natural form. Under this condition, the stationary density of PISGD is proportional to $\exp (- S (\{x_j\}_j; \widetilde{\lambda}) )$, without additional prefactors on the exponent.

\subsection{Instantons: Classical Saddles that Reproduce Quantum Tunneling}
\label{sec:instantons}

Standard SGD on $V$ escapes a local minimum by thermally climbing over the barrier: the noise must supply enough energy to lift the iterate above the barrier top, at a cost set by the barrier height $H_V$.
PISGD escapes by a qualitatively different route.  Because its state is an entire worldline rather than a single point, it can cross the barrier gradually, with only a small fraction of the time slices above the barrier top at any moment.  The corresponding path in the landscape of $S$ passes through a saddle point known as the \emph{bounce}, or equivalently an \emph{instanton}--\emph{anti-instanton} pair.  Its action cost $B_V$ is an integrated quantity, the \emph{Agmon distance}~\cite{26e649f4-ac21-3f96-8d07-d5d51780f780,carmona1981pointwise,hislop2000exponential}, which after normalization can be far smaller than $H_V$.

PISGD is a purely classical algorithm, yet this classical saddle carries exactly the action that governs quantum tunneling in the semiclassical limit.  The escape rate therefore inherits the tunneling scaling, without quantum dynamics being simulated directly.

Bounces and instantons are well studied in quantum mechanics and quantum scalar field theories~\cite{Coleman:1977py,callan1977fate,vauinshteuin1982abc,andreassen2017precision} and they give the dominant exponential factor of the quantum tunneling rate.

We consider a potential $V$ on $\mathbb{R}^d$ with a local minimum at~$x^\bullet$ and a global minimum at~$x^\star$, separated by a barrier whose lowest crossing point is an index-one saddle~$x^\circ$~\footnote{The Hessian at an index-one saddle has only one negative eigenvalue with the rest of its eigenvalues positive.}; see Figure~\ref{fig:sgd_potential_barrier} for an illustration. The critical points of the continuum action~\eqref{eqn:euclidean_action}
are paths satisfying the Euler--Lagrange equation
\begin{equation}
    \frac{\mathrm{d}^2 x}{\mathrm{d}\tau^2}
    = \lambda^2 \nabla V(x),
\label{eqn:euler_lagrange}
\end{equation}
which is the equation of motion for a particle moving in the \emph{inverted} potential $-\lambda^2 V$. (The sign flip comes from the ``$+$'' in the Euclidean action.)
The constant paths $x(\tau)\equiv x^\bullet$ and $x(\tau)\equiv x^\star$ are trivial solutions and correspond to local minima of $S$: they represent the optimizer sitting permanently in one basin of~$V$. The constant path $x(\tau)\equiv x^\circ$ is also a trivial solution, a saddle point of $S$.

The inverted potential also admits a nontrivial classical solution known as the bounce we denote as $x_{\mathrm{cl}} (\tau)$, a saddle-point path in path space that mediates between the constant paths $x(\tau)\equiv x^\bullet$ and $x(\tau)\equiv x^\star$: a trajectory that starts
near the hilltop at $x^\bullet$, rolls down through the barrier region of the inverted potential, reaches a turning point $x_{\mathrm{tp}}$ on the turning surface $\{x : V(x) = V(x^\bullet)\}$ where it comes momentarily to rest, and then retraces its path back to $x^\bullet$ (see Figure~\ref{fig:main_fig}). This solution stays close to $x^\bullet$ except within an
imaginary-time window of width $\sim 1/(\lambda\sqrt{\mu_{\min}})$, where $\mu_{\min} > 0$ is the smallest eigenvalue of $\nabla^2 V(x^\bullet)$. The \emph{instanton} is the outgoing half of the bounce, from $x^\bullet$ to $x_{\mathrm{tp}}$, and the \emph{anti-instanton} is the returning half.

The action cost of the bounce reduces to a spatial integral. Eq.~\eqref{eqn:euler_lagrange} conserves the energy $E = \tfrac{1}{2}\abs{\diff x_{\mathrm{cl}}/\diff \tau}^2 - \lambda^2 V(x_{\mathrm{cl}})$, and on the infinite imaginary-time line the bounce is asymptotically at rest at $x^\bullet$, so $E = -\lambda^2 V(x^\bullet)$. Following the standard treatment, we work in this infinite-$T$ limit, where the conservation law gives the speed along the trajectory of the classical bounce solution,
\begin{equation}
    \abs{\frac{\diff x_{\mathrm{cl}}}{\diff \tau}} = \sqrt{2}\,\lambda\sqrt{V(x_{\mathrm{cl}})-V(x^\bullet)}
    .
\label{eqn:bounce_speed}
\end{equation}
Parametrizing the instanton by $\tau \in (-\infty, \tau_c]$ and the anti-instanton by $\tau \in [\tau_c, \infty)$ with $\tau_c \in \mathbb{R}$ the center of the bounce, $x_{\mathrm{cl}}(-\infty) = x^\bullet$ and $x_{\mathrm{cl}}(\tau_c) = x_{\mathrm{tp}}$, the action of the instanton (or the anti-instanton) relative to the constant path at $x^\bullet$ is the \emph{instanton action} $S_I$, and the bounce action is $B_V = 2 S_I$. Substituting~\eqref{eqn:bounce_speed} makes the two terms in the Lagrangian equal and gives
\begin{equation}
    S_I
    = \int_{-\infty}^{\tau_c} \diff \tau \! \left[ \frac{1}{2} \abs{\frac{\diff x_{\mathrm{cl}}}{\diff \tau}}^2 + \lambda^2 \left(V(x_{\mathrm{cl}}) - V(x^\bullet)\right) \right] = \sqrt{2}\,\lambda
      \int_{\gamma_{\mathrm{cl}}} \abs{\diff x}
        \!\sqrt{V(x) - V(x^\bullet)},
\label{eqn:instanton_action}
\end{equation}
where subtracting $V(x^\bullet)$ measures the action relative to the reference path $x(\tau) \equiv x^\bullet$ (also necessary to make the integral finite), $\gamma_{\mathrm{cl}}$ is the spatial curve traced by the instanton from $x^\bullet$ to $x_{\mathrm{tp}}$, and $\abs{\diff x}$ is the arclength element. Among all curves joining the two points, the instanton traces the one of least length in the weight $\sqrt{V - V(x^\bullet)}$, so
\begin{equation}
    S_I = \,\lambda \, \rho_{V(x^\bullet)}(x^\bullet, x_{\mathrm{tp}}), \qquad
    \rho_{E}(x_1, x_2) = \inf_{\gamma : x_1 \rightsquigarrow x_2} \int_{\gamma} \abs{\diff x} \sqrt{2 \max\lbrace V(x) - E, 0 \rbrace}
    ,
\label{eqn:agmon_distance}
\end{equation}
where in the definition of the \textit{Agmon distance}~\cite{26e649f4-ac21-3f96-8d07-d5d51780f780,carmona1981pointwise,hislop2000exponential} $\rho_E(x_1, x_2)$, the infimum is over piecewise-smooth curves joining $x_1$ and $x_2$. Thus $S_I$ is the Agmon distance between $x^\bullet$ and $x_{\mathrm{tp}}$ at reference energy $E = V(x^\bullet)$, rescaled by $\lambda$. The infimum is realized by the instanton, as can be seen from the Maupertuis principle: at fixed energy, the classical trajectory is a critical point of the abbreviated action, which is the weighted length appearing in~\eqref{eqn:agmon_distance}~\cite{landau1982mechanics}. In one dimension,~\eqref{eqn:instanton_action} reduces to $S_I = \sqrt{2}\lambda\int_{x^\bullet}^{x_{\mathrm{tp}}} \diff x \sqrt{V(x)-V(x^\bullet)}$ (assuming $x^\bullet < x_{\mathrm{tp}}$ without loss of generality).

A quantum particle escapes a metastable basin not by climbing the barrier but by tunneling through it, at a rate governed by the bounce action rather than the barrier height. This mechanism has itself been proposed as a basis for optimization~\cite{liu2023quantum}, with performance characteristics unlike those of SGD. We state the tunneling rate here, and the same action $B_V$ reappears in Section~\ref{sec:runtime} as the quantity controlling the escape rate of PISGD, a classical algorithm that inherits the tunneling mechanism.

\paragraph{Quantum tunneling rate.} Semiclassically, the tunneling rate $\Gamma$ from a local minimum $x^\bullet$ to the global minimum $x^\star$ across a potential barrier in between is~\cite{callan1977fate,vauinshteuin1982abc,schmid1986quasiclassical}
\begin{equation}
    \Gamma_{\mathrm{QT}} = \exp(- \lambda B_V) \sqrt{ \frac{\lambda B_V}{2 \pi} } \left(- \frac{\det' \left(- \partial_\tau^2 + \lambda^2 \nabla^2 V(x_{\mathrm{cl}}(\tau))\right)}{\det \left(- \partial_\tau^2 + \lambda^2 \nabla^2 V(x^\bullet)\right)} \right)^{-1/2}
\label{eqn:tunneling_rate}
\end{equation}
where $x_\mathrm{cl}$ is the bounce solution (the classical trajectory), $\nabla^2$ denotes the Hessian matrix, and the determinants are functional determinants\footnote{For a linear differential operator $\mathcal{M}$ with eigenvalues $\lbrace \mu_i \rbrace_i$, its functional determinant is $\det \mathcal{M} = \prod_i \mu_i$. This product typically diverges, and only the ratio of two such determinants is well defined: for operators $\mathcal{M}_1$ and $\mathcal{M}_2$ with eigenvalues $\lbrace \mu^{(1)}_i \rbrace_i$ and $\lbrace \mu^{(2)}_i \rbrace_i$ ordered consistently, $\det \mathcal{M}_1 / \det \mathcal{M}_2 = \prod_i \left( \mu^{(1)}_i / \mu^{(2)}_i \right)$, which can converge.} of the linear differential operators governing quadratic fluctuations about each solution. The prime on $\det'$ indicates that the zero eigenvalue is omitted; it arises from the time-translation invariance of the bounce, and its contribution is the prefactor $\sqrt{\lambda B_V / 2\pi}$. The overall minus sign inside the ratio accounts for the single negative eigenvalue of the fluctuation operator about the bounce, which makes the ratio positive.

We note that $B_V$ plays a dual role: it supplies the dominant exponential factor in the quantum tunneling rate, and it plays the same role in PISGD that the barrier height plays in classical thermal escape, \textit{e.g.}, stochastic gradient descent (SGD). The latter role correspondence underlies the analysis in the remainder of the paper.

The essential property of the bounce action is that $B_V$ depends on the full profile of $V$ between $x^\bullet$ and $x_{\mathrm{tp}}$, on both the width and the height of the barrier, whereas $H_V$ depends on the height alone. A tall, thin barrier can therefore have a much smaller Agmon distance than a low, broad one. This is the regime where PISGD is expected to outperform standard SGD.

\paragraph{From rate to escape time.} A rate is not yet the escape time, and the two are related by
a short argument. Semiclassically,
decay out of the metastable basin is a memoryless process since at each instant the escape is
equally likely regardless of how long the system has already waited. The survival
probability in the basin therefore decays exponentially (See~\cite{andreassen2017precision} as an introduction),
\begin{equation}
    P_{\mathrm{stay}} (t) = \exp \left( - \Gamma t \right)
    ,
\label{eqn:survival_probability}
\end{equation}
and, neglecting return events from the global minimum back across the barrier, the
first-passage time $t$ has density
\begin{equation}
    p (t) = - \frac{\diff P_{\mathrm{stay}} (t)}{\diff t} = \Gamma \exp \left( - \Gamma t \right)
    .
\label{eqn:first_passage_density}
\end{equation}
The escape time is thus exponentially distributed with rate $\Gamma$, so that
\begin{equation}
    \mathbb{E} [t] = \int_0^\infty \diff t \, t \, p(t) = \frac{1}{\Gamma}
    , \qquad
    \mathrm{median} [t] = \frac{\ln 2}{\Gamma}
    .
\label{eqn:hitting_time_moments}
\end{equation}
The escape time for quantum tunneling is thus estimated by inverting~\eqref{eqn:tunneling_rate},
\begin{equation}
    \mathcal{T}_{\mathrm{QT}} \sim \frac{1}{\Gamma_{\mathrm{QT}}} = \exp(\lambda B_V) \sqrt{ \frac{2 \pi}{\lambda B_V} } \left(- \frac{\det' \left(- \partial_\tau^2 + \lambda^2 \nabla^2 V(x_{\mathrm{cl}}(\tau))\right)}{\det \left(- \partial_\tau^2 + \lambda^2 \nabla^2 V(x^\bullet)\right)} \right)^{1/2}
    \sim \exp(\lambda B_V)
    ,
\label{eqn:T_QT}
\end{equation}
where the last estimate retains only the dominant exponential factor.

\subsection{PISGD and the Quantum Ground State}\label{sec:distribution_equivalence}

In the context of PISGD, we state the connection between the marginalized stationary distribution of the path integral and the quantum ground-state distribution, a fact studied in the physics literature~\cite{gattringer2009quantum}.

The stationary density of PISGD $\exp (- S (\{x_j\}_j; \widetilde{\lambda}) )$, marginalized to a single time slice $j = 0$ to give $p(x_0)$, asymptotically converges to the quantum ground-state density $\abs{\psi_0 (x_0)}^2$ in the continuum limit $a \rightarrow 0$ followed by the large-$T$ limit, assuming a nondegenerate ground state ($E_0 < E_1$):
\begin{equation}
    p(x_0) = \abs{\psi_0 (x_0)}^2 \left(1 + O\!\left(e^{-T(E_1 - E_0)}\right)\right),
\label{eqn:ground_state_correspondence}
\end{equation}
where $E_0 < E_1 \leq E_2 \leq \cdots$ are eigenvalues of the Hamiltonian $\hat{H} (\widetilde{\lambda})$ (implicitly depending on $\widetilde{\lambda}$).

The Euclidean-time evolution operator with a small lattice spacing $a$ ($a = T / N_T$) is
\begin{equation}
    \hat{U} (a) = \exp(- a \, \hat{H} (\widetilde{\lambda}) ) = \exp(- \frac{1}{2} a \,\widetilde{\lambda}^2  V (\hat{x}) ) \exp(- \frac{1}{2} a \, \hat{p}^2) \exp(- \frac{1}{2} a \, \widetilde{\lambda}^2 V (\hat{x}) ) (1 + O(a^2))
    ,
\end{equation}
so for any two points $x, x' \in \mathbb{R}^d$,
\begin{equation}
\begin{aligned}
    \mel{x'}{\hat{U} (a)}{x} = & \, \exp(- \frac{1}{2} a \, \widetilde{\lambda}^2 (V (x) + V(x')) ) \mel{x'}{\exp(- \frac{1}{2} a \, \hat{p}^2)}{x} (1 + O(a^2)) \\
    = & \, \exp(- \frac{1}{2} a \, \widetilde{\lambda}^2 (V (x) + V(x')) ) \int_{\mathbb{R}^d} \diff p \, \mel{x'}{\exp(- \frac{1}{2} a \, \hat{p}^2)}{p} \braket{p}{x} (1 + O(a^2)) \\
    = & \, \frac{1}{(2\pi)^d} \exp(- \frac{1}{2} a \, \widetilde{\lambda}^2 (V (x) + V(x')) ) \int_{\mathbb{R}^d} \diff p \, \exp(i p \cdot (x' - x)) \exp(- \frac{1}{2} a p^2) (1 + O(a^2)) \\
    = & \, \frac{1}{(2 \pi a)^{d/2}} \exp(- \frac{1}{2} a \, \widetilde{\lambda}^2 (V (x) + V(x')) ) \exp( - \frac{(x' - x)^2}{2 a}) (1 + O(a^2))
\end{aligned}
\end{equation}
where we have used the wavefunction of the momentum eigenstates $\ket{p}$ with $p \in \mathbb{R}^d$,
\begin{equation}
    \braket{x}{p} = \frac{1}{(2 \pi)^{d/2}} \exp(i p \cdot x)
    ,
\end{equation}
and the Gaussian integral $\int_{\mathbb{R}^d} \diff p \, \exp(i p \cdot y) \exp(- a p^2 / 2) = \left(2\pi / a\right)^{d/2} \exp\!\left(- y^2 / (2a)\right)$.

The diagonal entry of the quantum
density matrix $\hat{\rho}_T = \exp (- T \, \hat{H} (\widetilde{\lambda})) / Z_T$ is
\begin{equation}
\begin{aligned}
    Z_T \mel{x_0}{\hat{\rho}_T}{x_0} = & \, \mel{x_0}{\exp(- T \, \hat{H} (\widetilde{\lambda}))}{x_0} \\
    = & \, \lim_{N_T \rightarrow \infty} \int_{\mathbb{R}^{(N_T-1) d}} \diff x_1 \cdots \diff x_{N_T-1} \, \mel{x_0}{\hat{U}(a)}{x_1} \mel{x_1}{\hat{U}(a)}{x_2} \cdots \\ & \quad\quad\quad\quad \, \mel{x_{N_T-2}}{\hat{U}(a)}{x_{N_T-1}} \mel{x_{N_T-1}}{\hat{U}(a)}{x_0} (1 + O(N_T a^2))  \\
    = & \, \lim_{N_T \rightarrow \infty} \left( \frac{1}{2 \pi a} \right)^{N_T d / 2} \\
    & \quad\quad\quad\quad \int_{\mathbb{R}^{(N_T-1) d}} \diff x_1 \cdots \diff x_{N_T-1} \, \exp \left( - a \sum_{j=0}^{N_T-1} \left( \frac{\left(x_j - x_{j+1}\right)^2}{2 a^2} + \frac{1}{2} \widetilde{\lambda}^2 \left(V (x_j) + V(x_{j+1}) \right) \right) \right) \\
    = & \, \lim_{N_T \rightarrow \infty} \left( \frac{1}{2 \pi a} \right)^{N_T d / 2} \\
    & \quad\quad\quad\quad \int_{\mathbb{R}^{(N_T-1) d}} \diff x_1 \cdots \diff x_{N_T-1} \, \exp \left( - a \sum_{j=0}^{N_T-1} \left( \frac{\left(x_j - x_{j+1}\right)^2}{2 a^2} + \widetilde{\lambda}^2 V(x_j) \right) \right) \\
    = & \, \lim_{N_T \rightarrow \infty} \left( \frac{1}{2 \pi a} \right)^{N_T d / 2} \int \left.[\mathcal{D} x]\right|_{x(0) = x(T) = x_0} \, \exp \left(- S_E [x] \right)
\end{aligned}
\end{equation}
where in the second-to-last line, we have used $x_0 \equiv x_{N_T}$; in the last line, we formally denote the path integral measure for any path in the constrained space with $x(0) = x(T) = x_0$ as $\left.[\mathcal{D} x]\right|_{x(0) = x(T) = x_0}$, and the partition function is $Z_T$. In addition, we have used $\lim_{N_T \rightarrow \infty} O(N_T a^2) = T^2 \lim_{N_T \rightarrow \infty} O(1 / N_T) = 0$ with $T$ being a finite constant ($T = a N_T$).

The partition function $Z_T$ can similarly be written as 
\begin{equation}
\begin{aligned}
    Z_T
    = & \, \tr (\rho_T) \\
    = & \, \int_{\mathbb{R}^d} \diff x_0 \, \mel{x_0}{\exp(- T \, \hat{H} (\widetilde{\lambda}))}{x_0} \\
    = & \, \lim_{N_T \rightarrow \infty} \left( \frac{1}{2 \pi a} \right)^{N_T d / 2} \\
    & \quad\quad\quad\quad \int_{\mathbb{R}^{N_T d}} \diff x_0 \diff x_1 \cdots \diff x_{N_T-1} \, \exp \left( - a \sum_{j=0}^{N_T-1} \left( \frac{\left(x_j - x_{j+1}\right)^2}{2 a^2} + \frac{1}{2} \widetilde{\lambda}^2 \left(V (x_j) + V(x_{j+1}) \right) \right) \right) \\
    = & \, \lim_{N_T \rightarrow \infty} \left( \frac{1}{2 \pi a} \right)^{N_T d / 2} \int \mathcal{D} x \, \exp \left(- S_E [x] \right)
    .
\end{aligned}
\end{equation}

Formally, in terms of path integrals with constraints at the endpoints $\tau = 0$ and $\tau = T$, the density matrix diagonal element is
\begin{equation}
    \mel{x_0}{\hat{\rho}_T}{x_0} = \frac{ \int \left.[\mathcal{D} x]\right|_{x(0) = x(T) = x_0} \, \exp \left(- S_E [x] \right) }{Z_T} =  \frac{ \int \left.[\mathcal{D} x]\right|_{x(0) = x(T) = x_0} \, \exp \left(- S_E [x] \right) }{\int_{\mathbb{R}^d} \diff y \, \int \left.[\mathcal{D} x]\right|_{x(0) = x(T) = y} \, \exp \left(- S_E [x] \right)}
    ,
\end{equation}
which is the same as the marginal probability $p(x_0)$ from the PISGD stationary density with the limit $N_T \rightarrow \infty$.

Then we characterize $p(x_0)$ and associate it with the quantum ground-state probability
density as follows,
\begin{equation}
\begin{aligned}
    Z_T p(x_0) = & Z_T \mel{x_0}{\hat{\rho}_T}{x_0} \\
    = & \mel{x_0}{e^{- T \hat{H}}}{x_0} \\
    = & \sum_{k=0}^{\infty} e^{- T E_k} \braket{x_0}{k}\braket{k}{x_0} \\
    = & \sum_{k=0}^{\infty} e^{- T E_k} \abs{\psi_k (x_0)}^2 \\
    = & e^{- T E_0} \abs{\psi_0 (x_0)}^2 \left( 1 + O_T \left( e^{- T (E_1 - E_0)} \right) \right)
\end{aligned}
\end{equation}
where $\ket{k}$ is the $k$-th energy eigenstate of $\hat{H}$ with the normalized
wavefunction $\psi_k (x) = \braket{x}{k}$, ($\int_{\mathbb{R}^d} \diff x \, \abs{\psi_k (x)}^2 = 1$)
$\hat{H} \ket{k} = E_k \ket{k}$, $E_0 \leq E_1 \leq E_2 \leq \cdots$. In the asymptotic
analysis of large $T$, we have assumed there is no degeneracy in the ground state,
\textit{i.e.}, $E_0 < E_1$.

Since $p(x_0)$ and $\abs{\psi_0(x_0)}^2$ are both normalized to integrate to $1$ over $\mathbb{R}^d$, and their ratio converges pointwise to a constant as $T \rightarrow \infty$ by the preceding line, that constant must equal $1$, which gives~\eqref{eqn:ground_state_correspondence}, \textit{i.e.}, the marginal probability of $x_0$ under the stationary distribution of PISGD converges to the quantum ground-state probability density in the large-$T$ limit, after the continuum limit $a \rightarrow 0$.

\section{SGD Runtime Analysis Revisited}\label{sec:sgd_runtime}

Throughout this section we restate known results for a generic objective $f \colon \mathbb{R}^d \to \mathbb{R}$, and instantiate $f = \lambda V$ at the end to recover the scaling used in the rest of the paper. The stochastic gradient descent update rule is 
\begin{equation}
    x^{(k+1)} = x^{(k)} - s \nabla f (x^{(k)}) + \sigma \sqrt{s} \xi^{(k)}
\label{eqn:sgd_update_rules}
\end{equation}
where $s$ is the learning rate, $\sigma$ is the noise strength, and $\xi^{(k)} \sim \mathcal{N} (0, I_{d})$ is Gaussian noise\footnote{In the setting of machine learning~\cite{shi2023learning}, the noise term $\xi$ originates from minibatching of the input data, and $\sigma$ is taken proportional to $\sqrt{s}$. However, here for an optimization task with a directly given objective function rather than a loss function constructed from data, the noise strength $\sigma$ is a hyperparameter and can be chosen freely.}.

In the continuous-time limit $s \rightarrow 0$, Eq.~\eqref{eqn:sgd_update_rules} becomes Langevin dynamics. We restate and reuse the results of~\cite{shi2023learning}. Writing $t^{(k)} = k s$ for the time at the $k$-th step and modeling the iterates as samples $x^{(k)} = X (t^{(k)})$ of a sufficiently smooth curve $X(t)$, a first-order Taylor expansion gives
\begin{equation}
    x^{(k+1)} = X(t^{(k+1)}) = X(t^{(k)}) + \dot{X}(t^{(k)}) \, s + O(s^2).
\end{equation}
The noise $\xi^{(k)}$ can be written as 
\begin{equation}
    \sqrt{s} \xi^{(k)} = W(t^{(k+1)}) - W(t^{(k)}) = s \frac{\diff W(t^{(k)})}{\diff t} + O(s^2)
\end{equation}
where $W$ is a standard Brownian motion. Retaining only the leading-order terms in $s$ in Eq.~\eqref{eqn:sgd_update_rules} gives a stochastic differential equation (SDE)
\begin{equation}
    \diff X = - \nabla f(X) \diff t + \sigma \diff W,
\label{eqn:sgd_sde}
\end{equation}
also known as overdamped Langevin dynamics. We denote the probability density of $X(t)$ by $\rho(t, \cdot)$. The stationary distribution, also known as the Gibbs distribution, is 
\begin{equation}
    \mu = \frac{1}{Z} \exp(- \frac{2 f}{\sigma^2})
    ,
\end{equation}
where $Z = \int_{\mathbb{R}^d} \diff x \, \exp(-2 f / \sigma^2)$ normalizes the total probability to unity. We also define the norm associated with $\mu^{-1}$ as $\norm{g}_{\mu^{-1}} := \left(\int_{\mathbb{R}^d} \diff x \, g^2 \mu^{-1}\right)^{1/2}$ for a function $g$.

The dynamics of the density $\rho$ of the SDE Eq.~\eqref{eqn:sgd_sde} follows the Fokker-Planck equation,
\begin{equation}
    \frac{\partial \rho}{\partial t} = \nabla \cdot \left( \rho \nabla f \right) + \frac{\sigma^2}{2} \Delta \rho.
\label{eqn:sgd_fokker_planck}
\end{equation}
Eq.~\eqref{eqn:sgd_fokker_planck} can be reformulated as an imaginary-time Schr\"{o}dinger equation\footnote{This Schr\"{o}dinger equation and the attendant quantum-mechanical language describe SGD itself. They are distinct from the quantum tunneling and path-integral discussion in the rest of the paper.} as 
\begin{equation}
    \frac{\partial \psi}{\partial t} = \frac{\sigma^2}{2} \Delta \psi - \left( \frac{\norm{\nabla f}^2}{2 \sigma^2} - \frac{\Delta f}{2} \right) \psi = - \frac{1}{2 \sigma^2} \Delta_f \psi
\end{equation}
where $\psi$ is a wavefunction-like function defined as 
\begin{equation}
    \psi(t, \cdot) = \frac{\rho(t, \cdot)}{\sqrt{\mu}},
\end{equation}
and $\Delta_f$ is the Witten Laplacian
\begin{equation}
    \Delta_f = - \sigma^4 \Delta + \norm{\nabla f}^2 - \sigma^2 \Delta f
\label{eqn:witten-laplacian}
\end{equation}
with eigenvalues $0 = \zeta_0 < \zeta_1 \leq \zeta_2 \leq \cdots$.
When the stationary distribution is achieved, $\rho(t, \cdot) = \mu$, the analogue of the squared wavefunction becomes the stationary density: $\psi(t, \cdot)^2 = \mu$.

As shown in~\cite{shi2023learning}, $\rho(t, \cdot)$ converges to $\mu$ as $t$ increases, since
\begin{equation}
    \norm{\rho(t, \cdot) - \mu}_{\mu^{-1}} \leq \exp(- \Gamma_{\mathrm{SGD}} t) \norm{\rho(0, \cdot) - \mu}_{\mu^{-1}}
\end{equation}
where $\Gamma_{\mathrm{SGD}}$ is the exponential decay constant for SGD, given by the spectral gap $\zeta_1$ of the Witten Laplacian $\Delta_f$ (since $\zeta_0 = 0$), as $\Gamma_{\mathrm{SGD}} = \zeta_1 / 2$.

\paragraph{SGD decay rate and escape runtime.} As references~\cite{shi2023learning,herau2011tunnel} show, when a local minimum $x^\bullet$ and an index-one saddle point $x^\circ$ of $f$ are involved, the spectral gap is governed by the barrier height $H_f = f(x^\circ) - f(x^\bullet)$, up to a prefactor determined by the curvature at the two critical points. We also note that this is a well-known result, a version of the Kramers escape rate discussed in the physics literature~\cite{hanggi1990reaction}.

Instantiating $f = \lambda V$, matching the scaling of the potential term in the PISGD action~\eqref{eqn:discrete_action_lambda2}, this reads
\begin{equation}
    \Gamma_{\mathrm{SGD}} = \frac{\zeta_1}{2} = \exp(- \frac{2 \lambda H_V}{\sigma^2}) \left( \frac{\lambda \abs{\eta_V (x^\circ)}}{2 \pi} \left(-  \frac{\det \nabla^2 V(x^\circ)}{\det \nabla^2 V(x^\bullet)} \right)^{-1/2} + o(\sqrt{\sigma}) \right)
    ,
\label{eqn:sgd_decay_rate}
\end{equation}
where $H_V = V(x^\circ) - V(x^\bullet)$ is the barrier height of $V$ and $\eta_V (x^\circ)$ is the unique negative eigenvalue of the Hessian $\nabla^2 V(x^\circ)$. Only the barrier height appears in the exponent, and it enters linearly in $\lambda$.

Whenever the overdamped Langevin dynamics~\eqref{eqn:sgd_sde} describes SGD, the decay constant~\footnote{As with quantum tunneling in the semiclassical limit, treated in Section~\ref{sec:instantons}, the decay for SGD is exponential~\cite{shi2023learning}, so the escape time $t$ is exponentially distributed with rate $\Gamma_{\mathrm{SGD}}$, giving $\mathbb{E}[t] = 1 / \Gamma_{\mathrm{SGD}}$ and $\mathrm{median}[t] = \ln 2 / \Gamma_{\mathrm{SGD}}$.} sets the runtime. The characteristic escape time is $\Gamma_{\mathrm{SGD}}^{-1}$ in continuous time, and since each iteration advances the time by $s$, the runtime measured in iterations is
\begin{equation}
    \mathcal{T}_{\mathrm{SGD}} \sim \frac{1}{s \, \Gamma_{\mathrm{SGD}}} \sim \exp\!\left(\frac{2 \lambda H_V}{\sigma^2}\right) = \exp\!\left(\widetilde{\lambda}_{\mathrm{SGD}} H_V\right)
    ,
\label{eqn:sgd_runtime}
\end{equation}
with $\widetilde{\lambda}_{\mathrm{SGD}} := 2 \lambda / \sigma^2$, where the second relation retains only the dominant exponential factor. The barrier height $H_V$ therefore controls the runtime of SGD, with the prefactors contributing polynomially.

\subsection{Fixed-Noise SGD Convergence Requirement}
\label{sec:fixed-noise_sgd_convergence}

Although~\cite{shi2023learning} is primarily concerned with learning-rate decay and the noise-level decay of $\sigma$ inherited from it, analogous to the temperature annealing of simulated annealing, we consider the more restricted fixed-noise setting. This separates the underlying escape mechanism from the annealing schedule and gives a cleaner picture, although in practical optimization the interplay between the two can have further effects.%

An obstacle for noise-driven SGD in the optimization task~\eqref{eq:nonconvex-optimization-problem} is that $\sigma$ controls two competing effects. Increasing $\sigma$ speeds up escape from local minima, but it also widens the stationary distribution $\mu$, making it unlikely to draw a sample within the prescribed radius $\varepsilon$ of the global minimum $x^\star$. Using a large noise level first and a smaller one later to settle near the solution would resolve the tension, but designing such an annealing schedule without prior knowledge of the function can itself be difficult.

The stationary distribution of SGD, $\mu = (1/Z) \exp(-2 \lambda V / \sigma^2)$, shows that a large $\lambda$ plays a concentrating role. In the semiclassical regime of large $\lambda$, for a sufficiently smooth $V$ with a nondegenerate global minimum, $\mu$ is dominated by the neighborhood of $x^\star$ and is approximately Gaussian,
\begin{equation}
    \mu (x) \approx \frac{1}{Z} \exp(- \frac{2 \lambda}{\sigma^2} \left( V(x^\star) + \frac{1}{2} (x - x^\star)^{\top} \nabla^2 V(x^\star) (x - x^\star) \right))
    ,
\label{eqn:stationary_distribution_quadratic}
\end{equation}
where higher-order terms of the Taylor expansion have been omitted, and the Hessian $\nabla^2 V(x^\star)$ is assumed positive definite.

As Appendix~\ref{sec:convergence_from_gaussians} discusses, when the dimension $d$ is
large, a greater value of $\lambda$ is required for the probability of sampling in the
$L^2$-ball $B(x^\star, \varepsilon)$ to remain bounded away from zero rather than
decreasing exponentially in $d$. Applying the result of
Appendix~\ref{sec:convergence_from_gaussians} to
\eqref{eqn:stationary_distribution_quadratic}, whose covariance is the inverse of $P = (2
\lambda / \sigma^2) \nabla^2 V(x^\star)$, gives the threshold
\begin{equation}
    \widetilde{\lambda}_{\mathrm{SGD}} = \frac{2 \lambda}{\sigma^2} \gtrsim \frac{1}{\varepsilon^2} \tr \left[ \left( \nabla^2 V(x^\star) \right)^{-1} \right]
    .
\label{eqn:lambda_requirement_sgd}
\end{equation}
For a separable potential whose components all have curvature $(\omega^\star)^2$ at the
global minimum, $\tr [ ( \nabla^2 V(x^\star) )^{-1} ] = d / (\omega^\star)^2$ and the
requirement scales as
\begin{equation}
    \widetilde{\lambda}_{\mathrm{SGD}} \gtrsim \frac{1}{\varepsilon^2} \frac{d}{(\omega^\star)^2}
    .
\label{eqn:optimal_lambda_tilde_sgd}
\end{equation}

\paragraph{SGD optimal escape time.}
Since the runtime~\eqref{eqn:sgd_runtime} grows with $\widetilde{\lambda}_{\mathrm{SGD}}$
while the concentration requirement~\eqref{eqn:lambda_requirement_sgd} bounds it from
below, the best available choice is to saturate the threshold. Substituting
\eqref{eqn:lambda_requirement_sgd} into \eqref{eqn:sgd_runtime}, the optimal runtime to
escape from the local minimum is
\begin{equation}
    \mathcal{T}_{\mathrm{SGD}}^{\mathrm{opt}} \sim \exp\!\left( \frac{1}{\varepsilon^2} \tr \left[ \left( \nabla^2 V(x^\star) \right)^{-1} \right] H_V\right)
    .
\label{eqn:sgd_optimal_runtime}
\end{equation}
Further, in the isotropic case $\nabla^2 V(x^\star) = (\omega^\star)^2 I_d$, equivalently
using~\eqref{eqn:optimal_lambda_tilde_sgd},
\begin{equation}
    \mathcal{T}_{\mathrm{SGD}}^{\mathrm{opt}} \sim \exp\!\left( \frac{1}{\varepsilon^2} \frac{d}{(\omega^\star)^2} H_V\right)
    .
\label{eqn:sgd_optimal_runtime_isotropic}
\end{equation}
Hence for SGD the escape runtime is exponential in $d$, with the exponent proportional to
the barrier height $H_V$.

\subsection{Beyond a Single Barrier}
\label{sec:sgd_beyond_a_single_barrier}

The discussion above focuses on the case of a single potential barrier, with a local minimum $x^\bullet$, a saddle point $x^\circ$, and a global minimum $x^\star$. In a more general potential, there can be multiple local minima, multiple barriers, and multiple saddle points in between.

A typical class of problems to consider is separable potentials written as $V(x) = \sum_{i=0}^{d-1} V_i (x_i)$, where each component $V_i$ is a function of the single coordinate $x_i$ alone, and each $V_i$ itself can have the structure of a local minimum $x_i^\bullet$, a barrier top $x_i^\circ$, and a global minimum $x_i^\star$. For the total potential $V$, the global minimum is $x^\star = (x_0^\star, x_1^\star, \ldots, x_{d-1}^\star)$. The escape time in this case can similarly be analyzed through the spectral gap of the Witten Laplacian, instantiating~\eqref{eqn:witten-laplacian} with $f = V$ as
\begin{equation}
\begin{aligned}
    \Delta_V = &\, - \sigma^4 \Delta + \norm{\nabla V}^2 - \sigma^2 \Delta V \\
    = &\, - \sigma^4 \Delta + \sum_{i=0}^{d-1} \abs{V_i'(x_i)}^2 - \sigma^2 \sum_{i=0}^{d-1} V_i''(x_i) \\
    = &\, \sum_{i=0}^{d-1} \left( - \sigma^4 \partial_{x_i}^2 + \abs{V_i'(x_i)}^2 - \sigma^2 V_i''(x_i) \right) \\
    = & \, \sum_{i=0}^{d-1} \Delta_{V_i}
    .
\end{aligned}
\label{eqn:separable_witten_laplacian}
\end{equation}
The Witten Laplacian $\Delta_V$ is therefore the sum of the single-coordinate Witten Laplacians $\Delta_{V_i}$, each acting on its own coordinate $x_i$. Since these commute, the spectrum of $\Delta_V$ consists of the sums of the individual eigenvalues, and as every $\Delta_{V_i}$ has ground eigenvalue zero, the spectral gap of $\Delta_V$ is the minimum of the $d$ spectral gaps of the $\Delta_{V_i}$~\cite{herman2025mechanisms}. The worst-case escape rate to the global minimum is thus dominated by the slowest direction of escape among all $d$ dimensions, which in general occurs when the initial point of optimization is at $(x_0^\bullet, x_1^\bullet, \ldots, x_{d-1}^\bullet)$~\footnote{In general, an arbitrary initial point $x^{(0)} \in \mathbb{R}^d$ of optimization is allowed, and we assume that a simple gradient descent with a $d$-independent computational cost can be performed so that the initial point is always at a local minimum, which is utilized also by the convexity-honing algorithm from~\cite{herman2025mechanisms}.}. Throughout this paper, we focus on the worst-case analysis.

The same conclusion follows from a continuous-time Markov birth--death picture~\cite{feller1939grundlagen}, in which each step corresponds to a jump from $x_i^\bullet$ to $x_i^\star$ at a rate given by the decay rate in the direction of $x_i$. The overall rate from $(x_0^\bullet, x_1^\bullet, \ldots, x_{d-1}^\bullet)$ to $(x_0^\star, x_1^\star, \ldots, x_{d-1}^\star)$ is then dominated by the slowest of the $d$ directions.

Consequently the decay rate, which is proportional to the spectral gap, obeys
\begin{equation}
    \Gamma_{\mathrm{SGD}} \sim \min_{0 \leq i \leq d-1} \Gamma_{V_i, \mathrm{SGD}}
    ,
\label{eqn:sgd_decay_rate_separable}
\end{equation}
where $\Gamma_{V_i, \mathrm{SGD}}$ is the decay rate of the one-dimensional problem in the coordinate $x_i$ with potential $V_i$. Since runtimes are inversely proportional to decay rates, the worst-case runtime to escape to $x^\star$ is
\begin{equation}
    \mathcal{T}_{\mathrm{SGD}} \sim \exp\!\left( \widetilde{\lambda}_{\mathrm{SGD}} \max_{0 \leq i \leq d-1} H_{V_i} \right)
    ,
\label{eqn:sgd_runtime_separable}
\end{equation}
where $H_{V_i} = V_i(x_i^\circ) - V_i(x_i^\bullet)$ is the barrier height of the $i$-th component.

For the requirement of convergence near $x^\star$, the same condition as~\eqref{eqn:lambda_requirement_sgd} applies and reduces to
\begin{equation}
    \widetilde{\lambda}_{\mathrm{SGD}} \gtrsim \frac{1}{\varepsilon^2} \tr \left[ \left( \nabla^2 V(x^\star) \right)^{-1} \right] = \frac{1}{\varepsilon^2} \sum_{i=0}^{d-1} \frac{1}{V_i''(x_{i}^\star)}
    ;
\label{eqn:lambda_requirement_sgd_separable}
\end{equation}
and for the special case where the $V_i$ are all identical with $V_i''(x_i^\star) = (\omega^\star)^2$, further to
\begin{equation}
    \widetilde{\lambda}_{\mathrm{SGD}} \gtrsim \frac{1}{\varepsilon^2} \frac{d}{(\omega^\star)^2}
    .
\label{eqn:lambda_requirement_sgd_separable_isotropic}
\end{equation}

Therefore, for a separable potential, the optimal runtime to escape from the worst case to $x^\star$ with the saturated choice of $\widetilde{\lambda}_{\mathrm{SGD}}$ is
\begin{equation}
    \mathcal{T}_{\mathrm{SGD}}^{\mathrm{opt}} \sim \exp\!\left( \frac{1}{\varepsilon^2} \left(\sum_{i=0}^{d-1} \frac{1}{V_i''(x_{i}^\star)}\right) \max_i H_{V_i} \right)
    ,
\label{eqn:sgd_optimal_runtime_separable}
\end{equation}
using~\eqref{eqn:lambda_requirement_sgd_separable}. Further, when the $V_i$ are all identical, so that $\nabla^2 V(x^\star) = (\omega^\star)^2 I_d$, this reduces to
\begin{equation}
    \mathcal{T}_{\mathrm{SGD}}^{\mathrm{opt}} \sim \exp\!\left( \frac{1}{\varepsilon^2} \frac{d}{(\omega^\star)^2} H_V\right)
    ,
\label{eqn:sgd_optimal_runtime_isotropic_separable}
\end{equation}
where $H_V$ is the common value of $H_{V_i}$.

For the separable potentials above, once separability and the separating coordinates are known, optimizing each dimension individually makes the problem easy. However, as~\cite{leng2023quantumnonconvex,herman2025mechanisms} point out, a rotation hides this information, while Langevin dynamics depends on the potential only through its gradient and is therefore covariant (equivariant) under rotations of the coordinates. Strictly speaking we should then consider separable potentials in rotated coordinates, so that the separating directions are hidden. By covariance, however, a rotated instance is exactly as hard for the algorithm as the unrotated one, so doing so leaves the results unchanged and only blurs the presentation. Throughout this paper we therefore work with explicitly separable potentials without an artificial rotation, while never using separability or the separating coordinates in the optimization itself.

Beyond separable potentials, an explicit analysis becomes considerably more challenging. We postpone it to Section~\ref{sec:nonseparable_potentials}, where we treat a special case of nonseparable potentials explicitly, for both SGD and PISGD.

\section{PISGD Runtime Analysis}\label{sec:runtime}

PISGD is the SGD update rule applied to the discretized action $S (\{x_j\}_j;\lambda)$ in place of the objective function $f$, so one might expect the decay rate to follow directly from Eq.~\eqref{eqn:sgd_decay_rate}. However, two caveats exist as follows.

First, Eq.~\eqref{eqn:sgd_decay_rate} assumes that the Hessian at the saddle to be crossed is index one, with a unique negative eigenvalue and no zero eigenvalues. For PISGD the lowest saddle of $S (\{x_j\}_j;\lambda)$ is the bounce, whose fluctuation operator (a functional double derivative, analogous to the Hessian) has an exact zero eigenvalue: translating the center $\tau_c$ of the bounce in imaginary time costs no action, so the bounce comes in a one-parameter family rather than in isolation. The Hessian is therefore degenerate, and the determinant in the prefactor of~\eqref{eqn:sgd_decay_rate} vanishes.

Second, on a finite lattice PISGD optimizes over the finite-dimensional space $\mathbb{R}^{d \times N_T}$, but the analysis is most naturally carried out in the continuum limit $a \rightarrow 0$, where the configuration space becomes infinite-dimensional and the fluctuation determinants must be interpreted as functional determinants.

We therefore follow Langer's theory~\cite{langer1969statistical,hong2022introduction}, which computes the Kramers escape rate in the presence of zero modes at the saddle by treating the associated collective coordinate as a volume factor rather than a Gaussian integral. We first review and state the main result of Langer's theory, then apply it to PISGD.

A closely related line of work treats stochastic partial differential equations (SPDEs)
with one-dimensional potentials~\cite{berglund2013sharp,barret2015sharp}, which
corresponds to PISGD with a one-dimensional potential $V$ in the doubly continuous limit,
both in iteration time and in imaginary time.

\subsection{Kramers' Escape Rates from Langer's Theory}

Langer's theory obtains the escape rate as a \emph{flux over population} ratio: one constructs the steady state that is fed at the local minimum $x^\bullet$ and absorbed beyond the barrier, integrates the resulting probability current across the dividing surface through the saddle $x^\circ$, and divides by the population in the well. Near $x^\circ$ the steady state is Gaussian in the transverse directions, so the flux reduces to a product of Gaussian integrals, one per fluctuation mode.

This decomposition is what accommodates a zero mode. Writing the eigenvalues of $\nabla^2 f(x^\circ)$ as $\eta < 0 = \nu_1 < \nu_2 \leq \cdots$, the transverse mode with $\nu_1 = 0$ contributes the extent $L$ of the flat direction rather than the Gaussian integral produced by a positive mode. Removing that mode from the determinant, denoted $\det'$, and replacing it by the factor $L$, the escape rate for the overdamped Langevin dynamics~\eqref{eqn:sgd_sde} is
\begin{equation}
    \Gamma = \frac{\abs{\eta}\, L}{(2 \pi)^{3/2}} \sqrt{\frac{2}{\sigma^2}} \, \exp(- \frac{2 H_f}{\sigma^2}) \left( - \frac{{\det}' \nabla^2 f(x^\circ)}{\det \nabla^2 f(x^\bullet)} \right)^{-1/2}
    ,
\label{eqn:langer_rate}
\end{equation}
where $H_f = f(x^\circ) - f(x^\bullet)$ as usual. If instead $\nabla^2 f(x^\circ)$ has no zero eigenvalue, the same computation retains the corresponding Gaussian integral, $\det'$ becomes the full determinant, and~\eqref{eqn:langer_rate} reduces to~\eqref{eqn:sgd_decay_rate}. The derivation in Appendix~\ref{sec:langer_theory} is restated from the modern exposition of~\cite{hong2022introduction}.

\subsection{PISGD Analyzed by Langer's Theory}

One caveat in applying Langer's theory to PISGD is that the theory is formulated on a
finite-dimensional configuration space, whereas continuous imaginary time makes the path
space infinite-dimensional. The algorithmic implementation of PISGD, however, always
discretizes imaginary time into $N_T$ slices and therefore operates in finite dimension,
where the theory applies directly. We accordingly treat the continuum expressions below
as the $a \rightarrow 0$ limit of the finite-dimensional result, which is the limit the
implementation approaches as the lattice is refined.

Applying the theory requires two substitutions. The objective $f$ is replaced by the
action $S(\{x_j\}_j; \lambda)$, and the configuration space $\mathbb{R}^d$ is replaced by
the path space of maps $\mathbb{R} \rightarrow \mathbb{R}^d$. The bounce solution, constituted from an instanton--anti-instanton pair, is the saddle point in the path space, and its zero mode is the imaginary-time translation mode, which supplies a factor proportional to the extent $T$ of imaginary time in place of $L$. Correspondingly, the
Hessians in~\eqref{eqn:langer_rate} become fluctuation operators of $S$, their
determinants become functional determinants, and the barrier height $H_f$ becomes the
bounce action $B_V$.

\subsubsection{Functional Determinants of Fluctuation Operators}

Fluctuation operators of the action $S$ in the continuous imaginary time case can be understood as the $a \rightarrow 0$ limit of the Hessian of the discretized version of $S$ defined in~\eqref{eqn:discrete_action_lambda2}
\begin{align}
    S \!\left(\{x_j\}_j;\lambda\right)
    = a \sum_{j=0}^{N_T-1} \left(
        \frac{1}{2a^2}\left(x_j - x_{j+1}\right)^2
        + \lambda^2 V(x_j)
      \right).
\end{align}
Then the Hessian is
\begin{align}
    \frac{\partial^2}{\partial x_{k} \partial x_{l}} S \!\left(\{x_j\}_j;\lambda\right)
    = \frac{1}{a}
        \left(- \delta_{k, l-1} + 2 \delta_{k l} -  \delta_{k, l+1}\right)
        + a \lambda^2 \frac{\partial^2}{\partial x_{k} \partial x_{l}} V(x_{k})
\label{eqn:hesssian_S}
\end{align}
where $\delta_{ij}$ is the Kronecker delta. In the limit of small $a$,
\begin{align}
    \frac{1}{a} \frac{\partial^2}{\partial x_{k} \partial x_{l}} S \!\left(\{x_j\}_j;\lambda\right)
    = \frac{1}{a^2}
        \left(- \delta_{k, l-1} + 2 \delta_{k l} -  \delta_{k, l+1}\right)
        + \lambda^2 \frac{\partial^2}{\partial x_{k} \partial x_{l}} V(x_{k}) \sim - \frac{\partial^2}{\partial \tau^2} + \lambda^2 \nabla^2 V(x(\tau))
    .
\end{align}
The differential operator on the right side is often referred to as the double functional derivative for the continuous-imaginary-time action $S$
\begin{equation}
    \frac{\delta^2 S}{\delta x^2} := - \frac{\partial^2}{\partial \tau^2} + \lambda^2 \nabla^2 V(x(\tau))
    ,
\label{eqn:fluctuation_operator}
\end{equation}
which is also called the \textit{fluctuation operator}~\cite{dunne2005beyond}.
Note that the limit $a \rightarrow 0$ of the Hessian~\eqref{eqn:hesssian_S} does not
reproduce $\delta^2 S / \delta x^2$ directly; an additional factor of $1/a$ is required, as
made explicit above.

\subsubsection{The Extent from the Zero Mode}

At first glance the extent $L$ in~\eqref{eqn:langer_rate} should simply be the extent $T$ of imaginary time, since the coordinate associated with imaginary-time translation is $\tau_c$, the center of the bounce, with $\int_{[0, T]} \diff \tau_c = T$. This requires care, however, because the integral in Langer's theory is performed in a specific coordinate system, and a linear change of coordinates contributes a Jacobian.

Callan and Coleman in~\cite{callan1977fate} discuss the orthonormal basis of the fluctuation operator. A path $x(\tau)$ can be expressed as a linear combination of the eigenmodes $\xi_n(\tau)$ of the fluctuation operator at $x_\mathrm{cl}$ as 
\begin{equation}
    x(\tau) = x_{\mathrm{cl}} (\tau) + \sum_{n=0}^\infty c_n \xi_n (\tau)
\end{equation}
where
\begin{equation}
    \left( - \frac{\partial^2}{\partial \tau^2} + \lambda^2 \nabla^2 V(x_{\mathrm{cl}}(\tau)) \right) \xi_n (\tau) = \nu_n \xi_n (\tau)
\end{equation}
and normalized as
\begin{equation}
    \int_{[0, T]} \diff \tau \, \xi_m (\tau) \xi_n (\tau) = \delta_{mn}
    .
\label{eqn:continuum_normalization}
\end{equation}
Then the coefficients $\lbrace c_n \rbrace_{n=0}^{\infty}$ are the coordinates with respect to an orthonormal basis that diagonalizes the fluctuation operator, and can serve as the coordinates in the Langer's theory analysis.

The center of the bounce $\tau_c$ is associated with the zero-mode coefficient $c_1$ as
\begin{equation}
    \xi_1 (\tau) = \left(\lambda B_V\right)^{-1/2} \frac{\diff x_{\mathrm{cl}}}{\diff \tau}
\end{equation}
and thus also~\cite{callan1977fate}
\begin{equation}
    \diff c_1 = \sqrt{\lambda B_V} \diff \tau_c
    ,
\end{equation}
where $\int_{[0,T]} \diff \tau \, (\diff x_{\mathrm{cl}} / \diff \tau)^2 = \lambda B_V$ for the zero-energy bounce.

A further subtlety is that the normalization~\eqref{eqn:continuum_normalization} is the one natural for the path-integral measure, whereas Langer's theory is carried out in the discrete coordinates $\{x_j\}_j$ with the flat measure $\prod_j \diff x_j$, in which the modes must be orthonormal with respect to $\sum_k \xi_m (\tau_k) \xi_n (\tau_k) = \delta_{mn}$. Since $\int \diff \tau = a \sum_k$, the two conventions differ by $\xi_n^{\mathrm{disc}} = \sqrt{a} \, \xi_n$, and correspondingly the coefficients by $c_1^{\mathrm{disc}} = c_1 / \sqrt{a}$. Therefore, the volume integral of the zero-mode coordinate does not give $T$ but instead
\begin{equation}
    \int \diff c_1^{\mathrm{disc}} = \frac{1}{\sqrt{a}} \sqrt{\lambda B_V} \int_{[0, T]}\diff \tau_c = \sqrt{\frac{\lambda B_V}{a}} \, T
    .
\label{eqn:zero_mode_extent}
\end{equation}

\subsubsection{PISGD Escape Rates and Runtimes}

Substituting the quantities for PISGD into~\eqref{eqn:langer_rate}, with the
extent~\eqref{eqn:zero_mode_extent} and the unique negative eigenvalue of the Hessian at the
saddle, gives
\begin{equation}
\begin{aligned}
    \Gamma_{\mathrm{PISGD}} = & \, \exp(- \frac{2 \lambda B_V}{\sigma^2}) \frac{\abs{\eta_{S(\lambda)}^{\mathrm{disc}}}\, \sqrt{\lambda B_V / a} \, T}{(2 \pi)^{3/2}} \sqrt{\frac{2}{\sigma^2}} \,  \left( - \frac{{\det}' \nabla^2 S \!\left(\{(x_{\mathrm{cl}})_j\}_j;\lambda\right)}{\det \nabla^2 S \!\left(\{(x^\bullet)_j\}_j;\lambda\right)} \right)^{-1/2} \\
    = & \, \exp(- \frac{2 \lambda B_V}{\sigma^2}) \, a \, \frac{\abs{\eta_{S(\lambda)}}\, \sqrt{B_V} T}{(2 \pi)^{3/2}} \sqrt{\frac{2 \lambda}{\sigma^2}} \,  \left( - \frac{{\det}' \delta^2 S \!\left[x_{\mathrm{cl}} ;\lambda\right]}{\det \delta^2 S \!\left[x^\bullet;\lambda\right]} \right)^{-1/2} \\
    = &\, \exp(- \frac{2 \lambda B_V}{\sigma^2}) \, a \, \frac{\abs{\eta_{S(\lambda)}}\, \sqrt{B_V} T}{(2 \pi)^{3/2}} \sqrt{\frac{2 \lambda}{\sigma^2}} \,  \left( - \frac{{\det}' \left( - \frac{\partial^2}{\partial \tau^2} + \lambda^2 \nabla^2 V(x_{\mathrm{cl}}(\tau))\right)}{\det \left( - \frac{\partial^2}{\partial \tau^2} + \lambda^2 \nabla^2 V(x^\bullet) \right)} \right)^{-1/2} \\
    = & \, \exp(- \widetilde{\lambda} B_V) \, a \, \left(\frac{\sigma^2}{2}\right)^{3} \, \frac{\abs{\eta_{S(\widetilde{\lambda})}}\, \sqrt{B_V} T}{(2 \pi)^{3/2}} \sqrt{\widetilde{\lambda}} \,  \left( - \frac{{\det}' \left( - \frac{\partial^2}{\partial \tau^2} + \widetilde{\lambda}^2 \nabla^2 V(x_{\mathrm{cl}}(\tau))\right)}{\det \left( - \frac{\partial^2}{\partial \tau^2} + \widetilde{\lambda}^2 \nabla^2 V(x^\bullet) \right)} \right)^{-1/2}
    ,
\end{aligned}
\label{eqn:Gamma_PISGD}
\end{equation}
Throughout, $\eta$ denotes the unique negative eigenvalue of the fluctuation operator at the saddle, with the subscript recording the action and the scaling parameter it is evaluated at, and the superscript whether it belongs to the discrete or the continuum problem: the first line uses the discrete $\eta_{S(\lambda)}^{\mathrm{disc}} = a \, \eta_{S(\lambda)}$ inherited from~\eqref{eqn:hesssian_S}, the middle lines its continuum counterpart $\eta_{S(\lambda)}$, and the last line the same continuum eigenvalue expressed at the normalized hyperparameter, $\eta_{S(\widetilde{\lambda})}$.
The last line is expressed in the normalized hyperparameter $\widetilde{\lambda} = 2 \lambda / \sigma^2$ of~\eqref{eqn:hyperparameter_normalization}, using that the eigenvalues of the fluctuation operator scale as $\lambda^2$~\footnote{Rescaling the potential as $V \to c V$ sends the bounce to $x_{\mathrm{cl}}(\sqrt{c} \, \tau)$ and each normalized mode of $- \partial_\tau^2 + \nabla^2 V(x_{\mathrm{cl}})$ to $c^{1/4} \xi_n (\sqrt{c} \, \tau)$, which multiplies every eigenvalue by $c$. The potential enters the action as $\lambda^2 V$, so $c = \lambda^2$.}. The prefactor then collects $(\sigma^2/2)^2$ from $\abs{\eta_{S(\lambda)}}$ and a further $(\sigma^2/2)$ from the determinant ratio.
The three sources of $a$ combine into the single factor: (1) the determinant ratio
contributes $a^{1/2}$, since the numerator omits one mode relative to the denominator; (2)
the explicit $\abs{\eta_{S(\lambda)}^{\mathrm{disc}}}$ contributes $a$; (3) the zero-mode
extent~\eqref{eqn:zero_mode_extent} contributes $a^{-1/2}$. The residual linear factor is
expected, as the PISGD gradient $\partial S / \partial x_j = a \, \delta S / \delta x$
advances the continuum time by $a$ per iteration, so that a rate measured per iteration
carries one factor of $a$ relative to the continuum rate.
The runtime is the inverse of the tunneling rate as
\begin{equation}
\begin{aligned}
    \mathcal{T}_{\mathrm{PISGD}} \sim \frac{1}{s \Gamma_{\mathrm{PISGD}}} = &\, \exp(\widetilde{\lambda} B_V) \, \frac{1}{s a} \left(\frac{2}{\sigma^2}\right)^{3} \frac{(2 \pi)^{3/2}}{\abs{\eta_{S(\widetilde{\lambda})}} \sqrt{\widetilde{\lambda} B_V} \, T} \left( - \frac{{\det}' \left( - \frac{\partial^2}{\partial \tau^2} + \widetilde{\lambda}^2 \nabla^2 V(x_{\mathrm{cl}}(\tau))\right)}{\det \left( - \frac{\partial^2}{\partial \tau^2} + \widetilde{\lambda}^2 \nabla^2 V(x^\bullet) \right)} \right)^{1/2} \\
    \sim &\, \exp(\widetilde{\lambda} B_V)
    .
\end{aligned}
\label{eqn:T_PISGD}
\end{equation}
As in the introduction, the exponential factor governing the escape time of PISGD is controlled by the bounce action $B_V$, as in the quantum tunneling case~\eqref{eqn:T_QT}, rather than by the barrier height $H_V$ that governs SGD in~\eqref{eqn:sgd_optimal_runtime}.

\subsubsection{Fixed-Noise Convergence Requirement}

PISGD faces the same tradeoff as SGD in Section~\ref{sec:fixed-noise_sgd_convergence}. A smaller $\widetilde{\lambda}$ makes escape faster, while a larger one concentrates the stationary distribution more tightly near $x^\star$. Here the noise strength $\sigma$ enters only through $\widetilde{\lambda} = 2 \lambda / \sigma^2$, so we again adopt the fixed-noise setting and treat $\widetilde{\lambda}$ as the single hyperparameter to be chosen.

The relevant stationary distribution is the one established in Section~\ref{sec:distribution_equivalence}: the marginal $p(x_0)$ of a single time slice converges to the quantum ground-state density $\abs{\psi_0(x_0)}^2$ of $\hat{H}(\widetilde{\lambda})$, up to the relative correction $O(e^{-T(E_1 - E_0)})$ recorded in~\eqref{eqn:ground_state_correspondence}. The concentration requirement is therefore a condition on the quantum ground-state probability distribution $\abs{\psi_0}^2$.

Similar to the analysis in Section~\ref{sec:fixed-noise_sgd_convergence}, in the semiclassical limit where $\widetilde{\lambda}$ is large, the PISGD stationary distribution approaches $\abs{\psi_0}^2$ and is dominated by the structure near the global minimum $x^\star$. There the potential is approximated by its quadratic expansion $V(x) \approx V(x^\star) + \frac{1}{2} (x - x^\star)^{\top} \nabla^2 V(x^\star) (x - x^\star)$, so that $\hat{H}(\widetilde{\lambda})$ in~\eqref{eqn:hamiltonian} reduces to a harmonic oscillator with matrix $\widetilde{\lambda} \sqrt{\nabla^2 V(x^\star)}$, whose ground state reads~\cite{srednicki1993entropy}
\begin{equation}
    \psi_0 (x) = \pi^{-d/4} \left( \det \left( \widetilde{\lambda} \sqrt{\nabla^2 V(x^\star)} \right) \right)^{1/4} \exp(- \frac{\widetilde{\lambda}}{2} (x - x^\star)^\top \sqrt{\nabla^2 V(x^\star)} \, (x - x^\star))
\label{eqn:harmonic_ground_state}
\end{equation}
where the square root of the Hessian is the matrix square root: since $x^\star$ is a minimum, $\nabla^2 V(x^\star)$ is symmetric positive definite and admits the spectral decomposition $\nabla^2 V(x^\star) = \sum_{i=1}^d \omega_i^2 \, u_i u_i^\top$ with orthonormal eigenvectors $\lbrace u_i \rbrace_i$ and $\omega_i > 0$, and $\sqrt{\nabla^2 V(x^\star)} := \sum_{i=1}^d \omega_i \, u_i u_i^\top$.

Therefore, in the semiclassical limit, the stationary distribution for $x_0$ in PISGD is 
\begin{equation}
    p(x_0) = \abs{\psi_0(x_0)}^2 = \pi^{-d/2} \left( \det \left( \widetilde{\lambda} \sqrt{\nabla^2 V(x^\star)} \right) \right)^{1/2} \exp(- \widetilde{\lambda} (x_0 - x^\star)^\top \sqrt{\nabla^2 V(x^\star)} \, (x_0 - x^\star))
    ,
\label{eqn:stationary_distribution_quadratic_pisgd}
\end{equation}
which is again a Gaussian measure, now with $2 \widetilde{\lambda} \sqrt{\nabla^2 V(x^\star)}$ in place of the $\widetilde{\lambda} \nabla^2 V(x^\star)$ of~\eqref{eqn:stationary_distribution_quadratic}.

The derivation for SGD in Section~\ref{sec:fixed-noise_sgd_convergence} then applies verbatim. Applying the result of
Appendix~\ref{sec:convergence_from_gaussians} to
\eqref{eqn:stationary_distribution_quadratic_pisgd}, whose covariance is the inverse of $P = 2 \widetilde{\lambda} \sqrt{\nabla^2 V(x^\star)}$, gives the threshold
\begin{equation}
    \widetilde{\lambda}_{\mathrm{PISGD}} \gtrsim \frac{1}{2 \varepsilon^2} \tr \left[ \left( \sqrt{\nabla^2 V(x^\star)} \right)^{-1} \right]
    ,
\label{eqn:lambda_requirement_pisgd}
\end{equation}
where the subscript is restored to distinguish the PISGD hyperparameter from that of SGD.
For a separable potential whose components all have curvature $(\omega^\star)^2$ at the
global minimum, $\tr [ ( \sqrt{\nabla^2 V(x^\star) } )^{-1} ] = d / \omega^\star$ and the
requirement scales as
\begin{equation}
    \widetilde{\lambda}_{\mathrm{PISGD}} \gtrsim \frac{1}{2 \varepsilon^2} \frac{d}{\omega^\star}
    ,
\label{eqn:optimal_lambda_tilde_pisgd}
\end{equation}
which carries the same linear growth in $d$ as~\eqref{eqn:optimal_lambda_tilde_sgd}, but depends on $\omega^\star$ rather than on $(\omega^\star)^2$.

\paragraph{PISGD optimal escape time.}
Since the runtime~\eqref{eqn:T_PISGD} grows with $\widetilde{\lambda}_{\mathrm{PISGD}}$
while the concentration requirement~\eqref{eqn:lambda_requirement_pisgd} bounds it from
below, the best available choice is to saturate the threshold. Substituting
\eqref{eqn:lambda_requirement_pisgd} into \eqref{eqn:T_PISGD}, the optimal runtime to
escape from the local minimum is
\begin{equation}
    \mathcal{T}_{\mathrm{PISGD}}^{\mathrm{opt}} \sim \exp\!\left( \frac{1}{2 \varepsilon^2} \tr \left[ \left( \sqrt{\nabla^2 V(x^\star)} \right)^{-1} \right] B_V\right)
    .
\label{eqn:pisgd_optimal_runtime}
\end{equation}
Further, in the isotropic case $\nabla^2 V(x^\star) = (\omega^\star)^2 I_d$, equivalently
using~\eqref{eqn:optimal_lambda_tilde_pisgd},
\begin{equation}
    \mathcal{T}_{\mathrm{PISGD}}^{\mathrm{opt}} \sim \exp\!\left( \frac{1}{2 \varepsilon^2} \frac{d}{\omega^\star} B_V\right)
    .
\label{eqn:pisgd_optimal_runtime_isotropic}
\end{equation}
Therefore, for PISGD the escape runtime is again exponential in $d$, but with the exponent proportional to
the bounce action $B_V$ rather than the barrier height $H_V$ of~\eqref{eqn:sgd_optimal_runtime_isotropic}.

\subsection{Beyond a Single Barrier}
\label{sec:pisgd_beyond_a_single_barrier}

As in Section~\ref{sec:sgd_beyond_a_single_barrier}, we view PISGD as an instantiation of SGD, albeit with more subtleties and caveats. For separable potentials $V (x) = \sum_{i=0}^{d-1} V_i (x_i)$, the PISGD decay rate from the worst initial point, in the convex region containing $(x_0^\bullet, x_1^\bullet, \ldots, x_{d-1}^\bullet)$, to $x^\star = (x_0^\star, x_1^\star, \ldots, x_{d-1}^\star)$ obeys
\begin{equation}
    \Gamma_{\mathrm{PISGD}} \sim \min_{0 \leq i \leq d-1} \Gamma_{V_i, \mathrm{PISGD}}
    ,
\label{eqn:pisgd_decay_rate_separable}
\end{equation}
where $\Gamma_{V_i, \mathrm{PISGD}}$ is the decay rate of the one-dimensional problem in the coordinate $x_i$ with potential $V_i$. Since runtimes to escape are inversely proportional to decay rates, the worst-case runtime to escape to $x^\star$ is
\begin{equation}
    \mathcal{T}_{\mathrm{PISGD}} \sim \exp\!\left( \widetilde{\lambda}_{\mathrm{PISGD}} \max_{0 \leq i \leq d-1} B_{V_i} \right)
    ,
\label{eqn:pisgd_runtime_separable}
\end{equation}
where $B_{V_i}$ is the bounce action associated with the one-dimensional potential $V_i$.

For the requirement of convergence near $x^\star$, the same condition as~\eqref{eqn:lambda_requirement_pisgd} applies and reduces to
\begin{equation}
    \widetilde{\lambda}_{\mathrm{PISGD}} \gtrsim \frac{1}{2 \varepsilon^2} \tr \left[ \left( \sqrt{\nabla^2 V(x^\star)} \right)^{-1} \right] = \frac{1}{2 \varepsilon^2} \sum_{i=0}^{d-1} \frac{1}{\sqrt{V_i''(x_{i}^\star)}}
    ;
\label{eqn:lambda_requirement_pisgd_separable}
\end{equation}
and for the special case where the $V_i$ are all identical with $V_i''(x_i^\star) = (\omega^\star)^2$, further to
\begin{equation}
    \widetilde{\lambda}_{\mathrm{PISGD}} \gtrsim \frac{1}{2 \varepsilon^2} \frac{d}{\omega^\star}
    .
\label{eqn:lambda_requirement_pisgd_separable_isotropic}
\end{equation}

Therefore, for a separable potential, the optimal worst-case runtime to escape to $x^\star$, with the saturated choice of $\widetilde{\lambda}_{\mathrm{PISGD}}$, is
\begin{equation}
    \mathcal{T}_{\mathrm{PISGD}}^{\mathrm{opt}} \sim \exp\!\left( \frac{1}{2 \varepsilon^2} \left( \sum_{i=0}^{d-1} \frac{1}{\sqrt{V_i''(x_{i}^\star)}} \right) \max_{0 \leq i \leq d-1} B_{V_i} \right)
    ,
\label{eqn:pisgd_optimal_runtime_separable}
\end{equation}
using~\eqref{eqn:lambda_requirement_pisgd_separable}. Further, when the $V_i$ are all identical, so that $\nabla^2 V(x^\star) = (\omega^\star)^2 I_d$, this reduces to
\begin{equation}
    \mathcal{T}_{\mathrm{PISGD}}^{\mathrm{opt}} \sim \exp\!\left( \frac{1}{2 \varepsilon^2} \frac{d}{\omega^\star} B_V\right)
    ,
\label{eqn:pisgd_optimal_runtime_isotropic_separable}
\end{equation}
where $B_V$ is the common value of $B_{V_i}$.

Nonseparable potentials are considerably more difficult to analyze. Section~\ref{sec:nonseparable_potentials} treats a special case explicitly, for both SGD and PISGD.

\subsection{Quantum Tunneling Revisited}

The subsections above concern PISGD, which has much in common with quantum tunneling. Without an annealing schedule, the quantum Hamiltonian is $\hat{H} = - \frac{1}{2} \Delta + \lambda^2 V(\hat{x})$, and the decay rate and escape time for a single barrier follow from~\eqref{eqn:tunneling_rate},
\begin{equation}
    \Gamma_{\mathrm{QT}} \sim \exp(- \lambda B_V)
    , \quad \mathcal{T}_{\mathrm{QT}} \sim \exp(\lambda B_V)
    .
\label{eqn:qt_rate_and_runtime}
\end{equation}
Suppose now that we require the ground-state wavefunction to be concentrated within the ball $B(x^\star, \varepsilon)$. In the semiclassical limit of large $\lambda$, the potential near $x^\star$ is dominated by its second-order Taylor expansion, and the ground-state probability density becomes
\begin{equation}
    p_{\mathrm{GS}}(x) = \abs{\psi_0(x)}^2 = \pi^{-d/2} \left( \det \left( \lambda \sqrt{\nabla^2 V(x^\star)} \right) \right)^{1/2} \exp(- \lambda (x - x^\star)^\top \sqrt{\nabla^2 V(x^\star)} \, (x - x^\star))
    ,
\label{eqn:ground_state_distribution_qm}
\end{equation}
which matches~\eqref{eqn:stationary_distribution_quadratic_pisgd}. Quantum tunneling therefore has the same concentration requirement as PISGD,
\begin{equation}
    \widetilde{\lambda}_{\mathrm{QT}} \gtrsim \frac{1}{2 \varepsilon^2} \tr \left[ \left( \sqrt{\nabla^2 V(x^\star)} \right)^{-1} \right]
    ,
\label{eqn:lambda_requirement_qt}
\end{equation}
where we write $\widetilde{\lambda}_{\mathrm{QT}}$ for $\lambda$, since in the quantum tunneling case there is no associated hyperparameter such as $\sigma$ to normalize against.

\paragraph{Quantum tunneling optimal escape time.}
Since the runtime~\eqref{eqn:qt_rate_and_runtime} grows with $\widetilde{\lambda}_{\mathrm{QT}}$
while the concentration requirement~\eqref{eqn:lambda_requirement_qt} bounds it from
below, the best available choice is to saturate the threshold. Substituting
\eqref{eqn:lambda_requirement_qt} into \eqref{eqn:qt_rate_and_runtime}, the optimal runtime to
escape from the local minimum is
\begin{equation}
    \mathcal{T}_{\mathrm{QT}}^{\mathrm{opt}} \sim \exp\!\left( \frac{1}{2 \varepsilon^2} \tr \left[ \left( \sqrt{\nabla^2 V(x^\star)} \right)^{-1} \right] B_V\right)
    .
\label{eqn:qt_optimal_runtime}
\end{equation}
Further, in the isotropic case $\nabla^2 V(x^\star) = (\omega^\star)^2 I_d$, this reduces to
\begin{equation}
    \mathcal{T}_{\mathrm{QT}}^{\mathrm{opt}} \sim \exp\!\left( \frac{1}{2 \varepsilon^2} \frac{d}{\omega^\star} B_V\right)
    .
\label{eqn:qt_optimal_runtime_isotropic}
\end{equation}
Quantum tunneling thus obeys exactly the same optimal escape scaling as PISGD in~\eqref{eqn:pisgd_optimal_runtime_isotropic} in the case of a single barrier.

For a separable potential $V(x) = \sum_{i=0}^{d-1} V_i(x_i)$, the Hamiltonian decomposes as $\hat{H} = \sum_{i=0}^{d-1} \hat{H}_i$ with $\hat{H}_i = -\frac{1}{2}\partial_{x_i}^2 + \lambda^2 V_i(\hat{x}_i)$, and the $\hat{H}_i$ act on distinct coordinates and therefore commute. The propagator factorizes accordingly, $e^{-i\hat{H}t} = \bigotimes_{i=0}^{d-1} e^{-i\hat{H}_i t}$, so an initial product state evolves as a product state and each coordinate tunnels independently of the others~\footnote{We assume throughout that quantum tunneling dynamics is initialized in a product state, localized at $(x_0^\bullet, x_1^\bullet, \ldots, x_{d-1}^\bullet)$ in the worst case, which is the natural analogue of initializing PISGD at a point. For an initial state that is not of product form the evolution does not factorize, and the argument below does not apply.}. The tunneling rate out of the worst-case initial state is then governed by the slowest coordinate,
\begin{equation}
    \Gamma_{\mathrm{QT}} \sim \min_{0 \leq i \leq d-1} \Gamma_{V_i, \mathrm{QT}}
    ,
\label{eqn:qt_decay_rate_separable}
\end{equation}
where $\Gamma_{V_i, \mathrm{QT}}$ is the tunneling rate of the one-dimensional problem generated by $\hat{H}_i$, that is, the rate~\eqref{eqn:qt_rate_and_runtime} for the single-coordinate potential $V_i$ with bounce action $B_{V_i}$. The corresponding worst-case escape time is
\begin{equation}
    \mathcal{T}_{\mathrm{QT}} \sim \exp\!\left( \widetilde{\lambda}_{\mathrm{QT}} \max_{0 \leq i \leq d-1} B_{V_i} \right)
    ,
\label{eqn:qt_runtime_separable}
\end{equation}
exactly as for PISGD in~\eqref{eqn:pisgd_decay_rate_separable} and~\eqref{eqn:pisgd_runtime_separable}. The concentration requirement~\eqref{eqn:lambda_requirement_qt} likewise reduces to $\widetilde{\lambda}_{\mathrm{QT}} \gtrsim \frac{1}{2\varepsilon^2} \sum_{i=0}^{d-1} 1/\sqrt{V_i''(x_i^\star)}$, so saturating it gives the optimal escape time
\begin{equation}
    \mathcal{T}_{\mathrm{QT}}^{\mathrm{opt}} \sim \exp\!\left( \frac{1}{2 \varepsilon^2} \sum_{i=0}^{d-1} \frac{1}{\sqrt{V_i''(x_{i}^\star)}} \max_i B_{V_i} \right)
    ,
\label{eqn:qt_optimal_runtime_separable}
\end{equation}
which for identical components with $V_i''(x_i^\star) = (\omega^\star)^2$ becomes
\begin{equation}
    \mathcal{T}_{\mathrm{QT}}^{\mathrm{opt}} \sim \exp\!\left( \frac{1}{2 \varepsilon^2} \frac{d}{\omega^\star} B_V\right)
    ,
\label{eqn:qt_optimal_runtime_isotropic_separable}
\end{equation}
matching~\eqref{eqn:pisgd_optimal_runtime_isotropic_separable}. Quantum tunneling and PISGD therefore agree at the level of the exponential factor for separable potentials as well, not only for a single barrier.

For nonseparable potentials, this agreement between quantum tunneling and PISGD need not persist. One scenario where it can fail is when genuine quantum tunneling benefits from coherence among multiple tunneling paths, while PISGD, as a quantum Monte Carlo method, suffers from a topological obstruction~\cite{hastings2013obstructions,andriyash2017can}~\footnote{No such obstruction arises for the specific nonseparable potential we treat in Section~\ref{sec:nonseparable_potentials}.}. Our analysis therefore does not rule out a potential quantum advantage of quantum-tunneling-based algorithms over classical ones, a point we have discussed in Section~\ref{sec:related_work}.

\subsection{Beyond Exponential Factors}
\label{sec:beyond_exponential_factors}

We have so far focused on the exponential factors, whose exponents are proportional to bounce actions in the case of PISGD, which dominate the decay rates and escape times. These factors are dimensionless and carry no unit of time, so on their own they fix only how the runtime scales, not the runtime itself. To consolidate the analysis and to compare with numerical results directly, we now retain the prefactors that accompany them, as recorded in~\eqref{eqn:Gamma_PISGD} and~\eqref{eqn:T_PISGD}.

We restrict to separable potentials, where each component is a one-dimensional problem and the prefactors remain tractable.

For a one-dimensional component, every ingredient of~\eqref{eqn:Gamma_PISGD} is explicit except the ratio of functional determinants and the negative eigenvalue $\eta_{S(\widetilde{\lambda})}$.
Both are properties of the fluctuation operator $-\partial_\tau^2 + \widetilde{\lambda}^2
V''(x_{\mathrm{cl}}(\tau))$ on the periodic interval, and both are obtained in
Appendix~\ref{app:gelfand_yaglom} by the Gel'fand--Yaglom
method~\cite{dunne2005beyond,dunne2008functional}, which evaluates a functional determinant ratio
from the solution of an initial-value problem without computing the spectra explicitly. The
calculation splits by parity, with the negative mode in the even sector and the
translational zero mode in the odd sector. The result, quoted
from~\eqref{eqn:gy_collected_asymptotic} at large $T$, is
\begin{equation}
    \frac{{\det}' \mathcal{M}}{\det \mathcal{M}_{\mathrm{free}}} \longrightarrow - \frac{\widetilde{\lambda} B_V \, \mathcal{Q} (\infty)}{2 \, \omega \, x_{\infty} \, x_{\mathrm{cl}}'' (0)}
    ,
\label{eqn:det_ratio_quoted}
\end{equation}
where $\omega^2 = \widetilde{\lambda}^2 V''(x^\bullet)$ is the curvature at the local
minimum, $x_\infty$ is the amplitude with which the bounce relaxes back to $x^\bullet$ away
from its center, as $x_{\mathrm{cl}} (\tau) \approx x^\bullet + x_{\infty} \exp(- \omega
\tau)$, and $\mathcal{Q}(\infty) < 0$ is the even-sector contribution, obtained by
solving a single ordinary differential equation numerically.

Substituting~\eqref{eqn:det_ratio_quoted} into the last line
of~\eqref{eqn:Gamma_PISGD}, the factors of $\widetilde{\lambda}$ and $B_V$ carried by the
zero mode cancel against those in the determinant ratio, leaving
\begin{equation}
    \Gamma_{\mathrm{PISGD}} = \exp(- \widetilde{\lambda} B_V) \, a \left(\frac{\sigma^2}{2}\right)^{3} \frac{\abs{\eta_{S(\widetilde{\lambda})}} \, T}{(2 \pi)^{3/2}} \left( \frac{2 \, \omega \, x_{\infty} \, x_{\mathrm{cl}}'' (0)}{\mathcal{Q} (\infty)} \right)^{1/2}
    ,
\label{eqn:Gamma_PISGD_explicit}
\end{equation}
and correspondingly
\begin{equation}
    \mathcal{T}_{\mathrm{PISGD}} \sim \frac{1}{s \, \Gamma_{\mathrm{PISGD}}} = \exp(\widetilde{\lambda} B_V) \, \frac{1}{s a} \left(\frac{2}{\sigma^2}\right)^{3} \frac{(2 \pi)^{3/2}}{\abs{\eta_{S(\widetilde{\lambda})}} \, T} \left( \frac{\mathcal{Q} (\infty)}{2 \, \omega \, x_{\infty} \, x_{\mathrm{cl}}'' (0)} \right)^{1/2}
    .
\label{eqn:T_PISGD_explicit}
\end{equation}
Every quantity on the right is now either a hyperparameter of the algorithm, namely
$\widetilde{\lambda}$, $\sigma$, $a$, $s$, and $T$, or a property of the bounce profile
that can be computed once for a given potential.
Thus every quantity on the right is either an algorithmic hyperparameter or a property
of the bounce profile that can be computed independently of the measured runtimes from numerics.

\section{Numerical Experiments}\label{sec:numerics}

Our numerical experiments are designed to focus on testing the theory and
illustrating the escape mechanism, rather than to identify the best practical
PISGD hyperparameters. The main quantity of interest is the exponential
dependence, $\exp(\widetilde{\lambda}_{\mathrm{SGD}}H_V)$ for SGD versus
$\exp(\widetilde{\lambda}_{\mathrm{PISGD}}B_V)$ for PISGD, which dominates the
runtime in the regimes studied.

Throughout this section, runtime is the first hitting time of the ball
$B(x^\star,\varepsilon)$ in~\eqref{eq:nonconvex-optimization-problem},
reported in continuous-time units $t_\varepsilon=s k_\varepsilon$, where
$k_\varepsilon$ is the number of iterations. For PISGD, one iteration updates
all $N_T$ slices, whose gradients can be evaluated in parallel. Moreover,
$N_T$ does not directly enter the dominant exponential factor. We therefore do not
multiply the runtime by $N_T$ when reporting the results.

\subsection{Separable Variant Quartic Potentials}\label{sec:numerics_variant_quartic_func}

Quartic double-well potentials are widely used to study quantum tunneling~\cite{Coleman:1977py,dunne2008functional}. Nevertheless, Appendix~\ref{sec:quartic_functions} shows that, in our concentration-adjusted optimization setting, the leading PISGD runtime for this family increases even faster than the square of the corresponding SGD runtime as $d\to\infty$. Such a comparison is not meaningful for the bare barrier measures alone: $H_V$ is an energy, whereas $B_V$ is an action, carrying different units. Saturating the concentration requirements yields~\eqref{eqn:main_results_separable}, which instead compares $B_V/\omega^\star$ with $2H_V/(\omega^\star)^2$. In the units used here~\footnote{With the standard convention $\hbar=1$ in physics.}, $\omega^\star$ has units of energy, so both normalized quantities have units of inverse energy.

To exhibit regimes in which PISGD outperforms SGD, we replace the quartic beyond its turning point by a quadratic branch with tunable curvature, matched so that the potential and its first derivative remain continuous. This construction leaves the barrier height and bounce action unchanged while making $\omega^\star$ independently tunable, so the PISGD advantage condition can be satisfied for suitable parameter choices.

For $x=(x_0,\ldots,x_{d-1})\in\mathbb{R}^d$, consider the separable potential
\begin{equation}
    V(x)=\sum_{i=0}^{d-1}V_i(x_i),
    \qquad V_i\equiv U \quad \text{for every } i,
\end{equation}
where $U$ is the common one-dimensional component. As in Appendix~\ref{sec:quartic_functions}, we denote its scalar argument by $z$ and define
\begin{equation}
    q_\alpha(z):=\frac{1}{2}z^2-\frac{1}{2}z^3+\frac{1}{8}\alpha z^4.
\end{equation}
The variant quartic component is
\begin{equation}
    U(z;\alpha,c,\bar{\omega}^\star)=
    \begin{cases}
        c\,q_\alpha(z),
        & z\leq z^{\mathrm{tp}}, \\
        c\,q_\alpha'(z^{\mathrm{tp}})(z-z^{\mathrm{tp}})
        +\dfrac{c}{2}(\bar{\omega}^\star)^2(z-z^{\mathrm{tp}})^2,
        & z>z^{\mathrm{tp}}.
    \end{cases}
\end{equation}
We take $0<\alpha<1$, $c>0$, and $\bar{\omega}^\star>0$. Here $\alpha$ controls the quartic shape, $c$ sets the overall scale~\footnote{For this benchmark, $\lambda$ enters the action through the product $\lambda^2c$. We therefore absorb it into $c$ and set $\lambda=1$ without loss of generality.}, and $\bar{\omega}^\star$ sets the curvature of the quadratic branch. The metastable minimum, saddle, and turning point are
\begin{equation}
    z^\bullet=0,
    \qquad
    z^\circ=\frac{3-\sqrt{9-8\alpha}}{2\alpha},
    \qquad
    z^{\mathrm{tp}}=\frac{2\left(1-\sqrt{1-\alpha}\right)}{\alpha},
\end{equation}
with
\begin{equation}
    U(z^\bullet)=U(z^{\mathrm{tp}})=0,
    \qquad
    U'(z^{\mathrm{tp}})=c\,q_\alpha'(z^{\mathrm{tp}})
    =-\frac{2c\sqrt{1-\alpha}\left(1-\sqrt{1-\alpha}\right)^2}{\alpha^2}.
\end{equation}
The global minimum lies on the quadratic branch at
\begin{equation}
\begin{aligned}
    z^\star
    &=z^{\mathrm{tp}}-\frac{q_\alpha'(z^{\mathrm{tp}})}{(\bar{\omega}^\star)^2} \\
    &=\frac{2\left(1-\sqrt{1-\alpha}\right)}{\alpha}
      +\frac{2\sqrt{1-\alpha}\left(1-\sqrt{1-\alpha}\right)^2}
      {(\bar{\omega}^\star)^2\alpha^2}, \\
    U(z^\star)
    &=-\frac{c\left[q_\alpha'(z^{\mathrm{tp}})\right]^2}
      {2(\bar{\omega}^\star)^2}
     =-\frac{2c(1-\alpha)\left(1-\sqrt{1-\alpha}\right)^4}
      {(\bar{\omega}^\star)^2\alpha^4}.
\end{aligned}
\end{equation}
Thus the global minimum of the full separable potential is $x^\star=(z^\star,\ldots,z^\star)$.
By construction,
\begin{equation}
    \bar{\omega}^\star:=\sqrt{\frac{U''(z^\star)}{c}},
    \qquad
    \omega^\star:=\sqrt{U''(z^\star)}=\sqrt{c}\,\bar{\omega}^\star,
    \qquad
    U''(z^\star)=(\omega^\star)^2=c(\bar{\omega}^\star)^2,
\end{equation}
where $\omega^\star$ is the curvature scale used in the previous sections. Because $q_\alpha(z^{\mathrm{tp}})=0$ and the two branches have the same slope at $z^{\mathrm{tp}}$, $U$ is continuously differentiable, although its second derivative generally has a discontinuous jump at the turning point.

To maintain the notation of the preceding analysis, $H_V$ and $B_V$ below denote, respectively, the barrier height and bounce action of the common one-dimensional component $U$. Equivalently, they are the common values of $H_{V_i}$ and $B_{V_i}$. The analytic expressions in Appendix~\ref{sec:quartic_functions} show that $B_V/(2\sqrt{c})$ and $H_V/c$ depend only on $\bar{\omega}^\star$ and $\alpha$, so the concentration-adjusted comparison is independent of $c$. Define the critical normalized curvature
\begin{equation}
    \bar{\omega}_{\mathrm{c}}^\star(\alpha)
    :=\frac{H_V(\alpha,c)/c}{B_V(\alpha,c)/(2\sqrt{c})}.
\end{equation}
The PISGD exponent is smaller when $\bar{\omega}^\star<\bar{\omega}_{\mathrm{c}}^\star(\alpha)$, whereas the SGD exponent is smaller when $\bar{\omega}^\star>\bar{\omega}_{\mathrm{c}}^\star(\alpha)$. Figure~\ref{fig:variant_quartic_pisgd_sgd_phase_diagram} partitions the parameter space into the corresponding PISGD- and SGD-favored regimes. For the standard pure quartic, the normalized curvature follows the curve
\begin{equation}
    \bar{\omega}_{\mathrm{q}}^\star(\alpha)
    :=\left(
        \frac{9-8\alpha+3\sqrt{9-8\alpha}}{4\alpha}
    \right)^{1/2},
\end{equation}
which lies entirely in the SGD-advantage region. Strictly speaking, the pure quartic is not a member of the piecewise family defined above because it has no quadratic branch, but the same advantage condition compares its $B_V$, $H_V$, and $\omega^\star$ still applies.

\begin{figure}[!htbp]
    \centering
    \includegraphics[width=\linewidth]{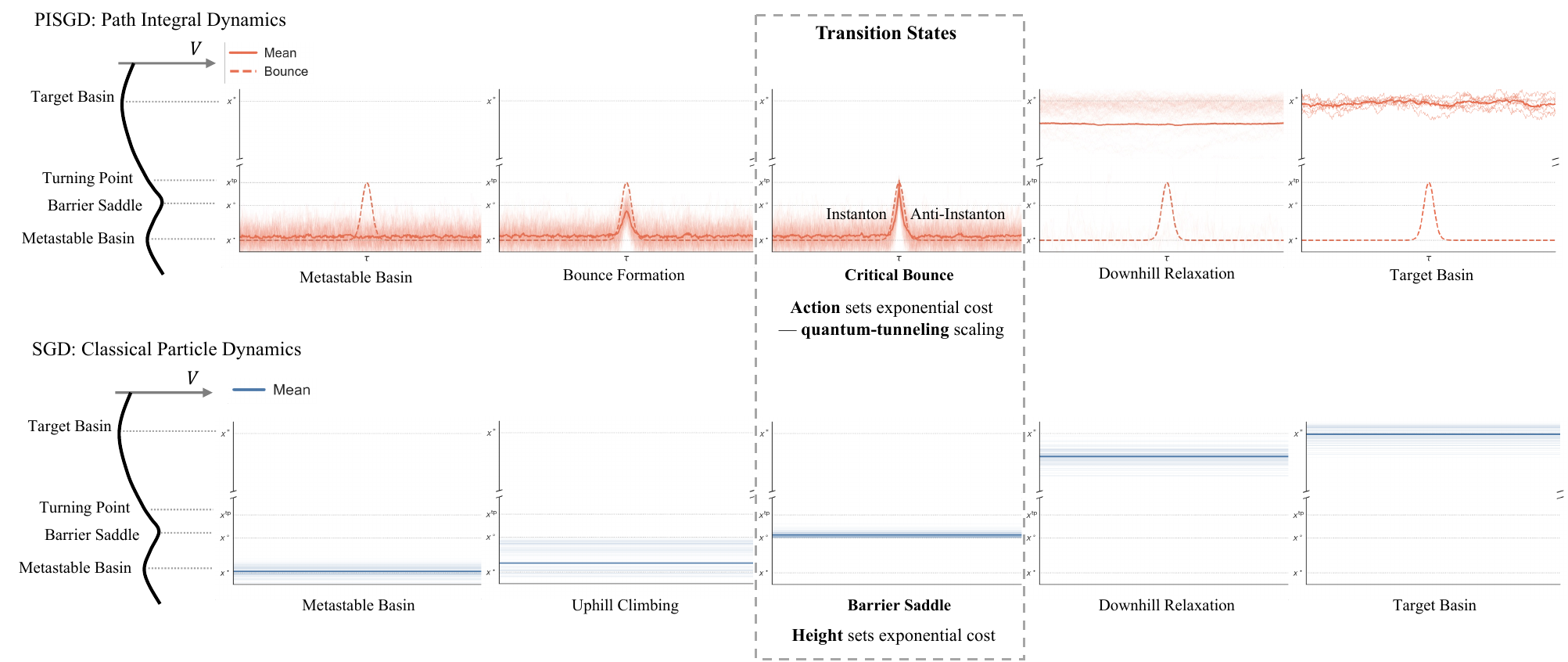}
    \caption{Event-aligned escape dynamics in numerical PISGD (upper row) and SGD (lower row) runs on the one-dimensional variant quartic potential with $\alpha = 0.9$, $c = 100$, and $\bar{\omega}^\star = 0.1$. Thin curves show the individual runs included in each panel, thick curves show their mean, and the dashed profile in the upper row is the bounce profile. In the central PISGD panel, the aligned mean path closely follows the bounce, displaying the predicted instanton--anti-instanton transition profile. The full path subsequently relaxes toward $x^\star$. By contrast, SGD crosses the barrier as a single scalar, plotted as a constant path. Because the runs are aligned to run-specific reference events, the panels compare escape mechanisms rather than elapsed runtimes. Appendix~\ref{app:escape_dynamics_figure} details the construction and statistics.}
    \label{fig:main_fig}
\end{figure}

\paragraph{Escape dynamics and numerical evidence for the bounce profile.}
Figure~\ref{fig:main_fig} shows event-aligned snapshots of PISGD and SGD trajectories from the numerical experiments. PISGD first develops a localized path fluctuation. At the PISGD transition event, defined here as the first iteration at which any time slice passes the turning point $x^{\mathrm{tp}}$, the mean of the aligned paths closely follows the theoretical bounce profile, an instanton--anti-instanton pair. This agreement provides numerical evidence for the predicted bounce profile in the PISGD dynamics and thus supports the proposed instanton-mediated escape mechanism. After the transition, the full path relaxes into the global basin. By contrast, SGD evolves a single point and escapes by passing over the saddle $x^\circ$. Because the runs are aligned at their respective reference events, the panels compare the two escape mechanisms rather than their elapsed runtimes. Appendix~\ref{app:escape_dynamics_figure} gives the complete simulation parameters, alignment procedure, panel offsets, sample counts, and crossing-time statistics, and discusses caveats associated with the event alignment.

\begin{figure}[!htbp]
    \centering
    \includegraphics[width=\linewidth]{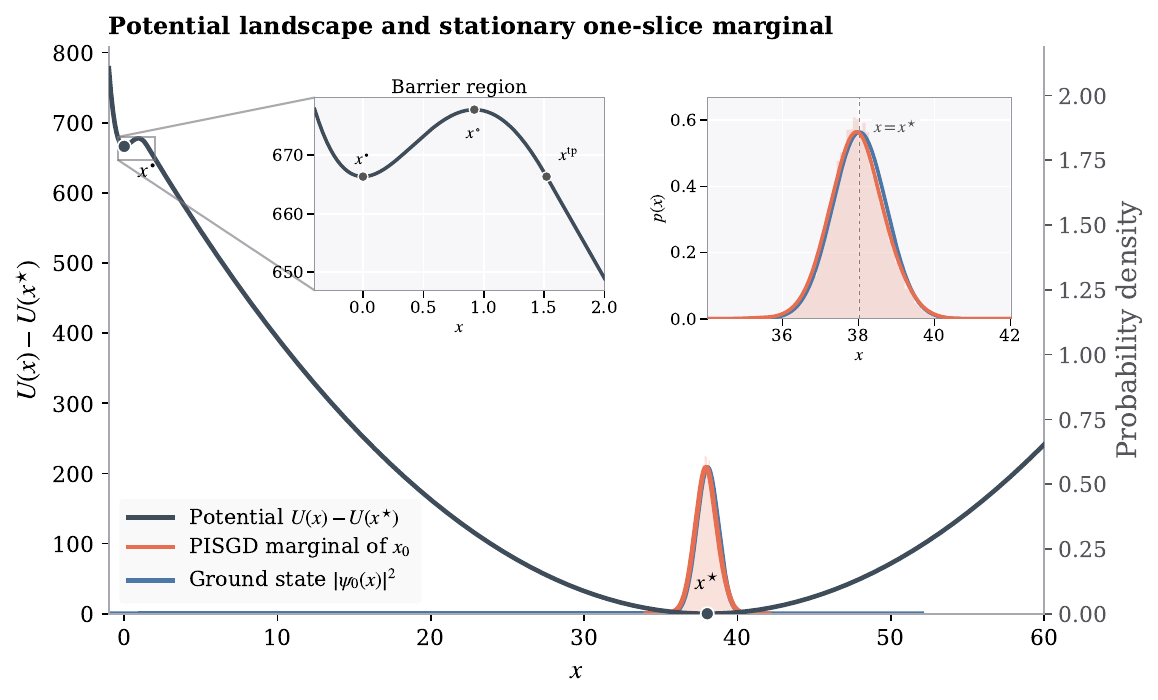}
    \caption{Numerical comparison of the PISGD one-slice marginal with the quantum ground-state density for the one-dimensional variant quartic potential with $\alpha=0.9$, $\bar{\omega}^\star=0.1$, and $c=100$. The main panel shows the single-component potential $U(x)-U(x^\star)$ on the left vertical axis. The right vertical axis corresponds to the density of the PISGD slice $x_0$ (orange histogram) from numerics, an Gaussian kernel-density estimate (orange curve), and the normalized ground-state density $\abs{\psi_0(x)}^2$ (blue curve).
    The left inset zooms into the metastable minimum $x^\bullet$, saddle $x^\circ$, and turning point $x^{\mathrm{tp}}$.
    The right inset zooms in for the density comparison near the global minimum.
    The observed agreement between the PISGD marginal and the quantum ground state is consistent with the theory~\eqref{eqn:ground_state_correspondence}.}
    \label{fig:pisgd_stationary_distribution_variant_quartic}
\end{figure}

\paragraph{PISGD stationary distribution.}
Section~\ref{sec:distribution_equivalence} shows that, in the continuum imaginary-time limit followed by the large-$T$ limit, the one-slice marginal of the PISGD stationary path measure converges to the quantum ground-state density~\eqref{eqn:ground_state_correspondence}. We test this correspondence numerically for the one-dimensional variant quartic potential at $\lambda=1$ and $\sigma^2=2$, for which $\widetilde{\lambda}_{\mathrm{PISGD}}=2\lambda/\sigma^2=1$. Starting with every slice at the metastable minimum $x^\bullet$, we run one long PISGD trajectory with $a=0.025$, $N_T=400$, $T=10$, and $s=0.01$. We discard the first $50{,}000$ iterations and retain the value of $x_0$ every $100$ iterations through iteration $999{,}900$, giving $9{,}500$ late-time samples. The comparison density $\abs{\psi_0(x)}^2$ is obtained independently by solving the one-dimensional Schr\"odinger eigenvalue problem~\eqref{eqn:hamiltonian} for the same potential.
Figure~\ref{fig:pisgd_stationary_distribution_variant_quartic} shows close agreement between the empirical PISGD marginal and the ground-state density in both location and width.

\begin{figure}[!htbp]
    \centering
    \begin{subfigure}[t]{0.49\linewidth}
        \centering
        \includegraphics[width=\linewidth]{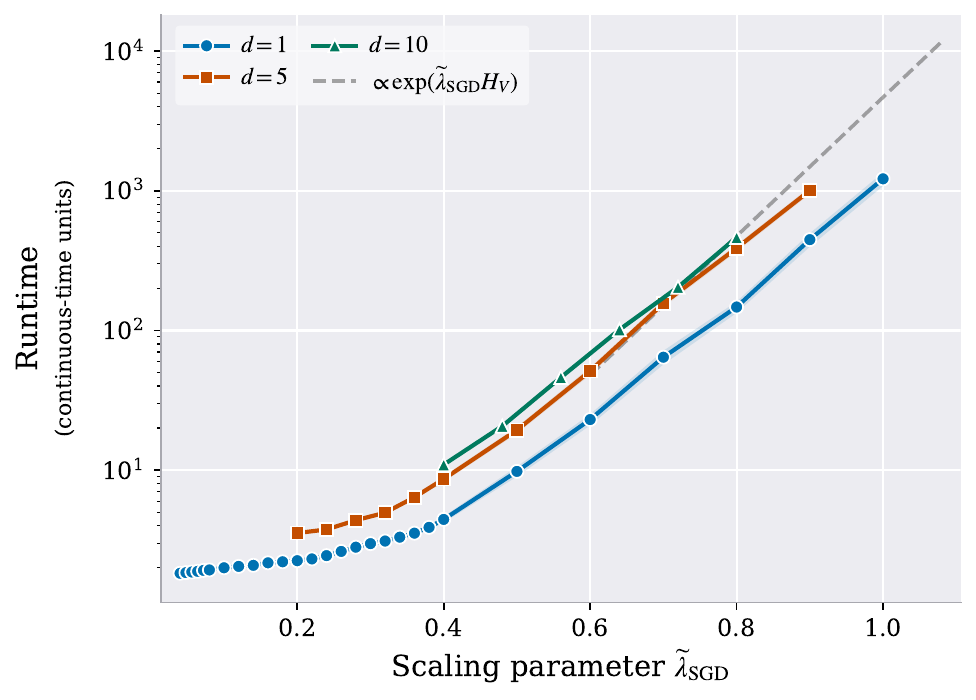}
        \caption{SGD: $\widetilde{\lambda}_{\mathrm{SGD}}\geq\widetilde{\lambda}_{\mathrm{conc},\mathrm{SGD}}(d)=0.04d$.}
        \label{fig:sgd_tilde_lambda_asymptotics}
    \end{subfigure}
    \hfill
    \begin{subfigure}[t]{0.49\linewidth}
        \centering
        \includegraphics[width=\linewidth]{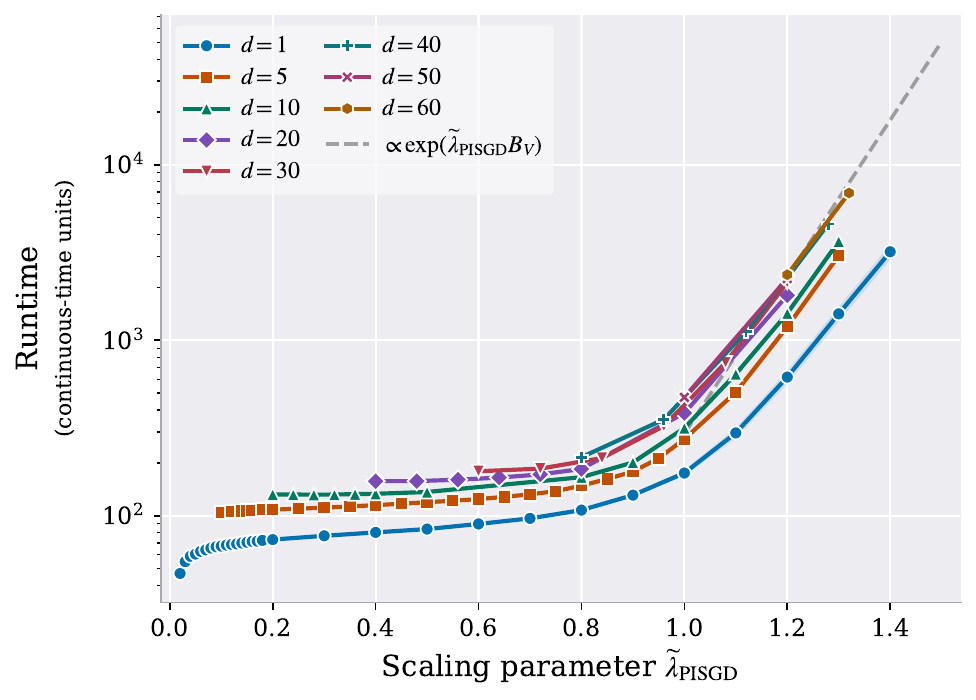}
        \caption{PISGD: $\widetilde{\lambda}_{\mathrm{PISGD}}\geq\widetilde{\lambda}_{\mathrm{conc},\mathrm{PISGD}}(d)=0.02d$.}
        \label{fig:pisgd_tilde_lambda_asymptotics}
    \end{subfigure}
    \caption{Mean runtime of (a) SGD and (b) PISGD on the $d$-dimensional separable variant quartic potential with $\alpha=0.9$, $\bar{\omega}^\star=0.1$, and $c=100$, shown as a function of the scaling parameters $\widetilde{\lambda}_{\mathrm{SGD}}=2/\sigma^2$ and $\widetilde{\lambda}_{\mathrm{PISGD}}=2/\sigma^2$  with $\lambda=1$.
    Runtime is measured in continuous-time units, \textit{i.e.}, the number of gradient steps multiplied by the learning rate $s=0.01$.
    All displayed points satisfy the corresponding concentration condition: Eq.~\eqref{eqn:optimal_lambda_tilde_sgd} for SGD and Eq.~\eqref{eqn:lambda_requirement_pisgd_separable_isotropic} for PISGD.
    The dashed lines indicate the predicted asymptotic exponential dependences $\mathcal{T}_{\mathrm{SGD}}\propto\exp(\widetilde{\lambda}_{\mathrm{SGD}}H_V)$ in (a) and $\mathcal{T}_{\mathrm{PISGD}}\propto\exp(\widetilde{\lambda}_{\mathrm{PISGD}}B_V)$ in (b). Their vertical offsets are chosen only to guide the eye.
    The shaded bands represent one standard error of the mean. The PISGD hyperparameters are the lattice spacing $a=0.025$, the number of imaginary-time slices $N_T=400$, and the imaginary-time length $T=N_Ta=10$.}
    \label{fig:mean_iterations_asymptotics_variant_quartic}
\end{figure}

\paragraph{Runtime with a scan of the scaling parameter $\widetilde{\lambda}$.}
For the runtime sweeps reported here, we define the runtime as the first-passage time of the reported iterate to the $L^2$-ball $B(x^\star,\varepsilon):=\{x\in\mathbb{R}^d:\lVert x-x^\star\rVert_2<\varepsilon\}$. Equivalently, the hitting iteration $k_\varepsilon$ is the first iteration at which the iterate enters this ball. For PISGD, the reported iterate is the slice $x_0$, as in Algorithm~\ref{alg:pisgd}. For SGD, it is the ordinary iterate. We set the target radius to $\varepsilon=5$.~\footnote{Although $\varepsilon=5$ may appear large as an optimization tolerance, the entire target ball lies within the quadratic region enclosing $x^\star$, and hence convex. We use this comparatively large radius because the concentration thresholds scale as $\varepsilon^{-2}$: decreasing $\varepsilon$ would require substantially larger values of $\widetilde{\lambda}$ and hence prohibitively long runtimes. The purpose of these experiments is to test the predicted escape mechanism and asymptotic scaling rather than to optimize practical stopping accuracy. The dependence on the target radius is examined in Appendix~\ref{app:ablation_target_radius}.}
We report the hitting time in continuous-time units, $t_\varepsilon=s k_\varepsilon$, rather than as a raw iteration count. This is the natural dynamical-time rescaling in the continuous-time limit $s\to0$; Appendix~\ref{app:ablation_learning_rate} examines the dependence on the finite learning rate.~\footnote{Each PISGD iteration evaluates $\nabla V$ on $N_T$ time slices, so a serial-work count would include an additional factor of $N_T$. These slice-wise evaluations can be parallelized, and we hold $N_T=400$ fixed across $d$ and $\widetilde{\lambda}$ in these sweeps. Including this factor would therefore change only the overall prefactor,  not the leading exponential dependence. The effects of the imaginary-time discretization $a$ and length $T$ are discussed in Appendices~\ref{app:ablation_lattice_spacing} and~\ref{app:ablation_imaginary_time_length}, respectively.}
Figure~\ref{fig:mean_iterations_asymptotics_variant_quartic} reports the mean runtimes for both SGD and PISGD. For sufficiently large values of their respective scaling parameters $\widetilde{\lambda}$, the runtimes approach the predicted leading exponential dependences $\mathcal{T}_{\mathrm{SGD}}\propto\exp(\widetilde{\lambda}_{\mathrm{SGD}}H_V)$ and $\mathcal{T}_{\mathrm{PISGD}}\propto\exp(\widetilde{\lambda}_{\mathrm{PISGD}}B_V)$.

\paragraph{Runtime across dimension.}
Figure~\ref{fig:comparison_variant_quartic} fixes each method at its concentration threshold, $\widetilde{\lambda}_{\mathrm{SGD}}=0.04d$ and $\widetilde{\lambda}_{\mathrm{PISGD}}=0.02d$, and varies $d$. Both runtimes eventually grow exponentially in $d$ with the rates predicted by~\eqref{eqn:main_results_separable}. Because $B_V/\omega^\star<2H_V/(\omega^\star)^2$ at this parameter setting, the PISGD has a scaling advantage over SGD.

\paragraph{Theoretical predictions of PISGD runtime.}
For we evaluate the full prefactor in
Eq.~\eqref{eqn:T_PISGD_explicit} at $\widetilde{\lambda}=1.2$, $\lambda=1$,
and hence $\sigma^2=5/3$.  Multiplying by $s$ gives back the runtime $t_{\varepsilon}$ in continuous-time units
\begin{equation}
    t_\varepsilon=s\mathcal{T}_{\mathrm{PISGD}}
    \sim\frac{K_{\mathrm{th}}}{aT},
    \qquad
    K_{\mathrm{th}}=A_{\mathrm{th}} \exp(\widetilde{\lambda}B_V),
\label{eqn:pisgd_continuous_time_prefactor}
\end{equation}
where the prefactor is
\begin{equation}
    A_{\mathrm{th}}
    :=\left(\frac{2}{\sigma^2}\right)^3
    \frac{(2\pi)^{3/2}}{\abs{\eta_{S(\widetilde{\lambda})}}}
    \left(
        \frac{\mathcal{Q}(\infty)}
        {2\omega x_\infty x_{\mathrm{cl}}''(0)}
    \right)^{1/2}.
\label{eqn:pisgd_A_th}
\end{equation}
The bounce initial-value problem and related calculation give the quantities above, independent of the numerical runs of PISGD itself. For the test potential in this section, the calculated values are
\begin{equation*}
\begin{gathered}
    B_V=10.287354,\quad \omega=12,\quad x_\infty=12.649111,\quad
    \mathcal{Q}(\infty)=-0.0288608,\\
    x_{\mathrm{cl}}''(0)=-52.569082,\qquad
    \eta_{S(\widetilde{\lambda})}=-57.843249,
\end{gathered}
\end{equation*}
so that
\begin{equation}
    A_{\mathrm{th}}=6.32724\times10^{-4},\qquad
    \exp(\!\widetilde{\lambda}B_V)=2.29768\times10^5,
    \qquad K_{\mathrm{th}}=145.379.
\label{eqn:theoretical_prediction_numbers}
\end{equation}
Thus the main setting $a=0.025$, $T=10$ has the theoretical prediction
$t_\varepsilon=K_{\mathrm{th}}/(aT)=581.518$, used in ablation study results Figures~\ref{fig:pisgd_lattice_spacing_ablation}--\ref{fig:pisgd_target_radius_ablation}.

\bigskip

Appendix~\ref{sec:pisgd_concentration_sweep} reports an additional sweep across the PISGD scaling parameter.

\subsection{Separable Spiky Potentials}\label{sec:numerics_spiky_func}

The advantage condition $B_V<2H_V/\omega^\star$ compares an action with a barrier height. A tall but thin barrier has a large height but a small action, because the bounce spends little imaginary time under it. We therefore use a spiky component, for which $B_V$ is much smaller than $2H_V/\omega^\star$, to show a more pronounced PISGD advantage than in the variant quartic case.

Consider the separable potential $V(x)=\sum_{i=0}^{d-1}U(x_i)$ with the common component
\begin{equation}
    U(z)=c\left(\frac{1}{2}(\bar{\omega}^\star)^2z^2+\frac{h}{\sqrt{2\pi w^2}}\exp\!\left(-\frac{(z-\mu)^2}{2w^2}\right)\right),
\label{eqn:spiky_component}
\end{equation}
a quadratic well with a Gaussian spike of width $w$ and weight $h$ centered at $\mu<0$. For $w\ll\abs{\mu}$, the spike splits the quadratic well into a metastable minimum $z^\bullet$ to its left and a global minimum $z^\star\approx0$, with $\omega^\star=\sqrt{U''(z^\star)}\approx\sqrt{c}\,\bar{\omega}^\star$. We take $w=0.25$, $h=50$, $\mu=-20$, $c=1$, and $\bar{\omega}^\star=1$, for which numerical evaluation gives
\begin{equation}
    H_V=64.14,\qquad B_V=15.07,\qquad \omega^\star=1.000,
\end{equation}
so that $2H_V/\omega^\star\approx 8.5\,B_V$. The exponent governing the SGD runtime is therefore about $8.5$ times larger than that of PISGD.

\paragraph{Runtime across dimension.}
We use the target radius $\varepsilon=5$ and set each method at its concentration threshold~\eqref{eqn:main_results_separable}, $\widetilde{\lambda}_{\mathrm{SGD}}=d/((\omega^\star)^2\varepsilon^2)=0.04d$ and $\widetilde{\lambda}_{\mathrm{PISGD}}=d/(2\omega^\star\varepsilon^2)=0.02d$. The predicted exponents then grow at the rates $H_V/((\omega^\star)^2\varepsilon^2)\approx2.57$ per dimension for SGD and $B_V/(2\omega^\star\varepsilon^2)\approx0.30$ per dimension for PISGD. SGD uses the learning rate $s=10^{-3}$. PISGD uses $s=10^{-2}$, $a=0.025$, and $T=2$ ($N_T=80$). Each point averages $100$ independent runs. Figure~\ref{fig:comparison_spike} shows that both runtimes approach the predicted exponential growth at large $d$.

\subsection{Nonseparable Potentials: Squared Sum of Quartics}
\label{sec:nonseparable_potentials}

To test beyond separable landscapes, we square a sum of identical quartic components. With $q_\alpha$ as in Section~\ref{sec:numerics_variant_quartic_func} and $0<\alpha<1$, let
\begin{equation}
    V_d(x)=c\bigg(b_d+\sum_{i=0}^{d-1}q_\alpha(x_i)\bigg)^2,
    \qquad x\in\mathbb{R}^d.
\label{eqn:squared_quartic_potential}
\end{equation}
The quartic $q_\alpha$ has a local minimum $z^\bullet=0$, a local maximum $z^\circ=(3-\sqrt{9-8\alpha})/(2\alpha)$, and a global minimum $z^\star=(3+\sqrt{9-8\alpha})/(2\alpha)$. We write $q^\circ:=q_\alpha(z^\circ)>0$ and $q^\star:=q_\alpha(z^\star)<0$, and note $q_\alpha(z^\bullet)=0$. The offset
\begin{equation}
    b_d:=\kappa_0-d\,q^\star,
    \qquad
    \kappa_0:=\frac{(\bar{\omega}^\star)^2}{2q_\alpha''(z^\star)},
\label{eqn:squared_quartic_offset}
\end{equation}
keeps the bracket in~\eqref{eqn:squared_quartic_potential} positive, at least $\kappa_0>0$. Squaring therefore leaves the $3^d$ stationary points $\{z^\bullet,z^\circ,z^\star\}^d$ and the ordering of their values unchanged. The global minimum is $x^\star=(z^\star,\ldots,z^\star)$, with $\nabla^2V_d(x^\star)=(\omega^\star)^2I_d$ and $\omega^\star=\sqrt{c}\,\bar{\omega}^\star$.
Unlike the separable case, the barrier for moving one coordinate depends on all other coordinates through the square.

\paragraph{Barrier heights and bounce actions.}
Appendix~\ref{app:squared_quartic_analysis} considers the transition of one coordinate from $z^\bullet$ to $z^\star$ while $n^\star\in\{0,\ldots,d-1\}$ of the other coordinates sit at $z^\star$ and the remaining $d-1-n^\star$ at $z^\bullet$. The barrier height and bounce action of this one-dimensional section are
\begin{equation}
\begin{aligned}
    H_V(n^\star)&=c\,q^\circ\left(q^\circ+2\kappa_0-2(d-n^\star)\,q^\star\right),\\
    B_V(n^\star)&=2\sqrt{2c}\int_{z^\bullet}^{z^{\mathrm{tp}}}\sqrt{q_\alpha(z)\left(q_\alpha(z)+2\kappa_0-2(d-n^\star)\,q^\star\right)}\,\diff z,
\end{aligned}
\label{eqn:squared_quartic_barriers}
\end{equation}
where $z^{\mathrm{tp}}=2(1-\sqrt{1-\alpha})/\alpha$ is the turning point of $q_\alpha$. Since $q^\star<0$, both decrease with $n^\star$. Their extreme values are
\begin{equation}
    H_V^{\max}:=H_V(0),\quad H_V^{\min}:=H_V(d-1),\qquad
    B_V^{\max}:=B_V(0),\quad B_V^{\min}:=B_V(d-1).
\label{eqn:squared_quartic_extremes}
\end{equation}
The maximal values grow with $d$, $H_V^{\max}$ as $O(d)$ and $B_V^{\max}$ as $O(\sqrt{d})$. The minimal values do not depend on $d$ and coincide with the one-dimensional values.

The Witten-Laplacian analysis of~\cite{Michel_2019,shi2023learning} shows that $H_V^{\max}$ sets the SGD spectral gap. For PISGD, no comparable theory is available, and we hypothesize by analogy that $B_V^{\max}$ plays the same role. Substituting these into the concentration-saturated scalings~\eqref{eqn:main_results_separable} gives
\begin{equation}
    \mathcal{T}_{\mathrm{SGD}}^{\mathrm{opt}}\sim\exp\!\left(\frac{d\,H_V^{\max}}{(\omega^\star)^2\varepsilon^2}\right),
    \qquad
    \mathcal{T}_{\mathrm{PISGD}}^{\mathrm{opt}}\sim\exp\!\left(\frac{d\,B_V^{\max}}{2\omega^\star\varepsilon^2}\right),
\label{eqn:squared_quartic_runtime}
\end{equation}
whose exponents grow as $O(d^2)$ and $O(d^{3/2})$, respectively. (Replacing the maximal values by $H_V^{\min}$ and $B_V^{\min}$ instead gives exponents linear in $d$, as in the separable case.)

\begin{figure}
    \centering
    \begin{subfigure}[t]{0.49\linewidth}
        \centering
        \includegraphics[width=\linewidth]{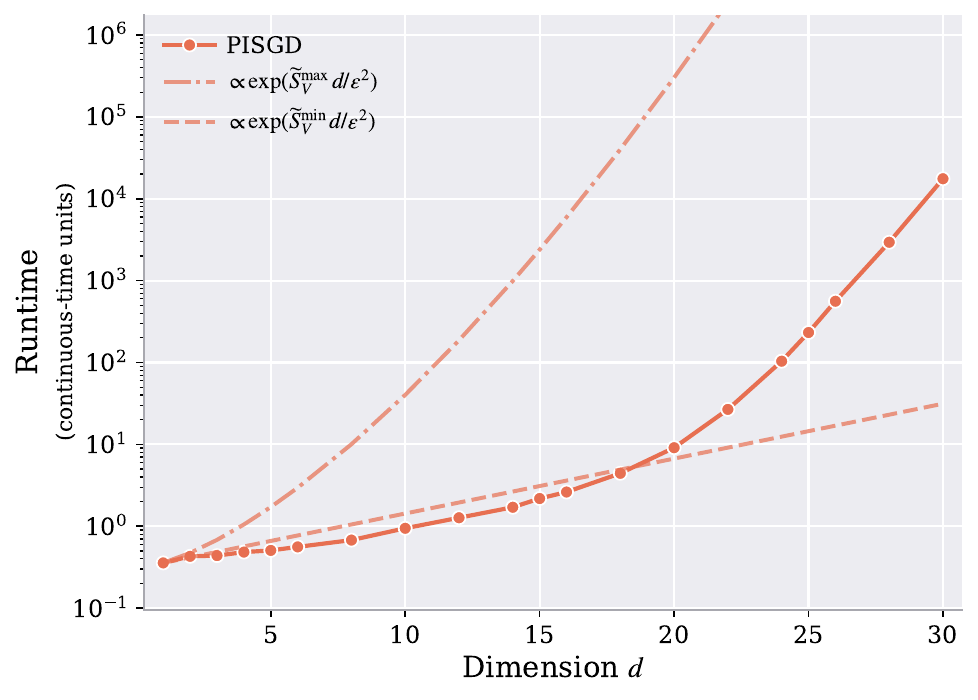}
        \caption{PISGD.}
        \label{fig:pisgd_squared_quartic}
    \end{subfigure}
    \hfill
    \begin{subfigure}[t]{0.49\linewidth}
        \centering
        \includegraphics[width=\linewidth]{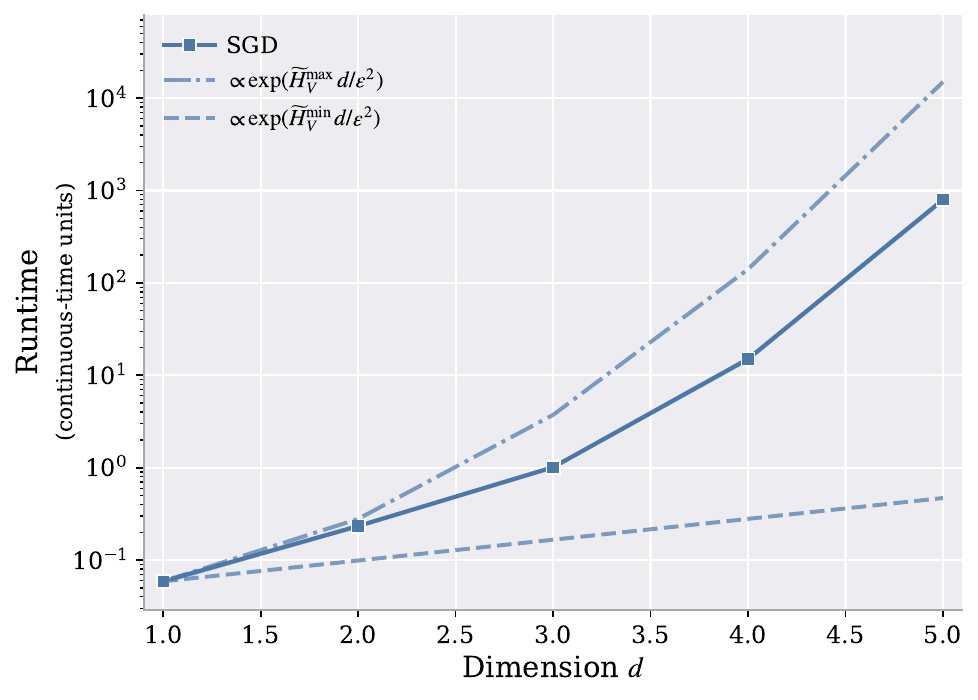}
        \caption{SGD.}
        \label{fig:sgd_squared_quartic}
    \end{subfigure}
    \caption{Mean runtime of (a) PISGD and (b) SGD on the squared quartic potential~\eqref{eqn:squared_quartic_potential} with $\alpha=0.3$, $c=1$, $\bar{\omega}^\star=1$, and $\varepsilon=5$, each set at its concentration threshold. Runtime is in continuous-time units. Error bars (small to be visible) denotes the standard error. Dash-dotted and dashed lines show~\eqref{eqn:squared_quartic_runtime} evaluated with the maximal ($B_V^{\max}=B_V(0)$, $H_V^{\max}=H_V(0)$) and the minimal ($B_V^{\min}=B_V(d-1)$, $H_V^{\min}=H_V(d-1)$) values of~\eqref{eqn:squared_quartic_barriers}, respectively. Both lines are anchored at the $d=1$ data point.}
    \label{fig:results_squared_quartic}
\end{figure}

\paragraph{Numerical results.}
We take $\alpha=0.3$, $c=1$, $\bar{\omega}^\star=1$, and $\varepsilon=5$, and set each method at its concentration threshold, $\widetilde{\lambda}_{\mathrm{SGD}}=0.04d$ and $\widetilde{\lambda}_{\mathrm{PISGD}}=0.02d$. Figure~\ref{fig:results_squared_quartic} compares the measured runtimes with~\eqref{eqn:squared_quartic_runtime} evaluated with the maximal and the minimal values. For both methods, the data lie between the two predictions asymptotically in large $d$. Figure~\ref{fig:comparison_squared_quartic} compares the two methods directly.

\section*{Acknowledgment}

We thank Rob Otter for executive support and valuable feedback on this project. J.S. also thanks Abhishek Som for helpful discussions. We further acknowledge our colleagues at the Global Technology Applied Research center of JPMorganChase.

\section*{Disclaimer}
This paper was prepared for informational purposes by the Global Technology Applied Research center of JPMorgan Chase \& Co. This paper is not a product of the Research Department of JPMorgan Chase \& Co. or its affiliates. Neither JPMorgan Chase \& Co. nor any of its affiliates makes any explicit or implied representation or warranty and none of them accept any liability in connection with this paper, including, without limitation, with respect to the completeness, accuracy, or reliability of the information contained herein and the potential legal, compliance, tax, or accounting effects thereof. This document is not intended as investment research or investment advice, or as a recommendation, offer, or solicitation for the purchase or sale of any security, financial instrument, financial product or service, or to be used in any way for evaluating the merits of participating in any transaction.

\section*{AI Use Disclosure}

Large language model tools were used interactively at all stages of the preparation of this manuscript: from the earliest exploratory research stages to the final exposition and writing steps. The final manuscript was substantially revised by the authors, who remain solely responsible for all mathematical claims, proofs, references, and conclusions.

\appendix

\section{Concentration of Gaussian Measures}\label{sec:convergence_from_gaussians}

\subsection{Isotropic Quadratic Functions}
\label{sec:isotropic_quadratic_functions}

We consider a stationary distribution
\begin{equation}
    \mu(x) = \frac{1}{Z} e^{-g(x)}
    , \qquad Z = \int \diff x \, e^{-g(x)}
    ,
\end{equation}
with $g: \mathbb{R}^d \rightarrow \mathbb{R}$, and take an isotropic quadratic function $g (x) = \frac{1}{2} \vartheta
\norm{x}^2$ with $\vartheta > 0$, so that $x \sim \mathcal{N} (0, (1/\vartheta) I_d)$.
Integrating in spherical coordinates and substituting $u = \frac{1}{2} \vartheta r^2$
reduces the probability of the ball $B(0, \varepsilon)$ to an incomplete gamma integral,
\begin{equation}
    P \left( \norm{x} < \varepsilon \right)
    = \frac{1}{\Gamma(d/2)} \int_0^{\frac{1}{2} \vartheta \varepsilon^2} \diff u \, u^{\frac{d}{2} - 1} e^{-u}
    = P \left(\frac{d}{2}, \frac{1}{2} \vartheta \varepsilon^2\right)
    ,
\end{equation}
where $P(s, x) = \gamma(s, x) / \Gamma(s)$ is the regularized gamma function, that is,
the cumulative distribution function of a gamma random variable with shape $s$ and scale
$1$. Equivalently, $\vartheta \norm{x}^2 \sim \chi^2_d$, and the above is its lower-tail
probability.

The threshold follows from the median $\nu(s)$ of this distribution, which
satisfies~\cite{choi1994medians}
\begin{equation}
    \nu(s) = s - \frac{1}{3} + \frac{8}{405 s} + \cdots
    .
\end{equation}
A finite probability requires $\frac{1}{2} \vartheta \varepsilon^2 \gtrsim \nu (d/2)$,
\textit{i.e.}, $\vartheta \varepsilon^2 \gtrsim d$. The transition is sharp in the large
$d$ limit, as the behavior on either side shows. For fixed
$\vartheta$ and $\varepsilon$ with $d \rightarrow \infty$,
\begin{equation}
    P \left(\frac{d}{2}, \frac{1}{2} \vartheta \varepsilon^2\right)
    = \frac{1}{\sqrt{\pi d}} \exp\left( -\frac{1}{2} \vartheta \varepsilon^2 + \frac{d}{2} \ln \left( \frac{e \vartheta \varepsilon^2}{d} \right) \right) \left( 1 + O(d^{-1}) \right)
    ,
\label{eqn:gamma_cdf_below_threshold}
\end{equation}
which is superexponentially small in $d$, the dominant factor being $\exp(- \frac{d}{2}
\ln (d / (e \vartheta \varepsilon^2)))$. The number of samples required to place at
least one point in $B(0, \varepsilon)$ is the reciprocal of this quantity. To make the probability finite for large $d$, $\vartheta \varepsilon^2$ needs to be $\Omega(d)$.

When $\vartheta \varepsilon^2$ is below the threshold $d$, $\vartheta \varepsilon^2 = d (1 - b)$ with $0 < b < 1$ fixed, the probability is
\begin{equation}
    P \left( \norm{x} < \varepsilon \right) = P \left(\frac{d}{2}, \frac{1}{2} \vartheta \varepsilon^2\right) \sim \frac{1}{b\sqrt{\pi d}}\;\exp\Big\{\frac{d}{2}\Big[b+\ln (1-b)\Big]\Big\}\left[1-\frac{1-2b}{6 b d}+O\left(\frac{1}{d^2}\right)\right]
    ,
\end{equation}
which is exponentially small in $d$ since $b+\ln (1-b)<0$.

When $\vartheta \varepsilon^2$ is above the threshold $d$, writing $\vartheta \varepsilon^2 = d (1 + b)$ with $b > 0$, the probability
obeys
\begin{equation}
    P \left( \norm{x} < \varepsilon \right) = P \left(\frac{d}{2}, \frac{d(1+b)}{2}\right) \sim 1 - \frac{1}{b \sqrt{\pi d}}\exp\left(-\frac{b^2 d}{4}\right) \left[1-\frac{2}{b^2 d}+O\left(\frac{1}{d^2}\right)\right]
    ,
\label{eqn:gamma_cdf_above_threshold}
\end{equation}
so for fixed $b$ the probability is
exponentially close to one.

Therefore, there is a very sharp threshold in order to achieve finite probability in the ball $B(0, \varepsilon)$, \textit{i.e.}, the condition
\begin{equation}
    \vartheta \varepsilon^2 > d \,
    ,
\end{equation}
needs to be met.

\subsection{Anisotropic Quadratic Functions}
\label{sec:convergence_from_gaussians_anisotropic}

The isotropic assumption above is not essential. We now take
\begin{equation}
    g(x) = \frac{1}{2} x^\top P x
\end{equation}
with $P$ symmetric positive definite, so that $x \sim \mathcal{N}(0, P^{-1})$. Writing
$0 < \vartheta_1 \leq \vartheta_2 \leq \cdots \leq \vartheta_d$ for the eigenvalues of
$P$ and passing to its eigenbasis, the squared $2$-norm becomes a weighted sum of
independent squared standard normals,
\begin{equation}
    \norm{x}^2 = \sum_{i=1}^d \frac{z_i^2}{\vartheta_i}
    , \qquad z_i \sim \mathcal{N}(0, 1) \text{ i.i.d.}
\label{eqn:weighted_chi_squared}
\end{equation}
Unlike the isotropic case, this generalized chi-squared variable has no closed-form
cumulative distribution function, so the exact evaluation in terms of the regularized
gamma function is no longer available. The location of the concentration threshold,
however, follows from~\eqref{eqn:weighted_chi_squared} directly.

Because the $z_i$ are independent with $\mathbb{E} (z_i^2) = 1$ and $\operatorname{Var}
(z_i^2) = 2$,
\begin{equation}
    \mathbb{E} (\norm{x}^2) = \sum_{i=1}^d \frac{1}{\vartheta_i} = \tr (P^{-1})
    , \qquad
    \operatorname{Var} (\norm{x}^2) = 2 \sum_{i=1}^d \frac{1}{\vartheta_i^2} = 2 \tr (P^{-2})
    .
\label{eqn:anisotropic_mean_variance}
\end{equation}
The event of interest is an event for this single scalar variable, since $\norm{x} <
\varepsilon$ if and only if $\norm{x}^2 < \varepsilon^2$. $\norm{x}^2$ concentrates, since its standard deviation $\sqrt{2 \tr (P^{-2})}$
is small compared with its mean $\tr (P^{-1})$ whenever no single eigenvalue dominates
the trace, so $\norm{x}^2$ takes values in a window of width $O (\sqrt{\tr (P^{-2})})$
about $\tr (P^{-1})$ with probability close to one. The threshold therefore sits at the mean and the transition has width $\sqrt{\tr
(P^{-2})}$, and the condition for $P(\norm{x} < \varepsilon)$ to be finite for large $d$ is
\begin{equation}
    \varepsilon^2 > \tr (P^{-1}) = \sum_{i=1}^d \frac{1}{\vartheta_i} \,
    .
\label{eqn:anisotropic_concentration_condition}
\end{equation}
For $P = \vartheta I_d$ we have $\tr (P^{-1}) = d / \vartheta$ and the isotropic
condition is recovered exactly.

The exponential smallness of the probability below the threshold can be made rigorous
without evaluating the distribution of~\eqref{eqn:weighted_chi_squared}. The tail bound
of~\cite{laurent2000adaptive} for weighted sums of squared normals, applied with weights
$1/\vartheta_i$, gives for any $t > 0$
\begin{equation}
    \mathbb{P} \left( \norm{x}^2 \leq \tr (P^{-1}) - 2 \sqrt{\tr (P^{-2}) \, t} \right) \leq e^{-t}
    .
\label{eqn:laurent_massart}
\end{equation}
Hence if $\varepsilon^2 \leq \tr (P^{-1}) - 2 \sqrt{\tr (P^{-2}) \, t}$, then
\begin{equation}
   \mathbb{P} (\norm{x} < \varepsilon) \leq e^{-t} \,.
\end{equation}

In the other direction of the threshold, The Chebyshev--Cantelli inequality applied to $\norm{x}^2$ with mean $\tr (P^{-1})$ and variance $2
\tr (P^{-2})$ gives
\begin{equation}
    \mathbb{P} \left( \norm{x}^2 \geq \tr (P^{-1}) + 2 \sqrt{\tr (P^{-2}) \, t} \right) \leq \frac{1}{1 + 2 t}
\label{eqn:cantelli}
\end{equation}
for any $t > 0$. Hence, if
$\varepsilon^2 \geq \tr (P^{-1}) + 2 \sqrt{\tr (P^{-2}) \, t}$, then
\begin{equation}
    \mathbb{P} \left( \norm{x} < \varepsilon \right) \geq 1 - \frac{1}{1 + 2 t}
    .
\label{eqn:anisotropic_lower_bound}
\end{equation}
A margin of a few multiples of $\sqrt{\tr (P^{-2})}$ above $\tr (P^{-1})$ therefore
already forces the ball probability above a constant, and that constant tends to one as
the margin grows (increasing $t$).

The two statements together place the
transition at~\eqref{eqn:anisotropic_concentration_condition}, with the width $\sqrt{\tr
(P^{-2})}$ reducing to the familiar result of $\sqrt{d} / \vartheta$ in the isotropic case of Appendix~\ref{sec:isotropic_quadratic_functions}.

\section{Langer's Theory for Kramers' Escape Rates with Zero Modes}
\label{sec:langer_theory}

This appendix derives~\eqref{eqn:langer_rate}, the escape rate when the fluctuation
operator at the bottleneck saddle has a zero mode. We follow the modern exposition
of~\cite{hong2022introduction} that reviews Langer's theory~\cite{langer1969statistical},
specialized throughout to the constant isotropic diffusion of the overdamped Langevin
dynamics~\eqref{eqn:sgd_sde}, which is the case relevant to SGD and PISGD. The argument
has three steps: solve for the steady state near the saddle, integrate its flux across
the dividing surface, and divide by the population of the well.

\paragraph{Setup.} We work with the dynamics~\eqref{eqn:sgd_sde}, $\diff X = - \nabla
f(X) \diff t + \sigma \diff W$, whose stationary distribution is $\mu = (1/Z) \exp(- 2 f
/ \sigma^2)$. The Fokker--Planck equation~\eqref{eqn:sgd_fokker_planck} is a continuity
equation $\partial_t \rho = - \nabla \cdot J$ with current
\begin{equation}
    J = - \frac{\sigma^2}{2} \, \mu \, \nabla \left( \frac{\rho}{\mu} \right)
    ,
\label{eqn:langer_current}
\end{equation}
so the isotropic diffusion constant is $\sigma^2 / 2$. Comparing with the Gibbs form $\mu \propto
e^{-\beta f}$ identifies the inverse temperature $\beta = 2 / \sigma^2$. We keep $\sigma$
explicit below.

Let $A$ be the basin of the local minimum $x^\bullet$ and $B$ the target region beyond
the barrier. Langer's construction replaces the true dynamics by the steady state
$\rho_{\mathrm{ss}}$ that is held at equilibrium in $A$ and absorbed in $B$,
\begin{equation}
    \nabla \cdot J[\rho_{\mathrm{ss}}] = 0 \quad \text{outside } A \cup B
    , \qquad
    \rho_{\mathrm{ss}} = \mu \ \text{ on } A
    , \qquad
    \rho_{\mathrm{ss}} = 0 \ \text{ on } B
    .
\label{eqn:langer_steady_state}
\end{equation}
The escape rate is then the flux through the dividing surface divided by the population
of $A$.

\paragraph{Steady state near the saddle.} Writing $q = \rho_{\mathrm{ss}} / \mu$, the
stationarity condition in~\eqref{eqn:langer_steady_state} becomes, by
\eqref{eqn:langer_current},
\begin{equation}
    \nabla \cdot \left( e^{- 2 f / \sigma^2} \nabla q \right) = 0
    ,
\label{eqn:committor_equation}
\end{equation}
with $q = 1$ on $A$ and $q = 0$ on $B$. Only a neighborhood of the saddle $x^\circ$
matters, since that is where the integrand of the flux is concentrated. Expanding to
quadratic order in the eigenbasis of $\nabla^2 f(x^\circ)$, with coordinates $y$ centered
at $x^\circ$ and eigenvalues $\eta < 0$ and $0 \leq \nu_1 \leq \nu_2 \leq \cdots \leq \nu_{d-1}$
as in Section~\ref{sec:runtime},
\begin{equation}
    f(x^\circ + y) \approx f(x^\circ) + \frac{1}{2} \eta \, y_0^2 + \sum_{j \geq 1} \frac{1}{2} \nu_j y_j^2
    .
\end{equation}
Seeking a solution of~\eqref{eqn:committor_equation} that depends only on the unstable
coordinate $y_0$ reduces it to the ordinary differential equation $q'' - (2 \eta / \sigma^2) y_0 q' = 0$,
whose solution matching $q \rightarrow 1$ as $y_0 \rightarrow - \infty$ and $q
\rightarrow 0$ as $y_0 \rightarrow + \infty$ is the Gaussian error integral
\begin{equation}
    q(y) = \sqrt{\frac{\abs{\eta}}{\pi \sigma^2}} \int_{y_0}^{\infty} \diff u \, \exp\left( - \frac{\abs{\eta} u^2}{\sigma^2} \right)
    .
\label{eqn:committor_solution}
\end{equation}
The escape is thus one-dimensional: the steady state varies only along the unstable
direction.

\paragraph{Flux.} Substituting~\eqref{eqn:committor_solution}
into~\eqref{eqn:langer_current} and integrating over the dividing surface $y_0 = 0$, the
transverse directions factorize into independent Gaussian integrals,
\begin{equation}
    \abs{J} = \frac{\sigma^2}{2} \, \mu(x^\circ) \sqrt{\frac{\abs{\eta}}{\pi \sigma^2}} \prod_{j \geq 1} \int \diff y_j \, \exp\left( - \frac{\nu_j y_j^2}{\sigma^2} \right)
    ,
\label{eqn:langer_flux}
\end{equation}
where the range of each $y_j$ is set by the extent of the saddle region. This factorized form is what accommodates a zero
mode. A direction with $\nu_j > 0$ contributes $\sqrt{\pi \sigma^2 / \nu_j}$, provided
its range is large compared with the Gaussian width $\sigma / \sqrt{\nu_j}$, and
collecting all such directions reproduces the determinant of $\nabla^2 f(x^\circ)$. A
direction with $\nu_j = 0$ has a flat integrand and instead contributes the extent $L$ of
that flat direction. Taking the zero mode to be unique, $0 = \nu_1 < \nu_2 \leq \cdots$,
as holds for the bounce in PISGD, we omit it from the determinant, write the result as
${\det}'$, and retain $L$ in its place. Several zero modes would be handled the same way,
one volume factor each.

\paragraph{Population and rate.} Since $\rho_{\mathrm{ss}} = \mu$ on $A$, the population
is $n = \int_A \diff x \, \mu(x)$, which in the same quadratic approximation about
$x^\bullet$ is a product of Gaussian integrals giving $\det \nabla^2 f(x^\bullet)$. In the
ratio $\Gamma = \abs{J} / n$ the two exponentials
combine into the barrier height $H_f = f(x^\circ) - f(x^\bullet)$, leaving
\begin{equation}
    \Gamma = \frac{\abs{\eta}\, L}{(2 \pi)^{3/2}} \sqrt{\frac{2}{\sigma^2}} \, \exp(- \frac{2 H_f}{\sigma^2}) \left( - \frac{{\det}' \nabla^2 f(x^\circ)}{\det \nabla^2 f(x^\bullet)} \right)^{-1/2}
    ,
\end{equation}
as quoted in the main text in~\eqref{eqn:langer_rate}.

\section{Functional Determinants by the Gel'fand--Yaglom Method}
\label{app:gelfand_yaglom}

This appendix evaluates the ratio of functional determinants entering the PISGD escape
rate~\eqref{eqn:Gamma_PISGD}, using the Gel'fand--Yaglom
method~\cite{dunne2005beyond,dunne2008functional,shen2023vacuum}. Throughout we work
with a single coordinate, as is sufficient for the separable potentials of
Section~\ref{sec:runtime}, and we write the fluctuation operators at the bounce and at the
local minimum as
\begin{equation}
    \mathcal{M} = - \frac{\diff^2}{\diff \tau^2} + \widetilde{\lambda}^2 V''(x_{\mathrm{cl}}(\tau))
    , \quad
    \mathcal{M}_{\mathrm{free}} = - \frac{\diff^2}{\diff \tau^2} + \widetilde{\lambda}^2 V''(x^\bullet)
    ,
\label{eqn:gy_operators}
\end{equation}
so that the quantity of interest is ${\det}' \mathcal{M} / \det \mathcal{M}_{\mathrm{free}}$.
The paths obey periodic boundary conditions on an interval of extent $T$. We adopt four
conventions that simplify the expressions below: the interval is centered at the origin,
so that it is $[-T/2, T/2]$; the bounce is centered at $\tau_c = 0$; the coordinate is
shifted so that $x^\bullet = 0$; and the stationary points are ordered as $0 = x^\bullet <
x^\circ < x^\star$. The last convention fixes the signs of the bounce data below: the
bounce excursion is positive, so $x_\infty > 0$, while its center is a maximum, so
$x_{\mathrm{cl}}''(0) < 0$. The bounce $x_{\mathrm{cl}}$ solves the Euler--Lagrange
equation
\begin{equation}
    \frac{\diff^2 x_{\mathrm{cl}}}{\diff \tau^2} - \widetilde{\lambda}^2 V'(x_{\mathrm{cl}}) = 0
    .
\label{eqn:gy_euler_lagrange}
\end{equation}

\subsection{Rescaling}

Write $\omega^2 = \widetilde{\lambda}^2 V''(x^\bullet)$ for the curvature at the local
minimum, and rescale imaginary time as $\tildetau := \omega \tau$ and $\widetilde{T} :=
\omega T$. The operators become
\begin{equation}
    \widetilde{\mathcal{M}} = \mathcal{M} / \omega^2 = - \frac{\diff^2}{\diff \tildetau^2} + 1 + \mathcal{V} (\tildetau)
    , \quad
    \widetilde{\mathcal{M}}_{\mathrm{free}} = \mathcal{M}_{\mathrm{free}} / \omega^2 = - \frac{\diff^2}{\diff \tildetau^2} + 1
    ,
\label{eqn:gy_rescaled_operators}
\end{equation}
with
\begin{equation}
    \mathcal{V} (\tildetau) := \frac{\widetilde{\lambda}^2 V''(x_{\mathrm{cl}}( \tildetau / \omega))}{\omega^2} - 1
    ,
\label{eqn:gy_potential}
\end{equation}
which decays as $\tildetau$ moves away from the bounce center $\tildetau_c = 0$. The
rescaling multiplies every eigenvalue by $\omega^2$, and since the reduced determinant in
the numerator omits one mode relative to the denominator, one factor survives,
\begin{equation}
    \frac{{\det}' \mathcal{M}}{\det \mathcal{M}_{\mathrm{free}}} = \frac{1}{\omega^2} \, \frac{{\det}' \widetilde{\mathcal{M}}}{\det \widetilde{\mathcal{M}}_{\mathrm{free}}}
    ,
\label{eqn:gy_rescaling_ratio}
\end{equation}
so it suffices to compute ${\det}' \widetilde{\mathcal{M}} / \det
\widetilde{\mathcal{M}}_{\mathrm{free}}$.

\subsection{Parity Decomposition}

Since the bounce is symmetric about its center, $\mathcal{V}(-\tildetau) =
\mathcal{V}(\tildetau)$, and the eigenfunctions of $\widetilde{\mathcal{M}}$ can be chosen
to be parity-even or parity-odd. The two sectors inherit different boundary conditions on
the half-interval $[0, \widetilde{T}/2]$.

In the even sector, $u(\tildetau) = u(-\tildetau)$ gives $u'(0) = 0$. At the endpoint,
oddness of the derivative gives $u'(\widetilde{T}/2) = - u'(-\widetilde{T}/2)$, while
periodicity gives $u'(\widetilde{T}/2) = u'(-\widetilde{T}/2)$; together these force
$u'(\widetilde{T}/2) = 0$. The even sector therefore obeys Neumann conditions $u'(0) =
u'(\widetilde{T}/2) = 0$.

In the odd sector, $u(\tildetau) = -u(-\tildetau)$ gives $u(0) = 0$, and the same pairing
of oddness with periodicity at the endpoint gives $u(\widetilde{T}/2) = 0$. The odd sector
therefore obeys Dirichlet conditions $u(0) = u(\widetilde{T}/2) = 0$.

This decomposition also separates the two special modes of the fluctuation operator. The negative mode is nodeless
and even, while the translational zero mode is proportional to $\diff x_{\mathrm{cl}}
/ \diff \tau$ and odd, with $\diff x_{\mathrm{cl}}
/ \diff \tau|_{\tau=0} = 0$. The determinant ratio factorizes as
\begin{equation}
    \frac{{\det}' \widetilde{\mathcal{M}}}{\det \widetilde{\mathcal{M}}_{\mathrm{free}}}
    = \frac{\det \widetilde{\mathcal{M}}^{\mathrm{even}}}{\det \widetilde{\mathcal{M}}^{\mathrm{even}}_{\mathrm{free}}}
    \cdot
    \frac{{\det}' \widetilde{\mathcal{M}}^{\mathrm{odd}}}{\det \widetilde{\mathcal{M}}^{\mathrm{odd}}_{\mathrm{free}}}
    ,
\label{eqn:gy_factorization}
\end{equation}
where $\det'$ excludes the zero mode.

\subsection{Even Sector}

The Gel'fand--Yaglom method replaces the boundary-value problem by an initial-value
problem. Let $u$ and $u^{\mathrm{free}}$ solve
\begin{equation}
    \widetilde{\mathcal{M}}^{\mathrm{even}} u = 0
    , \quad
    \widetilde{\mathcal{M}}^{\mathrm{even}}_{\mathrm{free}} u^{\mathrm{free}} = 0
    ,
\end{equation}
both with $u(0) = u^{\mathrm{free}}(0) = 1$ and $u'(0) = u^{\mathrm{free}\prime}(0) = 0$.
The determinant ratio is then read off at the endpoint,
\begin{equation}
    \frac{\det \widetilde{\mathcal{M}}^{\mathrm{even}}}{\det \widetilde{\mathcal{M}}^{\mathrm{even}}_{\mathrm{free}}} = \frac{u' (\widetilde{T} / 2)}{u^{\mathrm{free}\prime} (\widetilde{T} / 2)}
    ,
\label{eqn:gy_even_ratio}
\end{equation}
matching the Neumann conditions of the even sector. The free solution is
$u^{\mathrm{free}}(\tildetau) = \cosh(\tildetau)$.

Both $u'$ and $u^{\mathrm{free}\prime}$ grow exponentially at large $\widetilde{T}$, so it
is numerically preferable to track their ratio directly. Define
\begin{equation}
    \mathcal{Q} (\tildetau) = \frac{u' (\tildetau)}{u^{\mathrm{free}\prime} (\tildetau)}
    , \quad
    \mathcal{R} (\tildetau) = \frac{u (\tildetau)}{u^{\mathrm{free}} (\tildetau)}
    ,
\label{eqn:gy_QR}
\end{equation}
so that the answer is $\mathcal{Q}(\widetilde{T}/2)$. Differentiating $\mathcal{R}$ and
using $u^{\mathrm{free}\prime} / u^{\mathrm{free}} = \tanh(\tildetau)$ gives
\begin{equation}
    \mathcal{R}' (\tildetau) = \tanh (\tildetau) \left( \mathcal{Q} (\tildetau) - \mathcal{R} (\tildetau) \right)
    ,
\label{eqn:gy_R_prime}
\end{equation}
that is, $\mathcal{Q} = \mathcal{R} + \coth(\tildetau) \mathcal{R}'$. Differentiating once
more and using $u'' = (1 + \mathcal{V}) u$ together with $u^{\mathrm{free}\prime\prime} =
u^{\mathrm{free}}$ eliminates $\mathcal{Q}$ entirely,
\begin{equation}
    \mathcal{R}'' (\tildetau) + 2 \tanh (\tildetau) \mathcal{R}' (\tildetau) - \mathcal{V} (\tildetau) \mathcal{R} (\tildetau) = 0
    , \quad \mathcal{R} (0) = 1, \quad \mathcal{R}' (0) = 0
    .
\label{eqn:gy_R_ode}
\end{equation}
Solving~\eqref{eqn:gy_R_ode} and evaluating $\mathcal{Q}(\widetilde{T}/2)$ through
$\mathcal{Q} = \mathcal{R} + \coth(\tildetau) \mathcal{R}'$ yields the even-sector ratio.
It is negative for large but finite $\widetilde{T}$, which confirms that the bounce carries
exactly one unstable direction, as required for Langer's theory to apply.

The negative eigenvalue appears on its own in~\eqref{eqn:langer_rate}, not only inside the
determinant, so its value is needed separately. It is the lowest eigenvalue $\mu_0$ of the
Sturm--Liouville problem
\begin{equation}
    \left( - \frac{\diff^2}{\diff \tildetau^2} + 1 + \mathcal{V} (\tildetau) \right) u (\tildetau) = \mu \, u (\tildetau)
\end{equation}
on $[0, \widetilde{T}/2]$ with Neumann conditions, and $\eta_{S(\widetilde{\lambda})} =
\omega^2 \mu_0$ in the unscaled variables.

\subsection{Odd Sector and the Zero Mode}

The odd sector contains the imaginary-time translational zero mode, so $\det
\widetilde{\mathcal{M}}^{\mathrm{odd}} = 0$ and the ratio must be regularized. Following
the standard treatment~\cite{dunne2005beyond}, shift both operators by a small constant
$k^2$,
\begin{equation}
    \widetilde{\mathcal{M}}^{\mathrm{odd}} (k^2) = \widetilde{\mathcal{M}}^{\mathrm{odd}} + k^2 = - \frac{\diff^2}{\diff \tildetau^2} + 1 + k^2 + \mathcal{V} (\tildetau)
    ,
\label{eqn:gy_shifted_operator}
\end{equation}
\begin{equation}
    \widetilde{\mathcal{M}}^{\mathrm{odd}}_{\mathrm{free}} (k^2) = \widetilde{\mathcal{M}}^{\mathrm{odd}}_{\mathrm{free}} + k^2 = - \frac{\diff^2}{\diff \tildetau^2} + 1 + k^2
    ,
\label{eqn:gy_shifted_operator_free}
\end{equation}
so that every eigenvalue is raised by $k^2$ and in particular the zero mode acquires
eigenvalue $k^2$. We then consider
\begin{equation}
    \widetilde{\mathcal{M}}^{\mathrm{odd}} (k^2) u_{k^2} = 0
    , \quad
    \widetilde{\mathcal{M}}^{\mathrm{odd}}_{\mathrm{free}} (k^2) u^{\mathrm{free}}_{k^2} = 0
    ,
\end{equation}
now with Dirichlet-type initial data $u_{k^2}(0) = u^{\mathrm{free}}_{k^2}(0) = 0$ and
$u'_{k^2}(0) = u^{\mathrm{free}\prime}_{k^2}(0) = 1$. The free solutions are
\begin{equation}
    u^{\mathrm{free}}_{0} (\tildetau) = \sinh (\tildetau)
    , \quad
    u^{\mathrm{free}}_{k^2} (\tildetau) = \frac{\sinh ( \tildetau \sqrt{1 + k^2} )}{\sqrt{1 + k^2}}
    .
\end{equation}
Since one eigenvalue $k^2$ vanishes as $k^2 \to 0$,
\begin{equation}
    \frac{u_{k^2} (\widetilde{T} / 2)}{u^{\mathrm{free}}_{k^2} (\widetilde{T} / 2)} = \frac{\det \widetilde{\mathcal{M}}^{\mathrm{odd}} (k^2)}{\det \widetilde{\mathcal{M}}^{\mathrm{odd}}_{\mathrm{free}} (k^2)} \approx k^2 \, \frac{{\det}' \widetilde{\mathcal{M}}^{\mathrm{odd}}}{\det \widetilde{\mathcal{M}}^{\mathrm{odd}}_{\mathrm{free}}}
    ,
\label{eqn:gy_odd_regularized}
\end{equation}
so the reduced determinant is recovered as the coefficient of $k^2$.

To extract that coefficient, pair the shifted equation against the exact zero mode $u_0$ and
integrate by parts,
\begin{equation}
\begin{aligned}
    0 = & \, \int_0^{\widetilde{T} / 2} \diff \tildetau \, u_0 \left[ - u_{k^2}'' + \left( 1 + \mathcal{V} + k^2 \right) u_{k^2} \right] \\
    = & \, \left.\left( - u_0 u_{k^2}' + u_0' u_{k^2} \right)\right|_{\tildetau = 0}^{\widetilde{T} / 2} + k^2 \int_0^{\widetilde{T} / 2} \diff \tildetau \, u_0 u_{k^2} \\
    = & \, \left( - u_0(\widetilde{T} / 2) u_{k^2}'(\widetilde{T} / 2)+ u_0'(\widetilde{T} / 2) u_{k^2}(\widetilde{T} / 2) \right) + k^2 \int_0^{\widetilde{T} / 2} \diff \tildetau \, u_0 u_{k^2}
    ,
\end{aligned}
\label{eqn:gy_wronskian}
\end{equation}
where $\widetilde{\mathcal{M}}^{\mathrm{odd}}(k^2) u_{k^2} = 0$ and $\widetilde{\mathcal{M}}^{\mathrm{odd}} u_0 = 0$ have been used and the boundary terms
at the origin cancel. Equation~\eqref{eqn:gy_wronskian} can be reorganized into
\begin{equation}
    u_{k^2} (\widetilde{T} / 2) \left( 1 - \frac{u_0(\widetilde{T} / 2)}{u'_0(\widetilde{T} / 2)} \frac{u'_{k^2}(\widetilde{T} / 2)}{u_{k^2}(\widetilde{T} / 2)} \right) = k^2 \, \frac{\int_0^{\widetilde{T} / 2} \diff \tildetau \, u_0 (\tildetau) u_{k^2} (\tildetau)}{- u_0'(\widetilde{T} / 2)}
    ,
\label{eqn:gy_u_k2_reorganized}
\end{equation}
which exhibits $u_{k^2}(\widetilde{T}/2)$ as $O(k^2)$ once the bracket is shown to have a
finite nonzero limit.

The zero mode itself is fixed by the Euler--Lagrange
equation~\eqref{eqn:gy_euler_lagrange}. Differentiating it with respect to $\tau$ gives
$x_{\mathrm{cl}}''' = \widetilde{\lambda}^2 V''(x_{\mathrm{cl}}) \, x_{\mathrm{cl}}'$, so
$x_{\mathrm{cl}}'$ is annihilated by $\mathcal{M}$. Normalizing so that $u_0(0) = 0$ and
$u_0'(0) = 1$, matching the initial condition above, the zero mode in the rescaled variable is
\begin{equation}
    u_0 (\tildetau) = \omega \, \frac{x_{\mathrm{cl}}' (\tildetau / \omega)}{x_{\mathrm{cl}}'' (0)}
    ,
\label{eqn:gy_zero_mode}
\end{equation}
and one verifies directly that
\begin{equation}
\begin{aligned}
    \widetilde{\mathcal{M}} u_0 (\tildetau) = & \, - u_0'' (\tildetau) + \left( 1 + \mathcal{V} (\tildetau) \right) u_0 (\tildetau) \\
    = & \, - \frac{1}{\omega} \frac{x_{\mathrm{cl}}''' (\tildetau / \omega)}{x_{\mathrm{cl}}'' (0)} + \frac{\widetilde{\lambda}^2 V''(x_{\mathrm{cl}}(\tildetau / \omega))}{\omega^2} \, \omega \, \frac{x_{\mathrm{cl}}' (\tildetau / \omega)}{x_{\mathrm{cl}}'' (0)} \\
    = & \, - \frac{1}{\omega} \frac{1}{x_{\mathrm{cl}}'' (0)} \left( x_{\mathrm{cl}}''' (\tildetau / \omega) - \widetilde{\lambda}^2 V''(x_{\mathrm{cl}}(\tildetau / \omega)) \, x_{\mathrm{cl}}' (\tildetau / \omega) \right) \\
    = & \, 0
    ,
\end{aligned}
\label{eqn:gy_zero_mode_check}
\end{equation}
using~\eqref{eqn:gy_potential} for $\mathcal{V}$ and the differentiated Euler--Lagrange
equation in the last step.

It remains to evaluate the bracket in~\eqref{eqn:gy_u_k2_reorganized}, which requires the
asymptotics of the two solutions near the endpoint. For large but finite $\widetilde{T}$,
the bounce relaxes to the local minimum as
\begin{equation}
    x_{\mathrm{cl}} (\tildetau / \omega) \approx x_{\infty} \exp(- \tildetau)
    ,
\label{eqn:gy_bounce_asymptotics}
\end{equation}
where $x_\infty$ is a constant obtained by extrapolating the numerical profile.
By~\eqref{eqn:gy_zero_mode}, the zero mode inherits this decay,
\begin{equation}
    u_0 (\tildetau) \approx u_{\infty} \exp(- \tildetau) \propto \exp(- \tildetau)
    .
\label{eqn:gy_zero_mode_asymptotics}
\end{equation}

We now turn to the equations shifted by $k^2$. The free solution is
\begin{equation}
    u^{\mathrm{free}}_{k^2} (\tildetau) = \frac{\sinh(\sqrt{1 + k^2} \, \tildetau)}{\sqrt{1 + k^2}}
    .
\label{eqn:gy_free_shifted_solution}
\end{equation}
For the equation associated with $\widetilde{\mathcal{M}}^{\mathrm{odd}}(k^2)$, away from
the bounce region at large $\tildetau$ the potential term decays, $\mathcal{V}(\tildetau)
\to 0$, so this equation also becomes constant-coefficient, $u'' = (1 + k^2) u$, with the
two independent behaviors $\exp(\pm \sqrt{1 + k^2} \, \tildetau)$. The growing branch is
selected,
\begin{equation}
    u_{k^2} (\tildetau) \propto \exp(\sqrt{1 + k^2} \, \tildetau)
    ,
\label{eqn:gy_u_k2_asymptotics}
\end{equation}
matching the behavior of the free solution $u^{\mathrm{free}}_{k^2} (\tildetau) \sim
\exp(\sqrt{1 + k^2} \, \tildetau)$ and consistent
with~\eqref{eqn:gy_odd_regularized} being nonzero at order $k^2$.
Consequently $u_0'(\widetilde{T}/2) / u_0(\widetilde{T}/2) \to -1$ while $u'_{k^2}
(\widetilde{T}/2) / u_{k^2}(\widetilde{T}/2) \to +1$ as $\widetilde{T}$ grows, so the
bracket in~\eqref{eqn:gy_u_k2_reorganized} tends to $1 - (-1) = 2$ and
\begin{equation}
    u_{k^2} (\widetilde{T} / 2) = \frac{1}{2} k^2 \, \frac{\int_0^{\widetilde{T} / 2} \diff \tildetau \, u_0 (\tildetau) u_{k^2} (\tildetau)}{- u_0'(\widetilde{T} / 2)} \left( 1 + O(k^2) \right)
    .
\label{eqn:gy_u_k2_solved}
\end{equation}

Dividing by $u^{\mathrm{free}}_{k^2}(\widetilde{T}/2)$ and comparing
with~\eqref{eqn:gy_odd_regularized}, the factors of $k^2$ cancel and the limit $k^2 \to 0$
may be taken inside the integral, giving
\begin{equation}
    \frac{{\det}' \widetilde{\mathcal{M}}^{\mathrm{odd}}}{\det \widetilde{\mathcal{M}}^{\mathrm{odd}}_{\mathrm{free}}} = \frac{\int_0^{\widetilde{T} / 2} \diff \tildetau \, u_0 (\tildetau)^2}{- 2 \, u^{\mathrm{free}}_{0} (\widetilde{T} / 2) \, u_0'(\widetilde{T} / 2)}
    ,
\label{eqn:gy_odd_general}
\end{equation}
the $O(k^2)$ corrections dropping out because the left-hand side carries no $k^2$
dependence.

Equation~\eqref{eqn:gy_odd_general} can now be evaluated in terms of the bounce
alone. Using the zero mode~\eqref{eqn:gy_zero_mode} and substituting $\tau = \tildetau /
\omega$, the numerator is
\begin{equation}
    \int_0^{\widetilde{T} / 2} \diff \tildetau \, u_0 (\tildetau)^2 = \frac{\omega^3}{(x_{\mathrm{cl}}'' (0))^2} \int_0^{T/2} \diff \tau \, \left( \frac{\diff x_{\mathrm{cl}}}{\diff \tau} \right)^2 = \frac{\omega^3 \, \widetilde{\lambda} B_V}{2 \, (x_{\mathrm{cl}}'' (0))^2}
    ,
\label{eqn:gy_numerator}
\end{equation}
where the kinetic normalization $\int_{[-T/2,T/2]} \diff \tau \, (\diff x_{\mathrm{cl}} / \diff
\tau)^2 = \widetilde{\lambda} B_V$ of the zero-energy bounce has been used, together with
its symmetry about the center. The denominator is
\begin{equation}
    - 2 \, u^{\mathrm{free}}_{0} (\widetilde{T} / 2) \, u_0'(\widetilde{T} / 2) = - 2 \sinh \left( \frac{\omega T}{2} \right) \frac{x_{\mathrm{cl}}'' (T / 2)}{x_{\mathrm{cl}}'' (0)}
    .
\label{eqn:gy_denominator}
\end{equation}
Therefore
\begin{equation}
    \frac{{\det}' \widetilde{\mathcal{M}}^{\mathrm{odd}}}{\det \widetilde{\mathcal{M}}^{\mathrm{odd}}_{\mathrm{free}}} = - \frac{\omega^3 \, \widetilde{\lambda} B_V}{4 \, x_{\mathrm{cl}}'' (0) \, \sinh ( \omega T / 2 ) \, x_{\mathrm{cl}}'' (T / 2)}
    .
\label{eqn:gy_odd_closed_form}
\end{equation}

The apparent exponential growth of $\sinh(\omega T/2)$ is cancelled by the decay of the
bounce. Using $x_{\mathrm{cl}}''(\tau) \approx x_\infty \omega^2 \exp(- \omega \tau)$ at
large $\tau$,
\begin{equation}
    \sinh \left( \frac{\omega T}{2} \right) x_{\mathrm{cl}}'' \left( \frac{T}{2} \right) \approx \frac{\omega^2 x_{\infty}}{2}
    ,
\end{equation}
so~\eqref{eqn:gy_odd_closed_form} tends to a finite limit as $T$ grows. Combining it
with the even-sector ratio $\mathcal{Q}(\widetilde{T}/2)$
through~\eqref{eqn:gy_factorization} supplies the determinant ratio
in~\eqref{eqn:Gamma_PISGD}, and hence the full escape-rate prefactor.

\subsection{Collected Result}

Assembling the two sectors through~\eqref{eqn:gy_factorization} and restoring the
rescaling of~\eqref{eqn:gy_rescaling_ratio}, the determinant ratio entering the escape
rate~\eqref{eqn:Gamma_PISGD} is
\begin{equation}
\begin{aligned}
    \frac{{\det}' \mathcal{M}}{\det \mathcal{M}_{\mathrm{free}}}
    = & \, \frac{1}{\omega^2} \, \frac{\det \widetilde{\mathcal{M}}^{\mathrm{even}}}{\det \widetilde{\mathcal{M}}^{\mathrm{even}}_{\mathrm{free}}} \, \frac{{\det}' \widetilde{\mathcal{M}}^{\mathrm{odd}}}{\det \widetilde{\mathcal{M}}^{\mathrm{odd}}_{\mathrm{free}}} \\
    = & \, - \frac{\omega \, \widetilde{\lambda} B_V \, \mathcal{Q} (\widetilde{T} / 2)}{4 \, x_{\mathrm{cl}}'' (0) \, \sinh ( \omega T / 2 ) \, x_{\mathrm{cl}}'' (T / 2)}
    ,
\end{aligned}
\label{eqn:gy_collected}
\end{equation}
where $\mathcal{Q}(\widetilde{T} / 2)$ is obtained by solving~\eqref{eqn:gy_R_ode}
numerically and is itself negative, so the ratio as a whole is negative, as required for
the square root in~\eqref{eqn:Gamma_PISGD} to be real. In the large-$T$
regime relevant for the escape-rate analysis, the hyperbolic factor saturates as above and
\begin{equation}
    \frac{{\det}' \mathcal{M}}{\det \mathcal{M}_{\mathrm{free}}} \longrightarrow - \frac{\widetilde{\lambda} B_V \, \mathcal{Q} (\infty)}{2 \, \omega \, x_{\infty} \, x_{\mathrm{cl}}'' (0)}
    ,
\label{eqn:gy_collected_asymptotic}
\end{equation}
so every quantity entering the prefactor is fixed by the bounce profile together with a
single numerical solution of~\eqref{eqn:gy_R_ode}. The negative eigenvalue
$\eta_{S(\widetilde{\lambda})}$, which appears separately in~\eqref{eqn:Gamma_PISGD}, is
obtained from the Sturm--Liouville problem of the even sector.

\section{Separable Potentials with Quartic Functions}
\label{sec:quartic_functions}

A natural class of benchmarks is given by separable potentials of the form
\begin{equation*}
    V(x) = \sum_{i=0}^{d-1} V_i(x_i),
    \qquad V_i \equiv U \quad \text{for every } i,
\end{equation*}
where $U$ is the common one-dimensional asymmetric quartic double-well potential. We use $z$ for its scalar argument, so no coordinate index is needed in the one-dimensional analysis below.

To match the notation in the main text, we denote the barrier height and bounce action of $U$ by $H_V$ and $B_V$, respectively. Equivalently, these are the common values of $H_{V_i}$ and $B_{V_i}$.

A quartic function is the lowest-degree polynomial that can have two distinct local minima. A general quartic function has five coefficients. Four can be fixed by affine changes of the horizontal and vertical axes: a shift and rescaling of $z$, and a shift and rescaling of the function value, which will leave one dimensionless shape parameter $\alpha$ as follows. We use the canonical parameterization~\cite{dunne2005beyond,shen2023vacuum}
\begin{equation}
    U(z; \alpha, c) = c\,q_\alpha(z),
    \qquad
    q_\alpha(z) = \frac{1}{2}z^2 - \frac{1}{2}z^3 + \frac{1}{8}\alpha z^4,
\end{equation}
with $0<\alpha\leq 1$ and $c>0$. For $0<\alpha<1$, $U$ has a metastable local minimum, a saddle point, and a global minimum at
\begin{equation}
    z^\bullet = 0,
    \qquad
    z^\circ = \frac{3-\sqrt{9-8\alpha}}{2\alpha},
    \qquad
    z^\star = \frac{3+\sqrt{9-8\alpha}}{2\alpha},
\end{equation}
respectively. At $\alpha=1$, the two minima are degenerate. The first positive turning point satisfying $U(z^\mathrm{tp})=U(z^\bullet)$ is
\begin{equation}
    z^\mathrm{tp} = \frac{2\left(1-\sqrt{1-\alpha}\right)}{\alpha}.
\end{equation}

The barrier height is
\begin{equation}
    H_V = U(z^\circ)-U(z^\bullet)
    = c\,
    \frac{\left(3-\sqrt{9-8\alpha}\right)^2
    \left(4\alpha-3+\sqrt{9-8\alpha}\right)}{64\alpha^3}.
\end{equation}

The normalized curvature parameter at the global minimum is defined as
\begin{equation}
    (\bar{\omega}^\star)^2
    := \frac{U''(z^\star;\alpha,c)}{c}
    = \frac{9-8\alpha+3\sqrt{9-8\alpha}}{4\alpha},
\end{equation}
so $U''(z^\star;\alpha,c)=c(\bar{\omega}^\star)^2$. Therefore, the actual curvature scale used in the main text is
\begin{equation}
    \omega^\star:=\sqrt{U''(z^\star;\alpha,c)}
    =\sqrt{c}\,\bar{\omega}^\star,
    \qquad
    U''(z^\star;\alpha,c)=(\omega^\star)^2=c(\bar{\omega}^\star)^2.
\end{equation}

The bounce action is twice the Agmon distance from $z^\bullet$ to $z^\mathrm{tp}$ and is given by
\begin{equation}
\begin{aligned}
    B_V
    = ~& 2\sqrt{2}\,
            \int_{z^\bullet}^{z^\mathrm{tp}}
        \!\sqrt{U(z) - U(z^\bullet)}\;\mathrm{d}z \\
    = ~& 2\sqrt{2 c}\,
      \int_{0}^{z^\mathrm{tp}}
        \!\sqrt{q_\alpha(z)}\;\mathrm{d}z \\
    = ~& 2\sqrt{c}\left[
        \frac{2(3-2\alpha)}{3\alpha^{2}}
        -\frac{2(1-\alpha)}{\alpha^{5/2}}
        \ln\!\left(\frac{1+\sqrt{\alpha}}{\sqrt{1-\alpha}}\right)
        \right]
        .
\end{aligned}
\label{eqn:quartic_function_instanton_action}
\end{equation}
where the final expression at $\alpha=1$ is defined by continuity in terms of $\alpha$.

By~\eqref{eqn:main_results_separable}, PISGD has a smaller leading runtime exponent than SGD and therefore a scaling advantage in the considered setting when
\begin{equation}
    \frac{B_V}{2\sqrt{U''(z^\star)}}
    <
    \frac{H_V}{U''(z^\star)}.
\label{eqn:quartic_pisgd_advantage_condition}
\end{equation}
Both sides are invariant under the overall scale $c$. Therefore, the comparison depends only on $\alpha$.

To evaluate it, define the two rescaled quantities
\begin{equation}
    A(\alpha) := \frac{B_V}{2\sqrt{c}}
    = \sqrt{2}\int_0^{z^\mathrm{tp}}\!\sqrt{q_\alpha(z)}\,\mathrm{d}z,
    \qquad
    C(\alpha) := \frac{H_V}{\sqrt{cU''(z^\star)}}
    = \frac{H_V}{c\bar{\omega}^\star}.
\end{equation}
For $0<\alpha<1$, differentiating $A(\alpha)$ under the integral sign gives
\begin{equation}
\begin{aligned}
    \frac{\diff A}{\diff \alpha}
    ={}& \sqrt{2}
        \left(\frac{\diff}{\diff\alpha}
        \frac{2\left(1-\sqrt{1-\alpha}\right)}{\alpha}\right)
        \left.\sqrt{\frac{z^2}{2}-\frac{z^3}{2}+\frac{\alpha z^4}{8}}
        \right|_{z=z^\mathrm{tp}} \\
    &+ \sqrt{2}\int_0^{z^\mathrm{tp}}
        \frac{\partial}{\partial\alpha}
        \sqrt{\frac{z^2}{2}-\frac{z^3}{2}+\frac{\alpha z^4}{8}}\,\mathrm{d}z \\
    ={}& \frac{\sqrt{2}}{16}
        \int_0^{z^\mathrm{tp}}
        \frac{z^4}
        {\sqrt{\frac{z^2}{2}-\frac{z^3}{2}+\frac{\alpha z^4}{8}}}\,\mathrm{d}z
        >0.
\end{aligned}
\end{equation}
Here the first term vanishes because $q_\alpha(z^\mathrm{tp})=0$, while the remaining integrand is positive for $0<z<z^\mathrm{tp}$. Thus $A(\alpha)$ is increasing, with
\begin{equation}
    A(\alpha)
    > \lim_{\alpha\to0^+}A(\alpha)
    = \int_0^1 z\sqrt{1-z}\,\mathrm{d}z
    = \frac{4}{15}.
\label{eqn:quartic_action_lower_bound}
\end{equation}

The normalized barrier height is also increasing since direct differentiation gives
\begin{equation}
    \frac{\diff}{\diff\alpha}\frac{H_V}{c}
    = \frac{4}{
        81-72\alpha
        +(27-12\alpha)\sqrt{9-8\alpha}
        +8\alpha^2}
    > 0.
\end{equation}
Meanwhile,
\begin{equation}
    \frac{\mathrm{d}}{\mathrm{d}\alpha}(\bar{\omega}^\star)^2
    = -\frac{3\left(9-4\alpha+3\sqrt{9-8\alpha}\right)}
    {4\alpha^2\sqrt{9-8\alpha}}
    <0,
\end{equation}
so $\bar{\omega}^\star$ is decreasing. Consequently, $C(\alpha)=(H_V/c)/\bar{\omega}^\star$ is increasing and satisfies
\begin{equation}
    C(\alpha)
    \leq C(1)
    = \left.\frac{H_V}{c\bar{\omega}^\star}\right|_{\alpha=1}
    = \frac{1}{8}.
\label{eqn:quartic_barrier_upper_bound}
\end{equation}
Combining~\eqref{eqn:quartic_action_lower_bound} and~\eqref{eqn:quartic_barrier_upper_bound} yields
\begin{equation}
    \frac{\displaystyle B_V/\bigl(2\sqrt{U''(z^\star)}\bigr)}
         {\displaystyle H_V/U''(z^\star)}
    = \frac{A(\alpha)}{C(\alpha)}
    > \frac{32}{15}>2.
\label{eqn:quartic_runtime_exponent_ratio}
\end{equation}
The lower bound $32/15$ is not tight. In particular, the family of quartic functions lies on the opposite side of the PISGD advantage condition~\eqref{eqn:quartic_pisgd_advantage_condition}: for a $d$-dimensional separable potential with identical quartic components, the leading PISGD runtime exponent is more than twice the corresponding SGD exponent.

\section{Squared Sum of Quartics: Barrier and Action Analysis}\label{app:squared_quartic_analysis}

This appendix derives the barrier heights and bounce actions~\eqref{eqn:squared_quartic_barriers} used in Section~\ref{sec:nonseparable_potentials} for the potential $V_d$ in~\eqref{eqn:squared_quartic_potential}. Let $f(x):=b_d+\sum_{i=0}^{d-1}q_\alpha(x_i)$, so that $V_d=cf^2$ and $\nabla V_d=2cf\nabla f$. Because $f\geq\kappa_0>0$, the stationary points of $V_d$ coincide with those of $f$, namely the $3^d$ points $\{z^\bullet,z^\circ,z^\star\}^d$. Because squaring is monotone on positive values, the ordering of their values is also preserved. $V_d$ is invariant under permutations of the coordinates, so the value at a stationary point depends only on the numbers $n^\star$, $n^\circ$, and $n^\bullet$ of coordinates at $z^\star$, $z^\circ$, and $z^\bullet$, with $n^\star+n^\circ+n^\bullet=d$. We use $q_\alpha(z^\bullet)=0$ and $b_d=\kappa_0-d\,q^\star$ from~\eqref{eqn:squared_quartic_offset}.

\subsection{SGD: Spectral Gap of the Witten Laplacian}

We apply the labeling procedure of~\cite{Michel_2019,shi2023learning}. The index-1 saddles have $n^\circ=1$, and the local minima have $n^\circ=0$. Their values are
\begin{equation}
\begin{aligned}
    V_{\mathrm{sad}}(n^\star)&=c\left(\kappa_0-(d-n^\star)q^\star+q^\circ\right)^2,
    &\qquad 0&\leq n^\star\leq d-1,\\
    V_{\mathrm{min}}(n^\star)&=c\left(\kappa_0-(d-n^\star)q^\star\right)^2,
    &\qquad 0&\leq n^\star\leq d.
\end{aligned}
\end{equation}
Since $q^\star<0$, both decrease as $n^\star$ increases. In the notation of~\cite{shi2023learning}, the levels are therefore $\nu_0=\infty$ and $\nu_i=V_{\mathrm{sad}}(i-1)$ for $1\leq i\leq d$. The level $\nu_0$ is paired with the global minimum $x^\star$, which has $n^\star=d$. The sublevel set $\{V_d<\nu_1\}$ has a connected component that does not contain $x^\star$, and its minimum is $(z^\bullet,\ldots,z^\bullet)$, with $n^\star=0$. Proceeding inductively, $\nu_i$ is paired with the $\binom{d}{i-1}$ minima $x^\bullet_{i,j}$ with $n^\star=i-1$, whose values are $V_{\mathrm{min}}(i-1)$. The $2^d-1$ smallest positive eigenvalues of the Witten Laplacian are thus governed by the differences
\begin{equation}
    \nu_i-V_d(x^\bullet_{i,j})=H_V(i-1),
    \qquad
    H_V(n^\star):=V_{\mathrm{sad}}(n^\star)-V_{\mathrm{min}}(n^\star)
    =c\,q^\circ\left(q^\circ+2\kappa_0-2(d-n^\star)q^\star\right),
\end{equation}
each with multiplicity $\binom{d}{i-1}$. $H_V(n^\star)$ is also the barrier height of the one-dimensional section through a single coordinate, with the remaining coordinates fixed as above. It decreases with $n^\star$, so the spectral gap is set by its largest value, $H_V^{\max}=H_V(0)$. The smallest value $H_V^{\min}=H_V(d-1)$ does not depend on $d$.

\subsection{PISGD: Bounce Actions by Analogy}

No analogous spectral theory with mathematical rigor is available for the PISGD path measure. We therefore give a nonrigorous analysis by analogy with the SGD case. Consider tunneling of a single coordinate $x_i$ from $z^\bullet$ to $z^\star$ between nearest-neighbor minima, so that $n^\star$ changes by one. Among the other coordinates, $n^\star$ sit at $z^\star$ and $d-1-n^\star$ at $z^\bullet$. We assume that the bounce moves only along $x_i$.~\footnote{Moving other coordinates away from their minima increases both the potential along the path and the path length, and hence the action.} The one-dimensional section is
\begin{equation}
    U_{n^\star}(z)=c\left(q_\alpha(z)+\kappa_0-(d-n^\star)q^\star\right)^2,
\end{equation}
and its bounce action is
\begin{equation}
    B_V(n^\star):=2\sqrt{2}\int_{z^\bullet}^{z^{\mathrm{tp}}}\diff z \sqrt{U_{n^\star}(z)-U_{n^\star}(z^\bullet)}
    =2\sqrt{2c}\int_{z^\bullet}^{z^{\mathrm{tp}}} \diff z\sqrt{q_\alpha(z)\left(q_\alpha(z)+2\kappa_0-2(d-n^\star)q^\star\right)}.
\end{equation}
The turning point is the same as $z^{\mathrm{tp}}=2(1-\sqrt{1-\alpha})/\alpha$ of $q_\alpha$, which does not depend on $n^\star$. Like $H_V(n^\star)$, $B_V(n^\star)$ decreases with $n^\star$.
We therefore hypothesize that the PISGD spectral gap is governed by $B_V^{\max}=B_V(0)$.~\footnote{The analogy is not exact. The PISGD action landscape is infinite-dimensional, so finite-dimensional Morse theory does not apply directly. With periodic boundary conditions, the bounce is not an index-1 saddle, because of the zero mode from imaginary-time translations. Moreover, a subtlety is that the action barrier is measured relative to the constant path at $z^\bullet$, whereas the action difference between the constant paths at $z^\bullet$ and at $z^\star$ grows in proportion to $T$.}
The smallest value, $B_V^{\min}=B_V(d-1)$, does not depend on $d$. For large $d$, the integrand of $B_V(0)$ is $\sqrt{2d\abs{q^\star}q_\alpha(z)}\,(1+O(1/d))$, so $B_V^{\max}=O(\sqrt{d})$.

\section{Construction of the Escape-Dynamics Figure}
\label{app:escape_dynamics_figure}

This appendix describes the simulations and event alignment used to construct Figure~\ref{fig:main_fig}. The plotted configurations are snapshots from actual PISGD and SGD runs rather than schematic paths.

\subsection{Potential and Algorithm Parameters}

We use the one-dimensional variant quartic potential from Section~\ref{sec:numerics_variant_quartic_func} with
\begin{equation*}
    \alpha = 0.9, \qquad c = 100, \qquad \bar{\omega}^\star = 0.1.
\end{equation*}
Its relevant landmarks are
\begin{equation*}
    x^\bullet = 0, \qquad x^\circ \approx 0.921, \qquad x^{\mathrm{tp}} \approx 1.520, \qquad x^\star \approx 38.03.
\end{equation*}
This is the same potential used in the corresponding runtime experiments.

PISGD follows the update rule~\eqref{eqn:update_rule} with learning rate $s = 0.01$, noise strength $\sigma = \sqrt{2}$, lattice spacing $a = 0.025$, $N_T = 400$ imaginary-time slices, and imaginary-time length $T = N_T a = 10$. These parameters are already in the normalized convention~\eqref{eqn:hyperparameter_normalization}, with $\widetilde{\lambda} = 1$ and the overall scale of the potential carried by $c$.

For the SGD comparison, we use the same implementation with $N_T = 1$ and an immaterial lattice spacing $a = 1$. The kinetic term in~\eqref{eqn:gradient_action} then vanishes, and the update reduces to
\begin{equation}
    x^{(k+1)} = x^{(k)} - s V'(x^{(k)}) + \sqrt{2s}\,\xi^{(k)},
\label{eqn:figure_sgd_update}
\end{equation}
with learning rate $s = 0.01$ and $\widetilde{\lambda}_{\mathrm{SGD}} = 1$.

For each method, we perform $100$ independent runs with distinct random seeds. Every run is initialized at the local minimum, with $x_j^{(0)} = x^\bullet$ for every slice, and the configuration after each update is recorded.

\subsection{Stopping Criterion and Reference Events}

Let $k_\varepsilon$ be the first iteration satisfying
\begin{equation}
    \abs{x_0^{(k_\varepsilon)} - x^\star} < \varepsilon, \qquad \varepsilon = 0.01,
\label{eqn:figure_target_event}
\end{equation}
where the slice $x_0$ is used as the output, as in Algorithm~\ref{alg:pisgd}. After this event, the simulation continues for another $10^4$ iterations. All $100$ runs of both methods reach the target. We define two additional first-passage events:
\begin{itemize}
    \item The \emph{turning-point crossing} $k_{\mathrm{tp}}$ is the first iteration at which at least one time slice satisfies $x_j^{(k)} > x^{\mathrm{tp}}$. For PISGD, $j_{\mathrm{tp}}$ denotes the smallest slice index that crosses at this iteration.
    \item The \emph{saddle crossing} $k_\circ$ is the first iteration at which at least one time slice satisfies $x_j^{(k)} > x^\circ$. This event is used only for SGD, whose transition state is the saddle $x^\circ$. The transition state of PISGD is instead the bounce, which reaches $x^{\mathrm{tp}}$.
\end{itemize}

\subsection{Event Alignment and Panel Construction}

Escape times vary widely among runs, so snapshots taken at one absolute iteration would combine configurations at different stages of escape. Each panel therefore contains the configuration from every available run at iteration
\begin{equation*}
    k = k_{\mathrm{ref}} + \Delta k,
\end{equation*}
where $k_{\mathrm{ref}}$ is an event time for that run and $\Delta k$ is a fixed panel-dependent offset. Thin curves represent individual runs and the thick curve is their mean.

The periodic boundary condition makes the location of a PISGD bounce along imaginary time arbitrary. We remove this translational degree of freedom by cyclically shifting every PISGD configuration by $N_T/2-j_{\mathrm{tp}}$, placing the first slice that crosses $x^{\mathrm{tp}}$ at the center $\tau=0$ of the plotted axis. The dashed curve is the periodic bounce solution of~\eqref{eqn:euler_lagrange} on $[0,T]$, obtained by solving the boundary-value problem numerically and centered at the same point. Each SGD iterate is a scalar and is drawn as a constant path so that the two rows can share the same axes. The vertical axis is broken between $x^{\mathrm{tp}}$ and $x^\star$ because the global minimum lies far beyond the barrier region.

The upper row uses $k_{\mathrm{ref}}=k_{\mathrm{tp}}$ and offsets
\begin{equation*}
    \Delta k = -250,\ -20,\ 0,\ +25{,}000,\ +45{,}000.
\end{equation*}
The five PISGD panels in Figure~\ref{fig:main_fig} correspond to these offsets and are labeled, from left to right, \emph{metastable basin} (near $x^\bullet$), \emph{bounce formation}, \emph{critical bounce} (the transition state), \emph{downhill relaxation}, and \emph{target basin} (near $x^\star$). All $100$ runs contribute to each of the first three panels, whereas the $+25{,}000$ and $+45{,}000$ panels contain $99$ and $7$ runs, respectively. The smaller late-time sample sizes arise because each run ends $10^4$ iterations after reaching the target.

The lower row uses offsets
\begin{equation*}
    \Delta k = -1{,}000,\ -3,\ 0,\ +250,\ +10{,}000.
\end{equation*}
The offsets $-1{,}000$, $+250$, and $+10{,}000$ are measured relative to the turning-point crossing $k_{\mathrm{tp}}$, whereas $-3$ and $0$ are measured relative to the saddle crossing $k_\circ$. This mixed convention is partly a visualization choice: referencing the final two panels to $k_{\mathrm{tp}}$ aligns the runs at comparable post-crossing stages, whereas referencing them to $k_\circ$ would produce a visibly more dispersed ensemble.

The five SGD panels in Figure~\ref{fig:main_fig} correspond to these offsets and are labeled, from left to right, \emph{metastable basin} (near $x^\bullet$), \emph{uphill climbing}, \emph{barrier saddle} (the transition state), \emph{downhill relaxation}, and \emph{target basin} (near $x^\star$). All $100$ SGD runs contribute to every panel.

\section{Ablation Study}
\label{app:ablation_study}

This appendix collects auxiliary parameter sweeps used to assess effects of hyperparameters and the target threshold of PISGD.

\subsection{Concentration-Scale Sweep}
\label{sec:pisgd_concentration_sweep}

\begin{figure}[!htbp]
    \centering
    \includegraphics[width=0.75\textwidth]{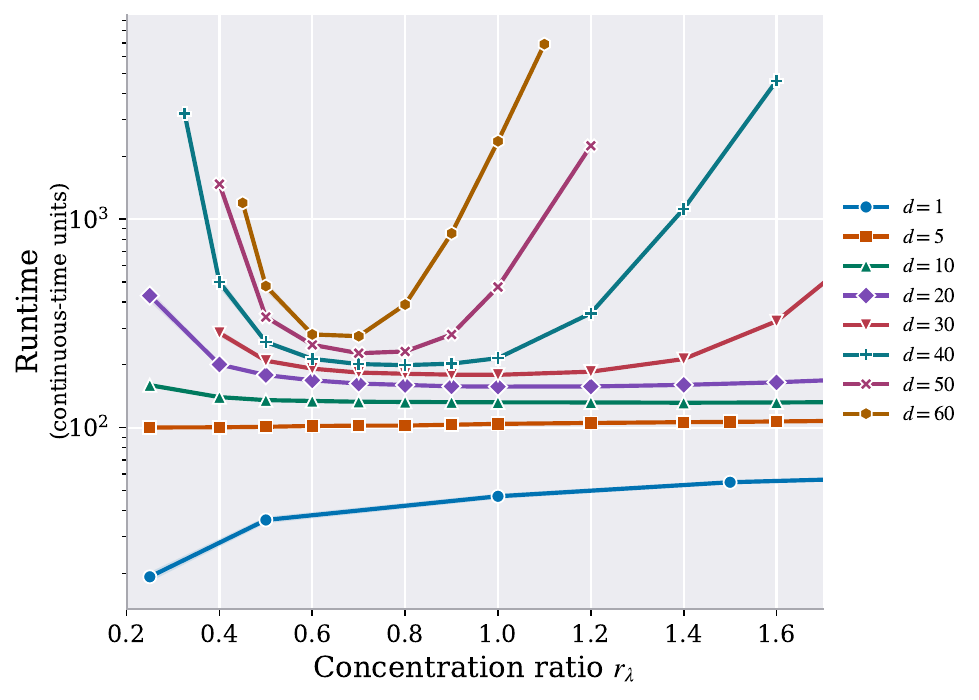}
    \caption{Mean PISGD runtime on the $d$-dimensional separable variant quartic potential of Section~\ref{sec:numerics_variant_quartic_func}, with $\alpha=0.9$, $\bar{\omega}^\star=0.1$, $c=100$, and $\varepsilon=5$. Runtime is in continuous-time units (gradient steps times the learning rate $s=0.01$); colors and marker shapes distinguish dimensions. The horizontal coordinate is the concentration ratio $r_\lambda=\widetilde{\lambda}/\widetilde{\lambda}_{\mathrm{conc},\mathrm{PISGD}}$, where $\widetilde{\lambda}=2/\sigma^2$ for $\lambda=1$ and $\widetilde{\lambda}_{\mathrm{conc},\mathrm{PISGD}}=0.02d$; thus $r_\lambda=1$ denotes the Gaussian concentration reference scale, not an independently tuned critical point. The transition from concentration-limited behavior at smaller $r_\lambda$ to escape-limited behavior at larger $r_\lambda$ becomes more pronounced as $d$ increases. Shaded bands denote the sample mean $\pm$ one estimated standard error, $s_{\mathrm{sample}}/\sqrt{n}$, with $s_{\mathrm{sample}}$ computed using one degree-of-freedom correction. The PISGD hyperparameters are lattice spacing $a=0.025$, number of imaginary-time slices $N_T=400$, and imaginary-time length $T=N_Ta=10$.}
    \label{fig:mean_iterations_to_solution_pisgd_critical_behavior_variant_quartic_alpha_0.9_omega_star_0.1_single-x}
\end{figure}

For PISGD, the concentration requirement in~\eqref{eqn:lambda_requirement_pisgd_separable} gives a reference value that depends on $d$. We denote the value obtained by saturating it by
\begin{equation}
    \widetilde{\lambda}_{\mathrm{conc},\mathrm{PISGD}}
    :=\frac{1}{2\varepsilon^2}
    \sum_{i=0}^{d-1}\frac{1}{\sqrt{V_i''(x_i^\star)}}.
\end{equation}
The subscript $\mathrm{conc}$ emphasizes that this is a concentration reference scale. For identical components with $V_i''(x_i^\star)=(\omega^\star)^2$, it reduces to
\begin{equation}
    \widetilde{\lambda}_{\mathrm{conc},\mathrm{PISGD}}
    =\frac{d}{2\varepsilon^2\omega^\star}.
\end{equation}
In the numerical setting below, $\omega^\star=\sqrt{c}\,\bar{\omega}^\star=1$ and $\varepsilon=5$, so $\widetilde{\lambda}_{\mathrm{conc},\mathrm{PISGD}}=0.02d$. For a compact horizontal-axis label, define the dimensionless concentration ratio
\begin{equation}
    r_\lambda
    :=\frac{\widetilde{\lambda}}
    {\widetilde{\lambda}_{\mathrm{conc},\mathrm{PISGD}}}.
\end{equation}
The results are displayed in Figure~\ref{fig:mean_iterations_to_solution_pisgd_critical_behavior_variant_quartic_alpha_0.9_omega_star_0.1_single-x}.

\subsection{Lattice Spacing \texorpdfstring{$a$}{a}}
\label{app:ablation_lattice_spacing}

The remaining ablations use the one-dimensional benchmark with $\alpha=0.9$,
$\bar{\omega}^\star=0.1$, $c=100$, $\lambda=1$, and
$\widetilde{\lambda}=1.2$ ($\sigma^2=5/3$).  Unless varied, we set
$a=0.025$, $T=10$, $s=0.01$, and $\varepsilon=5$.  Every path starts at
$x^\bullet=0$, and runtime is $t_\varepsilon=s k_\varepsilon$.  All displayed
means contain $100$ finite first hits of the $L^2$-ball $B(x^\star, \varepsilon)$.  All the runs are kept running until reaching the target.
For a scanned parameter $u\in\{a,T,s,\varepsilon\}$, we fit the mean runtime
to $\overline{t}_\varepsilon=C\,u^{p}$, both with $p$ free and with $p$ fixed to its predicted value. The
fixed-$p$ fit gives an empirical $K=aT\,\overline{t}_\varepsilon$, which we
compare with $K_{\mathrm{th}}=145.379$ from~\eqref{eqn:theoretical_prediction_numbers}.

\begin{figure}[!htbp]
    \centering
    \includegraphics[width=0.72\textwidth]{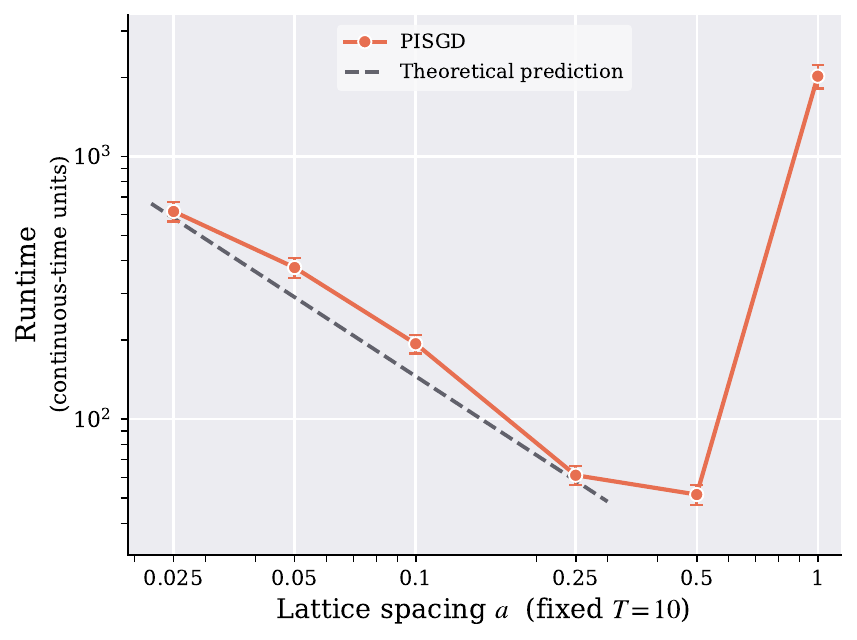}
    \caption{Mean continuous-time PISGD hitting time versus lattice spacing at
    $T=10$; error bars show one standard error over $100$ runs.  The dashed
    line is the theoretical prediction from
    Eq.~\eqref{eqn:pisgd_continuous_time_prefactor},
    $t_\varepsilon=14.5379/a$.}
    \label{fig:pisgd_lattice_spacing_ablation}
\end{figure}

At $T=10$ we scan
\begin{equation*}
    a\in\{1,0.5,0.25,0.1,0.05,0.025\},
    \qquad
    N_T=T/a\in\{10,20,40,100,200,400\}.
\end{equation*}
The results are displayed in Figure~\ref{fig:pisgd_lattice_spacing_ablation}.
We fit $\overline{t}_\varepsilon=C\,a^{p}$, for which
Eq.~\eqref{eqn:pisgd_continuous_time_prefactor} predicts $p=-1$.  Over
$a\leq0.25$, the free fit gives $p=-1.010$, and
fixing $p=-1$ gives $\overline{t}_\varepsilon=17.114/a$, \textit{i.e.}, $K=171.1$,
$17.7\%$ above $K_{\mathrm{th}}$.  The sharp increase at $a=1$ shows that the inverse
law $\propto a^{-1}$ is only valid for a fine lattice.

\subsection{Imaginary Time Length \texorpdfstring{$T$}{T}}
\label{app:ablation_imaginary_time_length}

\begin{figure}[!htbp]
    \centering
    \includegraphics[width=0.72\textwidth]{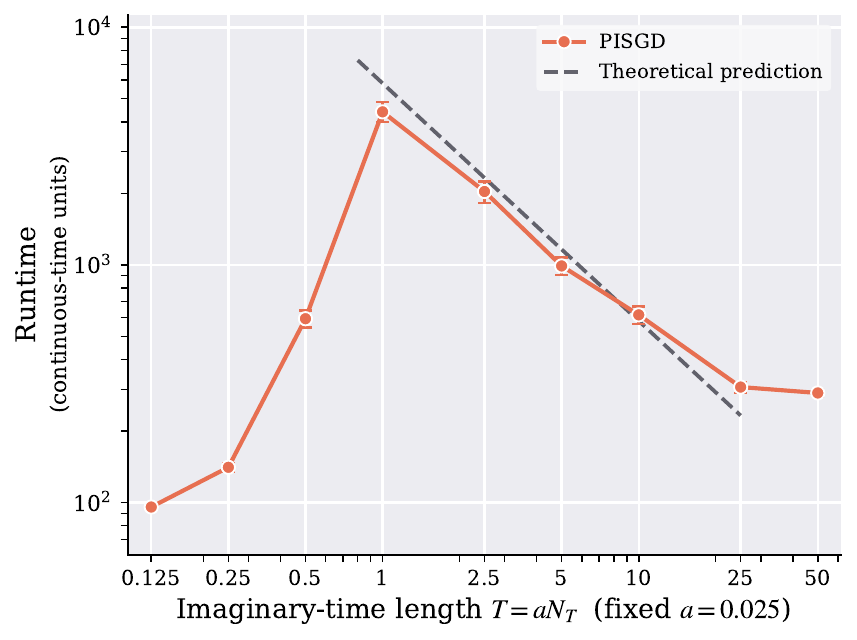}
    \caption{Mean continuous-time PISGD hitting time versus imaginary-time
    length at $a=0.025$. Error bars show one standard error over $100$ runs.
    The dashed line is the theoretical prediction from
    Eq.~\eqref{eqn:pisgd_continuous_time_prefactor},
    $t_\varepsilon=5{,}815.180/T$.}
    \label{fig:pisgd_imaginary_time_ablation}
\end{figure}

At $a=0.025$ we scan
\begin{equation*}
    T\in\{0.125,0.25,0.5,1,2.5,5,10,25,50\},
    \qquad
    N_T=T/a\in\{5,10,20,40,100,200,400,1000,2000\}.
\end{equation*}
The results are displayed in Figure~\ref{fig:pisgd_imaginary_time_ablation}.
We fit $\overline{t}_\varepsilon=C\,T^{p}$, for which
Eq.~\eqref{eqn:pisgd_continuous_time_prefactor} predicts $p=-1$.  Over
$1\leq T\leq10$, the free fit gives $p=-0.872$, and
fixing $p=-1$ gives $\overline{t}_\varepsilon=5{,}118/T$, \textit{i.e.}, $K=128.0$,
$12.0\%$ below $K_{\mathrm{th}}$.  The flattening at $T=25$ and $50$ shows
that the inverse law $\propto T^{-1}$ holds only for moderate $T$.
For $T\leq0.5$, the runtime instead drops well below its value at $T=10$.
This regime does not probe the tunneling mechanism of PISGD. When $T$ is
not much larger than the bounce, the path cannot accommodate a bounce, and PISGD reduces to SGD-like dynamics on $V$.  Its fast escape therefore may not give an advantage asymptotically.

\subsection{Learning Rate \texorpdfstring{$s$}{s}}
\label{app:ablation_learning_rate}

\begin{figure}[!htbp]
    \centering
    \includegraphics[width=0.72\textwidth]{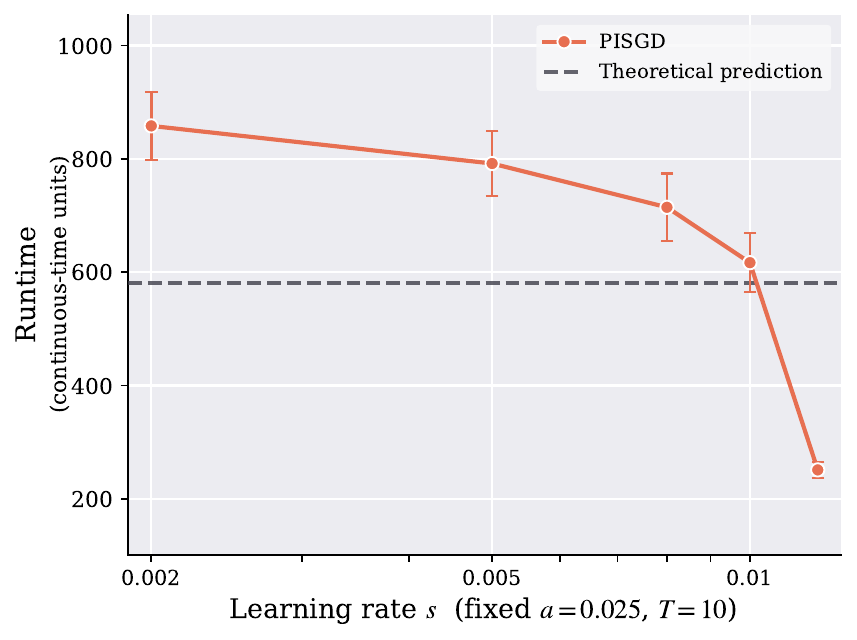}
    \caption{Mean continuous-time PISGD hitting time versus learning rate at
    $a=0.025$ and $T=10$. Error bars show one standard error over $100$ runs.
    The dashed line is the theoretical prediction from
    Eq.~\eqref{eqn:pisgd_continuous_time_prefactor},
    $t_\varepsilon=581.518$.}
    \label{fig:pisgd_learning_rate_ablation}
\end{figure}

At the central setting $a=0.025$, $T=10$, and $N_T=400$, we scan
\begin{equation*}
    s\in\{0.002,0.005,0.008,0.010,0.012\}.
\end{equation*}
$s=0.012$ is excluded from the fit.
The results are displayed in Figure~\ref{fig:pisgd_learning_rate_ablation}.
We fit $\overline{t}_\varepsilon=C\,s^{p}$, for which
Eq.~\eqref{eqn:pisgd_continuous_time_prefactor} predicts $p=0$.  Over
$s\leq0.01$, the free fit gives $p=-0.182$, and
fixing $p=0$ gives $\overline{t}_\varepsilon=739.7$, \textit{i.e.}, $K=184.9$,
$27.2\%$ above $K_{\mathrm{th}}$.  
A continuous-time plateau starts to appear at small $s$.

\subsection{Target Radius \texorpdfstring{$\varepsilon$}{epsilon}}
\label{app:ablation_target_radius}

\begin{figure}[!htbp]
    \centering
    \includegraphics[width=0.72\textwidth]{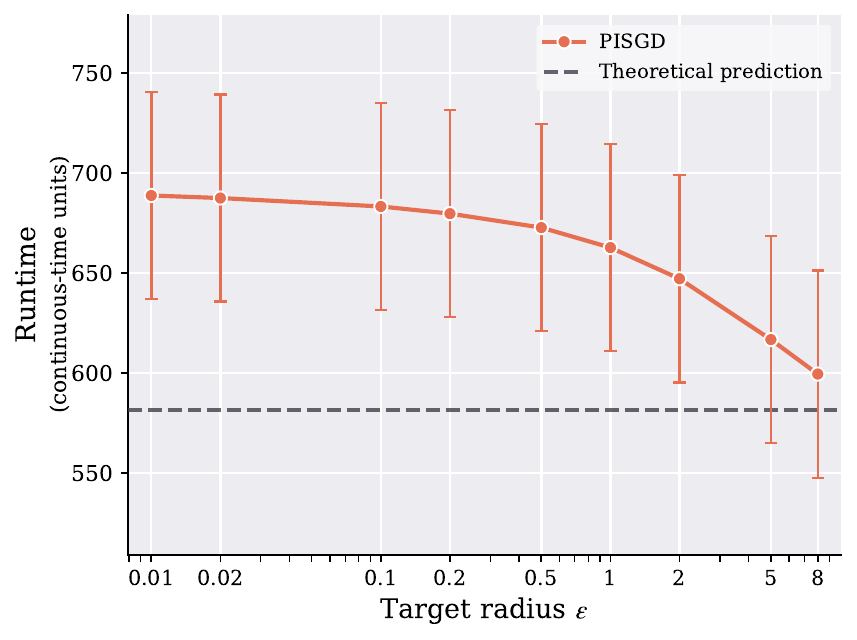}
    \caption{Mean continuous-time PISGD hitting time versus target radius at
    $a=0.025$, $T=10$, and $s=0.01$. Error bars show one standard error over
    $100$ runs.  The dashed line is the leading theoretical prediction from
    Eq.~\eqref{eqn:pisgd_continuous_time_prefactor},
    $t_\varepsilon=581.518$.}
    \label{fig:pisgd_target_radius_ablation}
\end{figure}

At the central setting we scan
\begin{equation*}
    \varepsilon\in\{0.01,0.02,0.1,0.2,0.5,1,2,5,8\}.
\end{equation*}
The concentration reference is
\begin{equation*}
    \widetilde{\lambda}_{\mathrm{conc},\mathrm{PISGD}}(\varepsilon)
    =\frac{1}{2\varepsilon^2},
\end{equation*}
so our central-setting value $\widetilde{\lambda}=1.2$ meets the concentration condition only for $\varepsilon\geq 1/\sqrt{2\widetilde{\lambda}}=1/\sqrt{2.4}\approx0.645$.
The results are displayed in Figure~\ref{fig:pisgd_target_radius_ablation}.
We fit $\overline{t}_\varepsilon=C\,\varepsilon^{p}$, for which
Eq.~\eqref{eqn:pisgd_continuous_time_prefactor} predicts $p=0$, since the
leading escape time does not depend on $\varepsilon$.  Over
$\varepsilon\in\{1,2,5,8\}$, the free fit gives $p=-0.049$, and
fixing $p=0$ gives $\overline{t}_\varepsilon=631.1$, \textit{i.e.},\ $K=157.8$,
$8.5\%$ above $K_{\mathrm{th}}$.  The increase for smaller $\varepsilon$
reflects the additional time needed to reach a smaller ball after escape.

\printbibliography

\end{document}